\documentclass[preview,10pt]{elsarticle}
\usepackage{amsmath,amsfonts,amssymb}
\usepackage[mathlines]{lineno} 
\usepackage{hyperref}
\usepackage{bm}
\usepackage{algorithm}
\usepackage{algorithmicx}
\usepackage[noend]{algpseudocode}
\let\Algorithm\algorithm
\usepackage{caption}
\usepackage{hyperref}
\usepackage[utf8]{inputenc}
\usepackage{lscape}
\usepackage{tikz}
\usepackage{float}
\usepackage{graphicx}
\usepackage{listings}%
\usepackage{mathtools}
\usepackage{wrapfig}
\usepackage{alltt}%
\usepackage{siunitx}
\usepackage{setspace}
\usepackage{eqparbox}
\usepackage{booktabs}
\usepackage{chemformula}
\usepackage[margin=2cm]{geometry} 

\renewcommand\algorithm[1][]{\Algorithm[#1]\setstretch{1.4}}

\usepackage[labelformat=simple]{subcaption} 

\DeclareCaptionLabelFormat{subcaptionlabel}{\normalfont(\textbf{#2}\normalfont)}
\newcommand{\half}{\scriptstyle\frac{1}{2}}

\newcommand{\wdiff}{{\bf{w}}_{\scriptscriptstyle \nabla^2}}
\newcommand{\fdiff}{\bm{f}^{\scriptscriptstyle \nabla^2}}

\usepackage{xcolor}
\definecolor{pyblue}{HTML}{1F77B4}
\definecolor{pyorange}{HTML}{FF7F0E}
\definecolor{pygreen}{HTML}{2CA02C}
\definecolor{pyred}{HTML}{D62728}

\journal{arXiv}
\biboptions{numbers,sort&compress}
\newcommand{\eg}{{\it e.g., }}
\newcommand{\ie}{{\it i.e., }}

\begin{document}
\begin{frontmatter}

\title{Solving the Discretised Multiphase Flow Equations with Interface Capturing on Structured Grids Using Machine Learning Libraries} 

\author[AMCG]{Boyang Chen}
\author[AMCG,IX]{Claire E. Heaney}
\author[UoA]{Jefferson L.M.A. Gomes\corref{correspond}}
\ead{jefferson.gomes@abdn.ac.uk}
\author[IX,Chem]{Omar K. Matar}
\author[AMCG,IX,DAL]{Christopher C. Pain}

\address[AMCG]{{Applied Modelling and Computation Group, Department of Earth Science and Engineering}, {Imperial College London},
{{London}, {SW7 2AZ} {UK}}}
\address[IX]{{Centre for AI-Physics Modelling, Imperial-X, White City Campus}, {Imperial College London},
{{London}, {W12 7SL} {UK}}}
\address[UoA]{{Fluid Mechanics Research Group, School of Engineering}, {University of Aberdeen},
{{Aberdeen}, {AB24 3FX} {UK}}}
\address[Chem]{{Department of Chemical Engineering}, {Imperial College London},
{{London}, {SW7 2AZ} {UK}}}
\address[DAL]{{Data Assimilation Laboratory, Data Science Institute}, {Imperial College London},
{{London}, {SW7 2AZ} {UK}}}

\cortext[correspond]{Corresponding author}


\begin{abstract}
This paper solves the discretised multiphase flow equations using tools and methods from machine-learning libraries. The idea comes from the observation that convolutional layers can be used to express a discretisation as a neural network whose weights are determined by the numerical method, rather than by training, and hence, we refer to this approach as Neural Networks for PDEs (NN4PDEs). To solve the discretised multiphase flow equations, a multigrid solver is implemented through a convolutional neural network with a U-Net architecture. Immiscible two-phase flow is modelled by the 3D incompressible Navier-Stokes equations with surface tension and advection of a volume fraction field, which describes the interface between the fluids. A new compressive algebraic volume-of-fluids method is introduced, based on a residual formulation using Petrov-Galerkin for accuracy and designed with NN4PDEs in mind. High-order finite-element based schemes are chosen to model a collapsing water column and a rising bubble. Results compare well with experimental data and other numerical results from the literature, demonstrating that, for the first time, finite element discretisations of multiphase flows can be solved using an approach based on (untrained) convolutional neural networks. A benefit of expressing numerical discretisations as neural networks is that the code can run, without modification, on CPUs, GPUs or the latest accelerators designed especially to run AI codes. 
\end{abstract}


\end{frontmatter}



\section{Introduction}
\subsection{Motivation}
Detailed numerical flow simulations provide significant insight into a wide range of sectors, including the environment~\citep{Dauxois2021}, energy~\citep{Hewitt} and food~\citep{Wilson2023} sectors. Over the past two decades, computational fluid dynamics (CFD) models, including multiphase flow models, have become critical for predicting environmental~\citep{atmos13081203} and industrial~\citep{XIANG2022105179} flow behaviour; conducting safety~\citep{VENKATESHWARAN2023104555} and environmental~\citep{WOODWARD2023163711} assessments; studying the effect of cloud formation on the climate~\citep{Grabowski2013}; analysing slugging in fluidised bed reactors~\citep{Ramirez2017} and pipelines~\citep{Osundare2022}; modelling food processing to reduce waste~\citep{Khan2018}; as well as being used to investigate fundamental multiphase turbulence~\citep{Ling_2018}. Our aim is to develop a method that can be used to simulate complex environmental and industrial processes involving immiscible multiphase flows and that has the potential to run on the latest GPUs and AI processors. The finite element discretisations and Petrov-Galerkin terms are all written as a neural network and solved with a neural network, giving the same result as if these methods had been programmed in a traditional way. We benchmark the proposed method with a collapsing water column problem and a rising bubble problem, running the code on a GPU.

\subsection{Background}
The so-called one-fluid approach has been widely used to simulate the flow of two or more immiscible fluids or phases~\citep{Prosperetti_chap3}. In this approach, a single model describes the flow throughout the domain and an interface marks the boundary between the different fluids or phases, across which, surface tension forces can be applied. Schemes for modelling the interface can be broadly classified as interface tracking or interface capturing. Although interface-tracking methods maintain a sharp interface and enable physical processes to be incorporated at the interface, these methods suffer from mass conservation issues, and struggle to model break-up and coalescence~\citep{Reddy2015,Mirjalili2017,Eikelder2021,Crialesi-Esposito2023}. Combining these methods with level-set methods has resulted in an improvement in their ability to conserve mass~\citep{Shin2017,Shin2018}. Interface-capturing methods treat interfaces as discontinuities in the material properties which are advected by a volume fraction field. In sharp-interface approaches, such as Volume-of-Fluid (VoF) methods~\citep{Rider1998,Scardovelli1999} or enriched element methods (including XFEM~\citep{Chessa2003} and Cut-Cell methods~\citep{Claus2019,Xie2020}), values of density and viscosity exhibit a discontinuity or jump at the interface, whereas for diffuse-interface approaches, such as the phase field method~\citep{Anderson1998}, values of density and viscosity are functions of the volume fraction field. VoF methods have good mass conservation properties, are able to maintain a sharp interface~\citep{Reddy2015,Montazeri2017} (provided there is sufficient resolution), and can handle coalescence and break-up~\citep{Estrem2020}. Phase-field methods diffuse the interface over a number of cells~\citep{Nochetto2016}, which improves stability but loses the sharpness of the interface. Level set methods can be considered as sharp~\citep{Mirjalili2017} or diffuse~\citep{Claus2019}, and, although the volume fraction field is smooth, conservation of mass can be problematic~\citep{Reddy2015,Eikelder2021}. For more details on many of the available approaches to modelling interfaces, see~\citep{Prosperetti,Elgeti2016,Mirjalili2017}, and for some recent work on the combination of interface-tracking and interface-capturing methods, see~\citep{Tryggvason}. In this paper, our goal is to model the break-up of immiscible two-phase flows, and we choose a VoF type approach for its ability in this regard. Following \citet{Pavlidis2014}, \citet{Pavlidis2016} and \citet{Obeysekara2021}, additional diffusion terms are included to keep the interface between the two fluids sharp: positive diffusion is added if oscillations are detected in the solution and negative diffusion is added if the solution is smooth. However, in this paper, oscillations are detected in a different manner to the methods presented previously~\citep{Pavlidis2014,Pavlidis2016,Obeysekara2021} and the amount of diffusion is calculated more conservatively to increase accuracy. 

The accuracy of CFD models strongly relies on the resolution of the mesh or grid, which becomes particularly challenging for multiphase flow problems due to the enormous range of length scales within the governing mechanisms~\citep{Dauxois2021,Wilson2023,Crialesi-Esposito2023}. As a consequence, computational costs are extremely high when simulating these processes. The use of high performance computing (HPC) resources can be beneficial, although, to obtain good performance, an efficient implementation of models, methods and algorithms is needed~\citep{Banchelli_2020}. Originally, HPC clusters consisted predominantly of central processing units (CPUs), however, Graphics Processing Units (GPUs), containing hundreds to thousands of processing cores, are now also available in such clusters~\citep{Niemeyer2014,Afzal2017}. In order to make use of GPUs, however, existing CFD models either need to be re-implemented in terms of graphical primitives, using specialist programming platforms such as CUDA, OpenCL and OpenACC~\citep{Memeti_2017,Araujo_2023}, or require implementation of communication protocols between program units across computational architectures~\citep{Lai_2020b,Zhu_2023}.   

\citet{Appleyard_2011} presented an early comparison between the performance of a single CPU and a single GPU when solving the level set equations describing an interface, and reported a difference of two orders of magnitude in favour of the GPU implementation.  Following this, a number of researchers developed hybrid CPU/GPU implementations of multiphase flow solvers~\citep{Codyer2012,Griebel,Reddy2015}. In all three cases, the computationally expensive pressure equation was solved on the GPUs and the other equations were solved on CPUs. For a variety of problems with up to 27~million grid points, speed-ups of 50~to 100 were obtained. 
\citet{Bryngelson2021} introduced the MFC code, a parallel multi-component, multiphase and multiscale flow solver that was extended by \citet{Radhakrishnan2023} to run on GPU systems. The solver uses high-order interface-capturing with diffuse-interface models to describe the interface between fluids. They ran their code on up to 576~GPUs to solve shock-induced collapse of air bubbles in the vicinity of a kidney stone with half a million grid points and on 960~GPUs to solve atomising droplets with 2~billion ($10^9$) grid points. \citet{Crialesi-Esposito2023} presented FluTAS, a solver for incompressible multiphase flows with heat transfer that can be run on multi-CPU and multi-GPU architectures. A direct method based on the Fast Fourier Transform is used to solve the pressure equation and the interface is represented by a volume-of-fluid method. Demonstrating their code on two-layer Rayleigh-B{\'e}nard convection and emulsions in homogeneous isotropic turbulence using up to 1~billion ($10^9$) grid points with up to 128~GPUs, they identify a communication burden which increased with the number of GPUs.

Not only is there significant work to be done in getting a code to run on a GPU system relating to re-implementing algorithms with graphical primitives, in addition, many of the afore-mentioned papers fine-tune their codes to obtain good performance, by assigning particular tasks to CPUs and others to GPUs. The level of computational expertise required can pose a barrier for many, however, a recent AI-based approach offers a way of circumventing this issue. Instead of using AI technologies to \emph{approximate} CFD models, as is often done, this recent approach solves discretised fluid dynamics systems \emph{exactly} (to within solver tolerances) using the functionality found within ML libraries~\citep{Zhao2020,Wang2022,Chen2023,Phillips2023,Phillips2022-progress}. This is because numerical discretisations can be expressed as discrete convolutions between a solution field and a stencil, the weights of which are determined explicitly by the numerical discretisation~\citep{Dong2017,Long2018}. Exactly the same operation is performed in the convolutional layer of a convolutional neural network (CNN)~\citep{Yamashita2018,Indolia_2018,tds_cnn} by a filter when it is convolved with an image. The weights of the filters of CNNs are usually determined by training, however, in this case, no training is needed as the weights are determined by the numerical discretisation. Hence, numerical discretisations can be expressed as CNNs using functionality from machine learning libraries. To solve the resulting systems, Jacobi methods can be implemented directly or multigrid methods can be implemented through a special neural network architecture known as a U-Net~\citep{Ronneberger2015}. Previous examples of solving discretised governing equations using machine learning techniques include \citet{Zhao2020}, who solved a finite difference discretisation of the Navier-Stokes equations using a neural network with pre-defined weights adopting an explicit solver for the pressure equation based on preconditioned conjugate gradients. The solutions of the neural network implementation were in good agreement with CFD benchmarks. \citet{Wang2022} developed a similar approach, and obtained linear weak scaling and a superlinear strong scaling for 2048~cores on TPUs (Google's accelerated tensor processing units) for CFD problems with up to a~billion ($10^9$) computational cells. \citet{Woo2023} developed a particle-in-cell multiphase solver to model particles immersed in a gas, porting parts of the MFiX code (Multiphase Flow with Interphase eXchanges) to TensorFlow. Some operations, such as particle initialisation, were handled by CPUs through Fortran code, but the computationally demanding calculations were handled by GPUs through python code and TensorFlow functions. \citet{Chen2023} implemented both finite difference and finite element discretisations through convolutional neural networks with a sawtooth multigrid method implemented within a U-Net architecture, to solve benchmark single-phase CFD problems. This method was extended to solve the neutron diffusion equations~\citep{Phillips2023} and radiation transport problems~\citep{Phillips2022-progress}. In the latter, a new convolutional finite element method (ConvFEM) was proposed, which was designed to simplify the discrete convolutions of higher-order Finite Element Methods (FEMs), so that they can be easily represented by convolutional layers in neural networks. Although applied to unstructured grids in this paper, by using graph neural networks, the NN4PDEs approach has been extended to unstructured meshes, where \citet{Li2024} apply the method to a Discontnuous Galerkin formulation for solving diffusion 
equations. 

\subsection{Contribution}
In this paper, we start from the NN4PDEs approach proposed by \citet{Chen2023}, which includes a single-phase CFD solver written as a neural network, and extend it to model immiscible two-phase flows. We solve the 3D incompressible Navier-Stokes equations with surface tension on structured grids and develop a type of compressive algebraic VOF method to capture the interface. A segregated pressure and velocity time-stepping method is used with explicit time-stepping schemes for volume fraction and velocity, and an implicit pressure update is used to enforce the incompressibility constraint. The latter is solved using a multigrid method with Jacobi smoothing through a U-Net architecture, as described in~\citet{Phillips2023}. Using convolutional neural networks, we solve the discretised system of equations and obtain the same answer (to within solver tolerance) as if the numerical methods had been coded in Fortran or C\verb!++!. The first advantage of this AI-based framework is that the code can be run on CPU- or GPU-based systems without having to make any changes or re-implement code using CUDA, OpenCL or OpenACC. The writers of ML libraries such as TensorFlow~\citep{tensorflow2015}, PyTorch~\citep{pytorch} and JAX~\citep{jax2018} have abstracted away platform-related code, meaning that it is very simple for the user to run their code on CPUs or GPUs. Furthermore, as the code uses functionality found in machine learning libraries, it can also be deployed on the latest energy-efficient AI accelerators such as those developed by GraphCore~\citep{GraphCore_IPU} and Cerebras~\citep{cerebras}, which have the potential for even greater computational speeds than GPUs. The \verb!C++! library Kokkos~\citep{Edwards2014} operates in a similar vein to ML libraries, abstracting away code relating to the low-level memory access needs for different GPU architectures, enabling users to run their \verb!C++! codes on CPUs and GPUs without writing in CUDA, OpenCL or OpenACC themselves. Kokkos has been used to simulate boiling in two-phase flow~\citep{Verdier2020} and two-phase-flow in a fluidised bed~\citep{Chattopadhyay2016}. A second advantage of expressing the discretised governing equations as neural networks is the potential for seamless integration with machine learning workflows such as surrogate models or digital twins. For example, having a CFD-based solver written as a neural network will streamline physics-informed approaches, which minimise the residual of the governing equations whilst training a neural network~\citep{Raissi_2019,Buhendwa2021}. In addition, the backpropagation functionality of machine learning libraries will allow the CFD model to be differentiated~\citep{Merrienboer_2018,Baydin_2018}, facilitating data assimilation, control and inversion. The NN4PDEs approach has already been successfully applied to discover subsurface geology by inverting measurements of electrical resistivity~\citep{AI4ERT}.

The contribution of this paper is to extend, to multiphase flow problems, the NN4PDEs approach of solving discretised PDEs by capitalising on the functionality found within machine-learning libraries. The numerical schemes used should all be suitable for implementation in the AI-based framework proposed here, so a compressive algebraic VoF method is developed which maintains a sharp interface between the two fluids. The remaining text is divided as follows: the methodology is described in Section~\ref{methods}; results for a series of collapsing water column problems and rising bubble problems are presented in Section~\ref{results}; and conclusions are drawn in Section~\ref{conclusion}.  

\section{Methodology}\label{methods}
In Section~\ref{sec:gov_eq}, we outline the governing equations: the advection equation for the interface between the fluids, and the equations conserving mass and momentum. The idea underlying the NN4PDEs approach is introduced in Section~\ref{sec:discretisation_AI4PDEs} and demonstrated for some simple cases. Here, we also explain the notation used to express discretisations through filters of convolutional layers which is fundamental to the NN4PDEs approach. In Section~\ref{sec:disc_gov_eq}, we describe the application of NN4PDEs to the equations governing multiphase flow, and give an outline of the interface capturing scheme that detects oscillatory behaviour and the non-linear Petrov-Galerkin method that attempts to diffuse away oscillations. The final solution algorithm is presented in Section~\ref{sec:solution_algorithm}. 

\subsection{Governing equations}\label{sec:gov_eq}
\subsubsection{Advection equation for the volume fraction field}\label{AD} 
The advection equation that governs the scalar volume fraction field~$C$ (or indicator field) is given by:
\begin{equation}
    \frac{\partial C}{\partial t} + \bm{q} \cdot \nabla C  
     =  0\,, \label{adv-diff-eqn} 
\end{equation}
where $\bm{q}=(u,v,w)^T$ represents the advection velocity. Here, two-phase flow is modelled, where a value of $C=1$ indicates pure liquid and a value of $C=0$ indicates a pure gas. The overall density, $\rho$, is: 
\begin{equation} \label{rho} 
\rho= C \rho_l + (1-C) \rho_g\,,
\end{equation}
with densities of the liquid and gas denoted by $\rho_l$ and $\rho_g$ respectively.

\subsubsection{Navier-Stokes equations}\label{NS}
The incompressible Navier-Stokes equations can be written as 
\begin{subequations}
\begin{eqnarray}
{\rho}\left( \frac{\partial \bm{q}}{\partial t} + \bm{q}\cdot \nabla \bm{q}\right) + 
\sigma \bm{q} - \nabla\cdot(\mu\nabla \bm{q}) &=& -\nabla p + \bm{s}_q +\bm{s}_t \,,
\label{mom-eqn} \\
\nabla  \cdot \bm{q} &=& {0} \,,
\label{NS-eqn-cty}
\end{eqnarray}
\end{subequations}
in which $p$ denotes the pressure, $\sigma$ is an absorption coefficient and $\mu$ is the dynamic viscosity. Buoyancy is represented by the term $\bm{s}_q= - \rho g \bm{e_z}$ where $g$ is the acceleration due to gravity and $\bm{e_z}$ is a unit vector in the upwards vertical direction. The effects of surface tension are represented through a source term, $\bm{s}_t$, given by 
\begin{equation} \label{surface-tension-eqn}
    \bm{s}_{t}=\sigma_t \kappa \nabla C \,,
\end{equation}
where $\sigma_t$ is the surface tension coefficient and the curvature is represented by~$\kappa$. The curvature is a continuous volumetric field which represents the effective normal to the liquid-gas interface and is defined as follows:
\begin{equation}
\kappa = \nabla \cdot \bm{n}_{t} \quad \text{where} \quad 
\bm{n}_{t}= \frac{\nabla C }{  \vert\vert\nabla C\vert\vert_2 }  \,.
\end{equation}
This formulation is sometimes referred to as continuous surface force~\citep{Mirjalili2017}. A consequence of Equations~\eqref{adv-diff-eqn},  \eqref{rho} and~\eqref{NS-eqn-cty}, is that the mass conservation equation holds:
\begin{equation}
\frac{\partial \rho}{\partial t} + \nabla\cdot (\rho \bm{q} )=0\,. 
\label{rho-mass} 
\end{equation}

Dirichlet or Neumann boundary conditions are applied as needed, depending on the example being investigated.

\subsection{Discretisation using convolutional layers} \label{sec:discretisation_AI4PDEs}
The NN4PDEs approach~\citep{Chen2023} is based on the idea that the matrix-vector multiplications associated with numerical discretisations can be written as discrete convolutions. Such operations are commonly used in the convolutional layers of neural networks, thus many numerical discretisations can be expressed using functions within machine learning libraries. See~\citep{Yamashita2018,Indolia_2018,tds_cnn} for more details about CNNs.

\subsubsection{Equivalence between numerical discretisations and discrete convolutions}
\begin{figure}[htbp]
\centering\scalebox{0.9}{\begin{tikzpicture}[scale=0.92]

\draw[gray, ultra thick] (2.5,-0.5) -- (6.5,-0.5) -- (6.5,6.5) -- (-0.5,6.5) -- (-0.5,-0.5) -- (2.5,-0.5); 
\foreach \y in {0,...,6}
    \node[] at (0,\y) {$C_{0\y}$};
\foreach \y in {0,...,6}    
    \node[] at (6,\y) {$C_{6\y}$}; 
\foreach \x in {1,...,5}
    \node[] at (\x,0) {$C_{\x 0}$};
\foreach \x in {1,...,5}    
    \node[] at (\x,6) {$C_{\x 6}$};     

\draw[gray, thick] (0.5,-0.5) -- (0.5,6.5);
\draw[gray, thick] (1.5,-0.5) -- (1.5,6.5);
\draw[gray, thick] (2.5,-0.5) -- (2.5,6.5);
\draw[gray, thick] (3.5,-0.5) -- (3.5,6.5);
\draw[gray, thick] (4.5,-0.5) -- (4.5,6.5);
\draw[gray, thick] (5.5,-0.5) -- (5.5,6.5);
\draw[gray, thick] (-0.5,5.5) -- (6.5,5.5);
\draw[gray, thick] (-0.5,4.5) -- (6.5,4.5);
\draw[gray, thick] (-0.5,3.5) -- (6.5,3.5);
\draw[gray, thick] (-0.5,2.5) -- (6.5,2.5);
\draw[gray, thick] (-0.5,1.5) -- (6.5,1.5);
\draw[gray, thick] (-0.5,0.5) -- (6.5,0.5);

\draw[pyorange, ultra thick, fill=pyorange!40!, fill opacity=0.2] (2.5,0.5) -- (5.5,0.5) -- (5.5,5.5) -- (0.5,5.5) -- (0.5,0.5) -- (2.5,0.5); 
\draw[pyorange, thick] (1.5,0.5) -- (1.5,5.5);
\draw[pyorange, thick] (2.5,0.5) -- (2.5,5.5);
\draw[pyorange, thick] (3.5,0.5) -- (3.5,5.5);
\draw[pyorange, thick] (4.5,0.5) -- (4.5,5.5);
\draw[pyorange, thick] (0.5,4.5) -- (5.5,4.5);
\draw[pyorange, thick] (0.5,3.5) -- (5.5,3.5);
\draw[pyorange, thick] (0.5,2.5) -- (5.5,2.5);
\draw[pyorange, thick] (0.5,1.5) -- (5.5,1.5);

\foreach \y in {1,...,5}
\foreach \x in {1,...,5}
        \node[font=\large] at (\x,\y) {$C_{\x\y}$}; 

\draw[pyblue, ultra thick, fill=pyblue!50!, fill opacity=0.2] (9.5,2.5) -- (11,2.5) -- (11,5.5) -- (8,5.5) -- (8,2.5) -- (9.5,2.5); 
\draw[pyblue, thick] (9,2.5) -- (9,5.5);
\draw[pyblue, thick] (10,2.5) -- (10,5.5);
\draw[pyblue, thick] (8,3.5) -- (11,3.5);
\draw[pyblue, thick] (8,4.5) -- (11,4.5);

\node[] at (8.5,3) {0}; 
\node[] at (8.5,4) {-1}; 
\node[] at (8.5,5) {0}; 
\node[] at (9.5,3) {-1}; 
\node[] at (9.5,4) {4}; 
\node[] at (9.5,5) {-1}; 
\node[] at (10.5,3) {0}; 
\node[] at (10.5,4) {-1}; 
\node[] at (10.5,5) {0}; 

\draw[pyblue, ultra thick] (2.5,1.6) -- (4.4,1.6) -- (4.4,4.4) -- (1.6,4.4) -- (1.6,1.6) -- (2.5,1.6); 
\draw[pyblue, ultra thick] (4.4,1.6) -- (11,2.5); 
\draw[pyblue, ultra thick] (1.6,4.4) -- (8,5.5);

\node[font=\huge] at (7.25,3.75) {*};

\draw[pyorange, ultra thick, fill=pyorange!40!, fill opacity=0.2]  (14,0.5) -- (17.5,0.5) -- (17.5,5.5) -- (12.5,5.5) -- (12.5,0.5) -- (14,0.5);  

\draw[pyorange, thick] (13.5,0.5) -- (13.5,5.5);
\draw[pyorange, thick] (14.5,0.5) -- (14.5,5.5);
\draw[pyorange, thick] (15.5,0.5) -- (15.5,5.5);
\draw[pyorange, thick] (16.5,0.5) -- (16.5,5.5);
\draw[pyorange, thick] (12.5,4.5) -- (17.5,4.5);
\draw[pyorange, thick] (12.5,3.5) -- (17.5,3.5);
\draw[pyorange, thick] (12.5,2.5) -- (17.5,2.5);
\draw[pyorange, thick] (12.5,1.5) -- (17.5,1.5);

\draw[pyorange, ultra thick, fill = pyorange!60!, fill opacity=0.2]  (15,2.5) -- (14.5,2.5) -- (14.5,3.5) -- (15.5,3.5) -- (15.5,2.5) -- (15,2.5); 
\node[font=\large] at (15,3) {$Y_{33}$}; 

\node[font=\small, left] at (17.5,-0.25) {$Y_{33}=4C_{33} - (C_{32}+C_{34}+C_{23}+C_{43})$};


\draw[gray, thick, rounded corners, fill=gray!50!, fill opacity=0.1] (9,-2.25) -- (17.5,-2.25) -- (17.5,-1.25) -- (-0.5,-1.25) -- (-0.5,-2.25) -- (9,-2.25) ;
\node[color=gray!65!black] at (3,-1.75) {\textsf{Pixel values of a 2D image}};
\node[color=gray!65!black] at (9.5,-1.75) {\textsf{Filter}};
\node[color=gray!65!black] at (15,-1.75) {\textsf{Feature map}};

\draw[gray, thick, rounded corners, fill=gray!50!, fill opacity=0.1] (9,7.25) -- (17.5,7.25) -- (17.5,8.25) -- (-0.5,8.25) -- (-0.5,7.25) -- (9,7.25) ;
\node[color=gray!65!black] at (3,7.75) {\textsf{Solution field on a grid}};
\node[color=gray!65!black] at (9.5,7.75) {\textsf{Central difference stencil}};
\node[color=gray!65!black] at (15,7.75) {\textsf{Diffusion discretisation}};

\end{tikzpicture}}
\caption{On the left (in orange) is a 5~by~5 image with pixel values $C_{ij}$ ($i,\,j\in\{1,\,2,\,\ldots,\,5\}$). A filter (in blue) of a convolutional layer, with weights as shown, is applied to the 3~by~3 part of the image centred around pixel $C_{33}$. The sum of the products of the weights and the pixel values determines the central value of the resulting feature map~$Y_{33}$ (in orange). Equivalently, on the left (in orange) is a 5~by~5 solution field $C_{ij}$ ($i,\,j\in\{1,\,2,\,\ldots,\,5\}$) to which is applied the stencil of the diffusion operator discretised by finite differences (in blue). This results in the value $Y_{33}$ (in orange). (The finite difference nodes are located at the centre of the cells.) Forming a halo around the inner domain are the white cells or ghost cells, through which boundary conditions can be applied.}
\label{fig:conv_filter_fd_equivalence}
\end{figure}
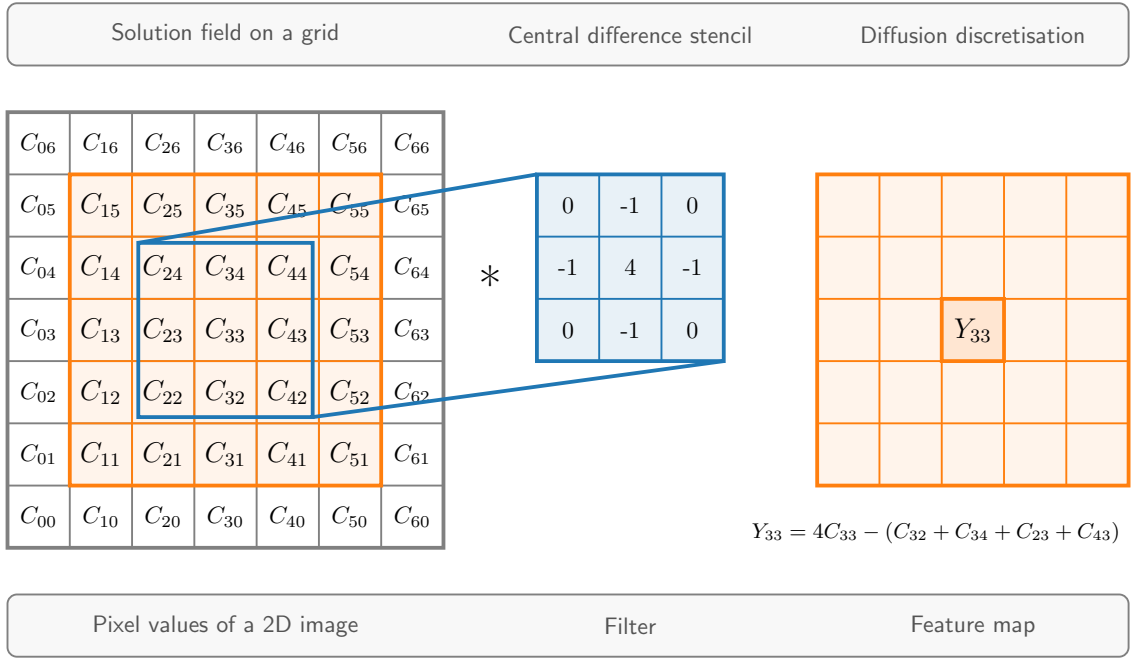

Although finite element discretisations are used in this paper, finite difference discretisation schemes have a simpler mathematical form, so an example of writing finite difference discretisations as discrete convolutions is presented here. For the discretised isotropic diffusion operator ($\nabla^2$) in 3D, a second-order finite difference scheme has the following 
representation as a $3\times3\times3$ convolutional filter: 
\begin{eqnarray}
{\wdiff}_{\cdot,\cdot,-1} = 
 \begin{bmatrix*}[c]
 0  & 0 & 0 \\
0 & \frac{-1}{(\Delta z)^2} & 0 \\
 0& 0 & 0
 \end{bmatrix*}, \ 
{\wdiff}_{\cdot,\cdot,0} =
 \begin{bmatrix*}[c]
 0
 & \frac{-1}{(\Delta y)^2} & 0 \\
\frac{-1}{(\Delta x)^2} & \frac{2}{(\Delta x)^2} +\frac{2}{(\Delta y)^2} + \frac{2}{(\Delta z)^2}& \frac{-1}{(\Delta x)^2} \\
 0 & \frac{-1}{(\Delta y)^2} & 0
 \end{bmatrix*}, 
{\wdiff}_{\cdot,\cdot,1} =
 \begin{bmatrix*}[c]
 0  & 0 & 0 \\
0 & \frac{-1}{(\Delta z)^2} & 0 \\
 0& 0 & 0
 \end{bmatrix*}\,,
\end{eqnarray}
where $\Delta x$, $\Delta y$ and $\Delta z$ denote the grid spacing between the nodes in the $x$, $y$ and $z$ directions. The rank~3 tensor is shown here in three `slices', labelled as~-1, 0 and~1. For more details on finite difference discretisations, the reader is referred to \citet{Fletcher,LingeLangtangen2017}. Considering this diffusion operator in 2D for a uniform grid ($\Delta x =1 = \Delta y$), Figure~\ref{fig:conv_filter_fd_equivalence} shows a stencil representing the discretised operator acting on a part of the solution field which calculates the value for cell $Y_{33}$. This application of the finite difference stencil is exactly equivalent to passing a filter, with weights as specified in Figure~\ref{fig:conv_filter_fd_equivalence}, over an image, as done in the convolutional layer of a convolutional neural network. In the left of this figure, a halo region can be seen, which is made up of ghost cells that are represented by the white cells (cells $ij$ where $i\in\{0,6\} ,\, j\in\{0,1,\ldots,6\}$ and $i\in\{0,1,\ldots,6\}\,, j\in\{0,6\}$), through which boundary conditions can be applied. In machine learning terminology, this is known as padding, and is used to obtain a feature map of the same dimension as that of the image to which the filter is applied. In this way, numerical discretisations can be represented as convolutional layers of neural networks. Systems arising from this can be solved directly with a Jacobi method or implicitly using a multigrid solver implemented through a U-Net architecture~\citep{Chen2023,Phillips2023}.

\subsubsection{Notation}\label{sec:notation}
The action of a discretisation stencil on a 3D tensor, $\bm{C}^n$, can be written as follows: 
\begin{equation}
\bm{f}(\bm{C}^n;{\bf{w}})\left\vert_{i,j,k}\right.
= \sum_{ii=-\ell}^\ell\sum_{jj=-\ell}^\ell\sum_{kk=-\ell}^\ell  \text{w}_{ii,jj,kk} \,C_{i+ii,j+jj,k+kk}^n 
\end{equation}
where $\bf{w}$ is of dimension $2\ell+1$ by $2\ell+1$ by $2\ell+1$ and represents the coefficients associated with the numerical discretisation. This expression is equivalent to convolving a filter of weights~${\bf w}$ with an image, as is done in a convolutional layer, see page~128 in chapter~3 of~\citep{Trefethen} and also~\citep{Long2018,Bishop}. This notation is used throughout to represent the application of a filter with weights~$\bf{w}$ to a tensor~$\bm{C}^n$ which holds the values of a solution field at the nodes or grid points. The discretised differential operator is expressed through the weights and is applied to the discretised field. Generally, the tensor will have a padding of~$\ell$ cells (through which boundary conditions are applied) and be of dimension $N_x+2\ell$~by $N_y+2\ell$ by~$N_z+2\ell$, but the output of the function can be written without the padding ($N_x$ by $N_y$ by $N_z$). The value of~$\ell$ depends on the discretisation applied, see Section~\ref{sec:bcs}.

\subsubsection{Finite element discretisation of first-order derivatives} 
Filters can also be used to represent finite element discretisations. The weights of the filter are calculated from integrals of products of basis functions $\left(N_{a}\right)$ and their derivatives in the usual manner, \eg 
\begin{equation}
\int\limits_{V} N_{a} \frac{\partial N_b}{\partial x}  dV, \hspace{.5cm} \forall \, b, \qquad \text{ or} \qquad \int\limits_{V}  N_{i,j,k}  \frac{\partial N_{i^{\prime},j^{\prime},k^{\prime}}}{\partial x} dV, \hspace{0.5cm} \forall \left(i^{\prime},j^{\prime},k^{\prime}\right)\,.
\end{equation}
in which the finite element nodes~$a$ and~$b$ are replaced with the equivalent tensor indices~$(i,j,k)$ and $\left(i^{\prime},j^{\prime},k^{\prime}\right)$ respectively. For every node away from the boundary and, with every element being of quadrilateral shape and of the same size (\ie regular structured quadrilateral elements), these matrix entries are the same for each node and written as the weights of a filter of a convolutional layer: 
\begin{equation}
\text{w}_{\text{x}\,ii,jj,kk}=\int\limits_{V}  N_{i,j,k} 
\frac{\partial
N_{i^{\prime},j^{\prime},k^{\prime}} }{\partial x} dV \quad \text{with} \quad
        ii=i^{\prime}-i,\ 
        jj=j^{\prime}-j,\ 
        kk=k^{\prime}-k\,.
\end{equation}
Due to the compact support of the FEM basis functions, these matrix entries are only non-zero for indices $ii\in\{-\ell,\ldots,\ell\}$, $jj\in\{-\ell,\ldots,\ell\}$, $kk\in\{-\ell,\ldots,\ell\}$ and they can be replaced by a~$2\ell+1$~by $2\ell+1$ by $2\ell+1$ tensor $
{\bf w}_{\bf x}$ or $\text{w}_{\text{x}\, ii,jj,kk}$. 
For linear elements $\ell=1$, quadratic elements $\ell=2$ and cubic elements $\ell=3$.  Thus, the weights of a linear FEM discretisation of the derivative with respect to~$x$,  ${\bf{w}}_{\bf{x}}$, are defined as 
\begin{equation}
{\bf{w}}_{{\bf{x}}{\cdot,\cdot,-1}} 
=
  \frac{m_l}{72\Delta x}
 \begin{bmatrix*}[c]
 \num{-1}  & 0 & \num{1} \\
 \num{-4} & 0 &  \num{4} \\
  \num{-1} & 0 &  \num{1}
 \end{bmatrix*},\  
{\bf{w}}_{{\bf{x}}{\cdot,\cdot,0}} =
 \frac{m_l}{72\Delta x}
 \begin{bmatrix*}[c]
  \num{-4} & 0 &  \num{4}  \\
 \num{-16}  & 0 &  \num{16} \\
  \num{-4} & 0 &  \num{4}
 \end{bmatrix*}, \ 
{\bf{w}}_{{\bf{x}}{\cdot,\cdot,1}} 
=
 \frac{m_l}{72\Delta x}
 \begin{bmatrix*}[c]
  \num{-1}  & 0 &  \num{1} \\
 \num{-4} & 0 &  \num{4} \\
  \num{-1} & 0 &  \num{1}
 \end{bmatrix*}\,,
\end{equation}
where $m_l = \Delta x \Delta y \Delta z$.  

A challenge arising during the design of quadratic or higher order FE methods within the NN4PDEs approach, is that the coefficients of the stencil are different at each node. For convolutional layers, the same filter is applied to each pixel in the image. In order to address this, a new FEM has been devised known as the Convolultional Finite Element Method (ConvFEM), which has the same coefficients at each node in the mesh or grid. This greatly simplifies the implementation of higher-order FE discretisations using the filters of convolutional layers on structured grids. The basic idea in forming the ConvFEM filters is to sum coefficients of the stencils and take the average (see \citet{Phillips2022-progress} for further details). A number of FE discretisations are applied in this paper, thus filters are tensors of shape of either $3\times3\times3$ for linear FE, $5\times5\times5$ for quadratic ConvFEM elements, or $7\times7\times7$ for cubic ConvFEM elements. These filters are listed in the GitHub repository by \citet{Phillips2022_RT_github}, along with the computer code that automatically generates them.

\subsubsection{Finite-element based discretisation of second-order derivatives}\label{Section:DiscDiffusion}
Second-order derivatives can be implemented in the same way as first-order derivatives, with the coefficients corresponding to the numerical discretisation making up the weights of the filters, ${\bf{w_{xx}}}$, ${\bf{w_{yy}}}$ and~${\bf{w_{zz}}}$. In this work, any stabilising terms (such as those derived from Petrov-Galerkin methods) also involve second-order spatial derivatives. Considering the second derivative in the $x$ direction of field $C^n$ with diffusion coefficient~$k_C^n$, its typical form can be expanded as three terms, all involving the Laplacian operator, 
\begin{equation}\label{k-Txx}
 \begin{array}{rl@{\hspace{2.5cm}}l}
  \displaystyle - \frac{\partial}{\partial x} \left({k_C^{n}} 
   \frac{\partial C^n}{\partial x}  \right) \!\!\!\! & \displaystyle = -\frac{1}{2}\left( \frac{\partial^2 (k_C^{n} {C}^n )}{\partial x^2}  + k_C^{n}  \frac{\partial^2 {C}^n}{\partial x^2}  - {C}^n  \frac{\partial^2 k_C^{n}}{\partial x^2}  \right) & \text{(Continuum Form)}\\[3mm]%
  &\multicolumn{2}{l}{\displaystyle \sim \phantom{-} \frac{1}{2}\left( \bm{f}(\bm{k_C}^{n}\odot\bm{C}^n;{\bf{w_{xx}}}) + \bm{k_C}^{n}\odot \bm{f}(\bm{C}^n;{\bf{w_{xx}}}) - \bm{C}^n\odot \bm{f}(\bm{k_C}^{n};{\bf{w_{xx}}}) \right)} \\[2mm] 
  & =:  \displaystyle  \phantom{-} \fdiff (\bm{C}^n,\bm{k_C}^{n};{\bf{w_{xx}}}) \,, &\text{(Discretised Form)}
\end{array}
\end{equation}
where ${\bf{w_{xx}}}$ contains weights associated with a particular discretisation of the second derivative (as yet unspecified), ${\bm k_C}^n$ is the tensor containing the diffusion coefficients at each node of the grid at time level $n$, and $\odot$ is the Hadamard product (also known as Schur product) which represents entry-wise multiplication. Equation~\eqref{k-Txx} presents a very convenient way of writing terms involving second derivatives of the form seen in the left-hand side of Equation~\eqref{k-Txx} for implementation using convolutional layers, as it only requires the use of a discretised Laplacian (or associated filter) to form the diffusion operator in the case of varying diffusion coefficients. For isotropic diffusion (in three directions) this equation becomes:
\begin{equation}\label{k-T}
 \begin{array}{rl@{\hspace{2.5cm}}l}
  \displaystyle - \nabla \cdot \left( k_C^n 
   \nabla {C}^n \right) & \displaystyle = -\frac{1}{2}\left( \nabla^2 ( k_C^n {C}^n ) + k_C^n  \nabla^2 {C}^n - {C}^n  \nabla^2 k_C^n \right) & \text{(Continuum Form)}\\[2mm]
  &\multicolumn{2}{l}{\displaystyle \sim  
  \phantom{-}\frac{1}{2}\left( \bm{f}(\bm{k_C}^n\odot\bm{C}^n;\wdiff) + \bm{k_C}^n\odot 
   \bm{f}(\bm{C}^n;\wdiff) - \bm{C}^n\odot \bm{f}(\bm{k_C}^n;\wdiff ) \right) } \\[2mm]
  & \displaystyle = \fdiff (\bm{C}^n,\bm{k_C}^n;\wdiff) \, , & \text{(Discretised Form)} 
  \end{array}
\end{equation}
in which the filter $\wdiff$ contains weights associated with a particular discretisation of isotropic diffusion. These expressions will be used, in the following sections, to represent viscous and stabilisation terms that are determined from a non-linear Petrov-Galerkin discretisation. 

\subsection{Discretisation of the governing equations using convolutional layers}\label{sec:disc_gov_eq}
We now discretise multiphase flow equations in space and time with FEM, and express these discretisations as convolutions using the notation introduced in Section~\ref{sec:notation}. 
\subsubsection{Advection equation for the volume fraction field} \label{adv-section}
The equation governing the volume fraction field, Equation~\eqref{adv-diff-eqn}, is discretised in time using forward Euler time-stepping:  
\begin{equation} \label{transport-indicator-disc-in-time}
    \frac{C^{n+1}-C^{n}}{\Delta t} + \bm{q}^{n+1}\cdot\nabla C^{n} 
    = 0\,,
\end{equation}
where $C^n$ is the volume fraction field at time level~$n$, $\bm{q}^n$ is the velocity field at time level~$n$ and $\Delta t$ is the time step. The forward Euler time-stepping method has been chosen because it has a negative diffusion coefficient in the truncation error which compresses the solution slightly~\citep{Pavlidis2014,Pavlidis2016}. Using the filters derived in Section~\ref{Section:DiscDiffusion}, Equation~\eqref{transport-indicator-disc-in-time} can be discretised in space as
\begin{eqnarray} \bm{r_C}^{n+\frac{1}{2}} &=& \bm{f}\left(\frac{\bm{C}^{n+1}-\bm{C}^{n} }{\Delta t}  ; {\bf{w_{ml}}} \right) 
+ \bm{\tilde{s}_C}^{n+\frac{1}{2}}\,, 
\label{Petrov-C} \\
\text{with}\quad \bm{\tilde{s}_C}^{n+\frac{1}{2}}
 &=&
 \bm{u}^{n+1}  \odot \bm{f}\left( \bm{C}^{n} ;{\bf{w_{x}}}\right)
+
 \bm{v}^{n+1}  \odot \bm{f}\left( \bm{C}^{n} ;{\bf{w_{y}}}\right)
+
 \bm{w}^{n+1}  \odot \bm{f}\left( \bm{C}^{n} ;{\bf{w_{z}}}\right)  \nonumber \\
&+&  
 \fdiff\left(\bm{C}^{n} ;\bm{k}_{\bm{C}}^{\bm{x}\,n} ,{\bf{w_{xx}^{PG}}}\right)
+\fdiff\left(\bm{C}^{n} ;\bm{k}_{\bm{C}}^{\bm{y}\,n} ,{\bf{w_{yy}^{PG}}}\right)
+\fdiff\left(\bm{C}^{n} ;\bm{k}_{\bm{C}}^{\bm{z}\,n} ,{\bf{w_{zz}^{PG}}}\right)\,,
\label{Petrov-Cs}
\end{eqnarray}
where $\bm{f}$ denotes the the convolution of a filter with the first argument of the function; $\bf{w_{ml}}$ represents a lumped mass tensor whose only non-zero entry corresponds to $i=0,\,j=0,\,k=0$ and has value $\Delta x \Delta y \Delta z$; $\bm{u}^{n+1}$, $\bm{v}^{n+1}$ and $\bm{w}^{n+1}$ are tensors which represent the three velocity components on the structured grid; the first-order discretised differential operators are represented by a convolutional layer with weights $\bf{w_{x}}$, $\bf{w_{y}}$ and $\bf{w_{z}}$. Finally, filters with weights from a linear $3\times3\times3$ FEM discretisation for calculating the second-order derivatives  are represented by ${\bf{w_{xx}^{PG}}}$, ${\bf{w_{yy}^{PG}}}$, and ${\bf{w_{zz}^{PG}}}$. These second derivative terms provide diffusion (based on the Petrov-Galerkin method) and are included to reduce the oscillations that can occur when central-difference or FEM-based discretisations are applied to an advection term (see Sections~\ref{Petrov-Galerkin} and~\ref{compress-adv-k}). 

\subsubsection{Navier-Stokes equations} 
The momentum equation, Equation~\eqref{mom-eqn}, is discretised in time using a two-level central difference scheme similar to the Crank-Nicolson time-stepping method, however $\sigma$ and the pressure terms are discretised using backward Euler time-stepping, resulting in: 
\begin{equation}\label{NS-eqn-disc-guess}
\begin{split}
\rho^{n} \left(\frac{\bm{\tilde{q}}^{n+1} -\bm{q}^{n} }{\Delta t} 
+ 
u^{n+\frac{1}{2}}\frac{\partial \bm{q}^{n+\frac{1}{2}} }{\partial x} +  v^{n+\frac{1}{2}}\frac{\partial \bm{q}^{n+\frac{1}{2}} }{\partial y} +  w^{n+\frac{1}{2}}\frac{\partial \bm{q}^{n+\frac{1}{2}} }{\partial z} 
\right) + \sigma \bm{\tilde q}^{n+1} 
- \nabla \cdot({\mu}\nabla \bm{q}^{n+\frac{1}{2}}) \\ = -\nabla p_{hs}^{n+1} -\nabla \tilde{p}_{nh}^{n+1} + \bm{s}_{q}^{n} 
+ \bm{s}_{t}^{n}\,. 
\end{split}
\end{equation}
The buoyancy and surface tension forces are represented by $\bm{s}_{q}^{n}$ and $\bm{s}_{t}^{n}$ respectively, and the pressure has been split into two terms: a term which balances hydrostatic and surface tension forces $\left(p_{hs}^{n+1}\right)$, and a non-hydrostatic non-surface-tension force component $\left(\tilde{p}_{nh}^{n+1}\right)$. The tilde sign above a variable indicates that we will use the best approximation currently available for that variable.  Equation~\eqref{NS-eqn-disc-guess} can then be discretised in space using a Petrov-Galerkin method~\citep{donea2003finite}, with mass lumping for the advection, time and absorption terms. Using the filter notation of Section~\ref{sec:notation}, this can be expressed as:
\begin{subequations}\label{Petrov}
\begin{eqnarray}
 \bm{r}_{u}^{n+\frac{1}{2}} =  \bm{f}\left( 
 \bm{\rho}^{n}\odot\left(\frac{\bm{\tilde u}^{n+1} -\bm{u}^n}{\Delta t}\right) +\bm{\sigma}\odot\bm{\tilde u}^{n+1}
 ; {\bf{w_{ml}} } \right) 
 + \bm{\tilde s}_{u}^{n+\frac{1}{2}}
 \;\;\; \text{with} 
 \nonumber \\
 \bm{\tilde s}_{u}^{n+\frac{1}{2}} =
 \bm{\rho}^{n}\odot\left(
 \bm{u}^{n+\frac{1}{2}} \odot \bm{f}( \bm{u}^{n+\frac{1}{2}};{\bf{w_{x}}} )
+
 \bm{v}^{n+\frac{1}{2}} \odot \bm{f}( \bm{u}^{n+\frac{1}{2}};{\bf{w_{y}}})
+
 \bm{w}^{n+\frac{1}{2}} \odot \bm{f}( \bm{u}^{n+\frac{1}{2}};{\bf{w_{z}}}) 
 \right) \label{Petrov-u} \\
+ 
 \fdiff (\bm{u}^{n+\frac{1}{2}},\bm{{k}}_u^{n+\frac{1}{2}};\wdiff )
 + \bm{f}( \bm{p_{hs}}^{n+1}+\bm{\tilde p_{nh}}^{n+1};{\bf{w_{x}}})
 -
 \bm{f}( 
 \bm{s}_{u}^{n} 
 +
 \bm{s}_{\bm{t}u}^{n}; {\bf{w_{m}}}) ,
 \nonumber \\[5mm]
 \bm{r}_{v}^{n+\frac{1}{2}} = \bm{f}\left(
 \bm{\rho}^{n}\odot\left( 
 \frac{\bm{v}^{n+1} - \bm{v}^n}{\Delta t} \right)
 +\bm{\sigma}\odot\bm{\tilde v}^{n+1}
 ; {\bf{w_{ml}}} \right) 
 + \bm{\tilde s}_{v}^{n+\frac{1}{2}}
 \;\;\; \text{with} 
 \nonumber \\
 \bm{\tilde s}_{v}^{n+\frac{1}{2}} =
 \bm{\rho}^{n}\odot\left(
 \bm{u}^{n+\frac{1}{2}} \odot \bm{f}( \bm{v}^{n+\frac{1}{2}};{\bf{w_{x}}})
+
 \bm{v}^{n+\frac{1}{2}} \odot \bm{f}( \bm{v}^{n+\frac{1}{2}};{\bf{w_{y}}})
+
 \bm{w}^{n+\frac{1}{2}} \odot \bm{f}( \bm{v}^{n+\frac{1}{2}};{\bf{w_{z}}}) \right) \label{Petrov-v} \\
+ 
 \fdiff (\bm{v}^{n+\frac{1}{2}},\bm{{k}}_v^{n+\frac{1}{2}};\wdiff )
 + \bm{f}( \bm{p_{hs}}^{n+1}+\bm{\tilde p_{nh}}^{n+1};{\bf{w_{y}}})
 -
 \bm{f}( \bm{s}_{v}^{n} 
 + 
 \bm{s}_{\bm{t}v}^{n}; {\bf{w_{m}}})
  ,
 \nonumber\\[5mm]
 \bm{r}_{w}^{n+\frac{1}{2}} = \bm{f}\left(
 \bm{\rho}^{n}\odot\left( 
 \frac{ \bm{\tilde w}^{n+1} -\bm{w}^n}{\Delta t} \right)
 +\bm{\sigma}\odot\bm{\tilde w}^{n+1}
 ; {\bf{w_{ml}}} \right) 
 + \bm{\tilde s}_{u}^{n+\frac{1}{2}}
 \;\;\; \text{with} 
 \nonumber \\
 \bm{\tilde s}_{w}^{n+\frac{1}{2}} = 
 \bm{\rho}^{n}\odot\left(
 \bm{u}^{n+\frac{1}{2}} \odot \bm{f}( \bm{w}^{n+\frac{1}{2}};{\bf{w_{x}}})
+
 \bm{v}^{n+\frac{1}{2}} \odot \bm{f}( \bm{w}^{n+\frac{1}{2}};{\bf{w_{y}}})
+
 \bm{w}^{n+\frac{1}{2}} \odot \bm{f}( \bm{w}^{n+\frac{1}{2}};{\bf{w_{z}}}) \right) \label{Petrov-w} \\
+  
 \fdiff (\bm{w}^{n+\frac{1}{2}},\bm{{k}}_w^{n+\frac{1}{2}};\wdiff)
 + \bm{f}( \bm{p_{hs}}^{n+1}+\bm{\tilde p_{nh}}^{n+1};{\bf{w_{z}}}) 
- \bm{f}( \bm{s}_{w}^{n}
+
 \bm{s}_{\bm{t}w}^{n} ;{\bf{w_{m}}}) ,
 \nonumber
\end{eqnarray}
\end{subequations}
in which ${\bf{w_{m}}}$ is the consistent mass filter. When solving the discretised Navier-Stokes equations the aim is to minimise the residual, that is, to obtain: $\bm{r}_{u}^{n+\frac{1}{2}}=\bm{0}$, $\bm{r}_{v}^{n+\frac{1}{2}}=\bm{0}$, $\bm{r}_{w}^{n+\frac{1}{2}}=\bm{0}$. These equations are solved using a two-step predictor-corrector time level approach, in order to obtain better estimates of the velocity variables at time level $n+\frac{1}{2}$.  In Equations~\eqref{Petrov}, the density at time level~$n$ is given by: 
\begin{equation}
\label{density-n} 
    \bm{\rho}^{n} = \rho_l \bm{C}^n + \rho_g (\bm{1}-\bm{C}^n)\,,
\end{equation}
where $\bm{1}$ is a tensor in which each entry is unity $\left(\text{that is } \bm{1}\left\vert\right. _{i,j,k}=1\right)$ and similarly, $\bm{0}$ represents the zero tensor (every entry is 0). There are three buoyancy tensors, one of which is non-zero, as determined by the direction of gravity
\begin{equation}\label{eq:gravity}
    \bm{s}_u^{n}=\bm{0}, \qquad 
    \bm{s}_v^{n}=\bm{0}, \qquad 
    \bm{s}_w^{n}= -g\,\bm{\rho}^{n}\,.
\end{equation}
On discretising the surface tension given in Equation~\eqref{surface-tension-eqn} we obtain: 
\begin{equation}
    \bm{s}_{\bm{t}u}^{n} =  \sigma_t \bm{\kappa}^{n} \odot \bm{C_x}^{n}, \qquad 
    \bm{s}_{\bm{t}v}^{n} =\sigma_t \bm{\kappa}^{n} \odot \bm{C_y}^{n}, \qquad 
    \bm{s}_{\bm{t}w}^{n} =\sigma_t \bm{\kappa}^{n} \odot \bm{C_z}^{n}\,,       
  \label{surface-tension-uvw}
\end{equation}
where the curvature tensor is represented by $\bm{\kappa}^{n}$. In order to calculate the curvature, we take the gradient of the normal to the interface using filters as follows
\begin{equation}
\bm{\kappa}^{n} 
= \bm{f}\left(
\bm{f}( \bm{n}_{\bm{t}u}^{n} ; {\bf{w_x}})
+ \bm{f}( \bm{n}_{\bm{t}v}^{n} ; {\bf{w_y}})
+ \bm{f}( \bm{n}_{\bm{t}w}^{n} ; {\bf{w_v}})
;{\bf{w_{ml}^{-1}}}\right) \,,
\label{surface-tension-k}
\end{equation} 
in which ${\bf{w_{ml}^{-1}}}$ is the inverse of the lumped mass filter and where the tensors representing the normal are given by 
\begin{equation}
    \begin{array}{rcl}
        \bm{n}_{\bm{t}u}^{n} &=&\bm{C_x}^{n} \oslash \left(\epsilon_t \bm{1}+ \bm{l}^{n}\right),\\
        \bm{n}_{\bm{t}v}^{n} &=& \bm{C_y}^{n} \oslash \left(\epsilon_t \bm{1}+ \bm{l}^{n}\right),\\
        \bm{n}_{\bm{t}w}^{n} &=& \bm{C_z}^{n} \oslash \left( \epsilon_t \bm{1}+ \bm{l}^{n} \right),
    \end{array}\text{ with} \quad {l}_{i,j,k}^{n} = \sqrt{  (C_{xi,j,k}^{n})^2 + (C_{yi,j,k}^{n})^2 + (C_{zi,j,k}^{n})^2 } \;\;\text{ and } \;\; \epsilon_t = \frac{10^{-7}}{h},
\end{equation}
where $h$ is the typical grid spacing and $\oslash$ is the Hadamard entry-wise division of equally sized 
tensors, \ie
\begin{equation*}
    \hat{\bm{c}}=\hat{\bm{a}}\oslash\hat{\bm{b}} \quad \Longleftrightarrow \quad \hat{c}_{i,j,k}=\frac{\hat{a}_{i,j,k}}{\hat{b}_{i,j,k}}.
\end{equation*}
Spurious currents can be generated when the capillary number is low~\citep{Inguva2022}, which is mitigated by introducing dissipation to the volume fraction field, done here through a Petrov-Galerkin formulation (see Section~\ref{Petrov-Galerkin}).

\subsubsection{Application of boundary conditions} \label{sec:bcs}
\begin{figure}[htp]
\centering
\subfloat[Symmetry boundary conditions]
{\includegraphics[scale=0.40,trim=45mm 20mm 98mm 15mm, clip]{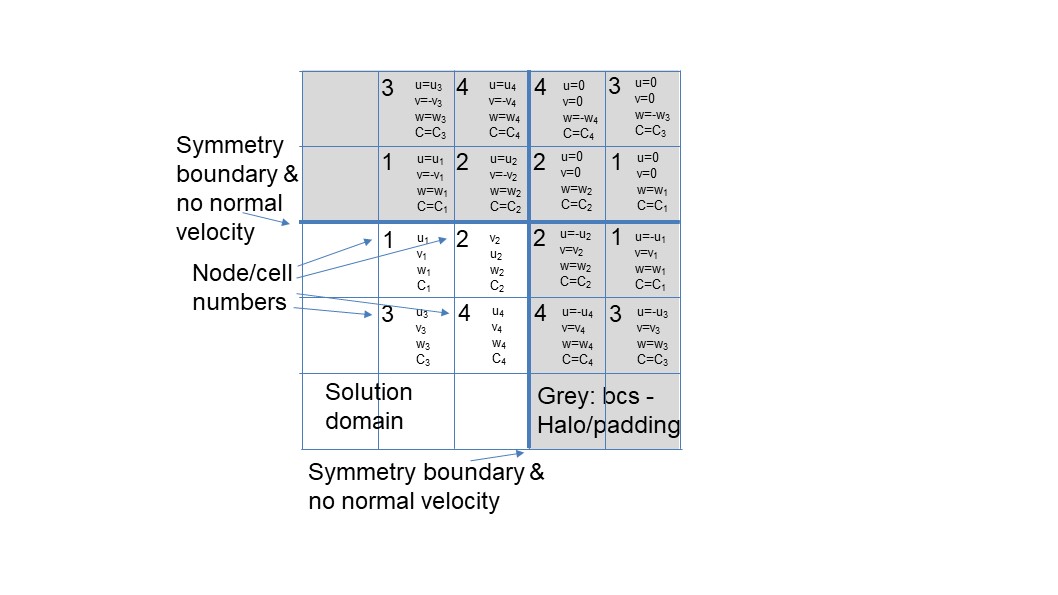} 
\label{fig:bc-pic1}}
\quad
\subfloat[Combined symmetry and wall boundary conditions  (zero velocity and Dirichlet $C=B$,  for volume fraction~$C$)]{\includegraphics[scale=0.40,trim=45mm 20mm 98mm 15mm, clip]{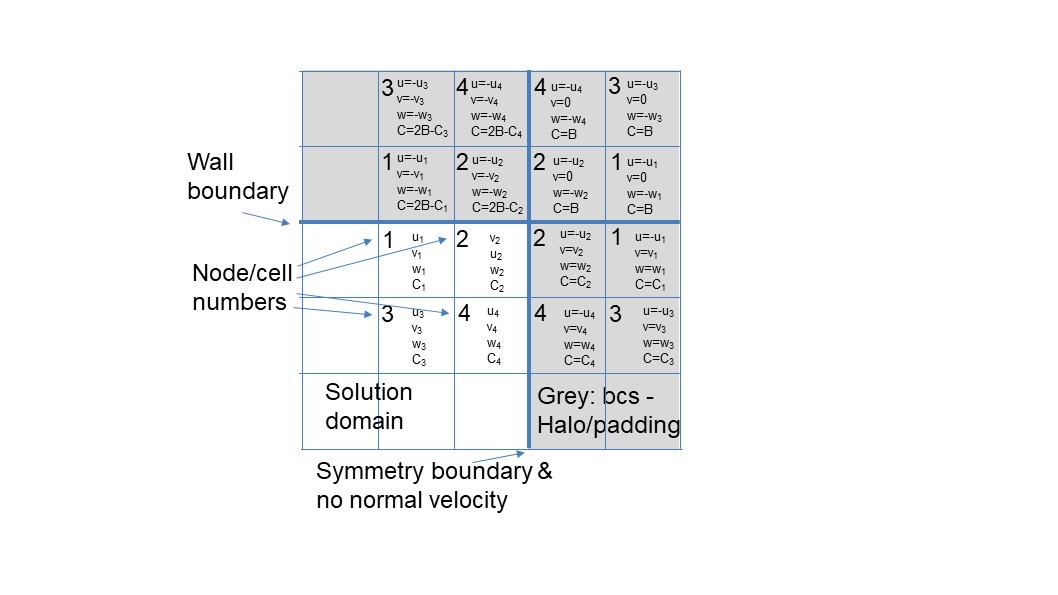}
\label{fig:bc-pic2}}
\quad
\subfloat[Wall boundary conditions (zero velocity and Dirichlet $C=B$)]{\includegraphics[scale=0.40,trim=45mm 20mm 98mm 15mm, clip]{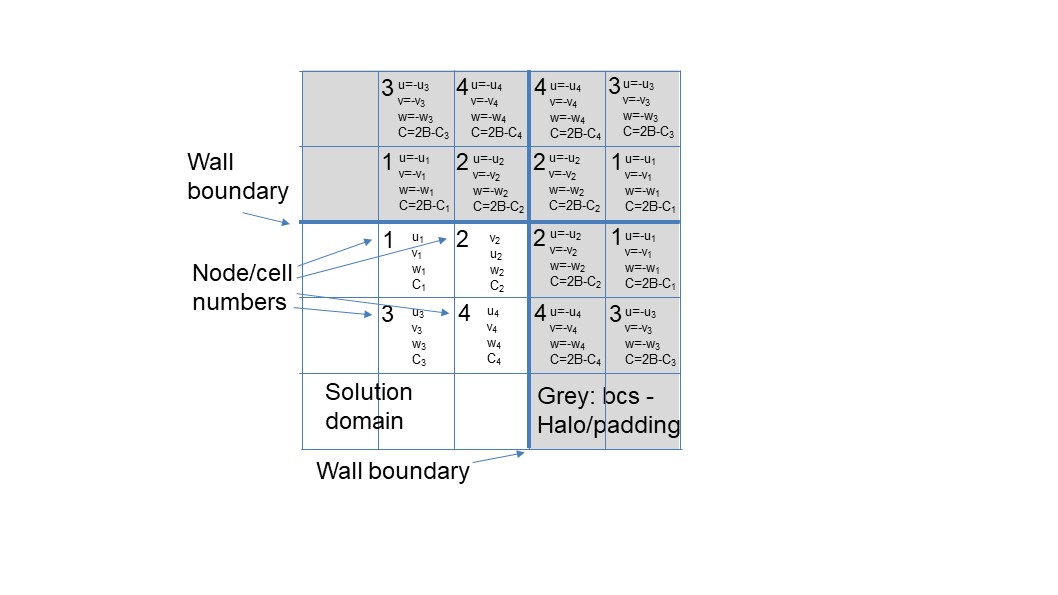}
\label{fig:bc-pic2}}
\caption{Diagrams showing how boundary conditions are applied through halo values when using the quadratic ConvFEM. Linear ConvFEM can be realised simply by ignoring the second layer of halo values. The indices~1, 2, 3 and~4 indicate the cell or node number within the solution domain. These same indices are used in the halo nodes to indicate where the information for forming the halo values is obtained. The basic idea behind the boundary conditions is to specify the value of the halo values such that when one interpolates these with the values inside the domain then one obtains the desired boundary conditions along the boundaries (indicated here by thick blue lines).}
\label{bc-applied} 
\end{figure}

The boundary conditions are applied through the halo nodes using padding (see Figure~\ref{fig:conv_filter_fd_equivalence}). This means that the discretisation scheme is the same wherever it is applied, whether this is next to the boundary or within the domain, thereby simplifying the implementation of the method. These halos are one node thick for linear elements, two nodes thick for quadratic elements and three nodes thick for cubic elements. Dirichlet boundary conditions are applied by setting the values of the halo nodes to equal the desired value. When applying a specified normal derivative, the value at the halo nodes is calculated by extrapolating from the nearest node within the domain using the specified gradient. Most commonly, a zero derivative is applied, in which case, this simplifies to copying the nearest nodal value within the domain to the halo nodes at least for linear ConvFEM. See Figure~\ref{bc-applied} for further details of how the boundary conditions are applied. 

Thus, for incoming flows one might specify the values of $C_{i,j,k}^n$ at the halo nodes based on the desired Dirichlet boundary conditions and for boundaries with outgoing flows one might apply a zero derivative boundary condition to $C_{i,j,k}^n$. For incoming velocities, the Dirichlet velocity components are specified at the halo nodes and for outgoing flows (outlet boundary), a zero normal-derivative boundary condition is applied by copying the nodal values at the boundary to the halo nodes for the velocity component normal to the boundary. For non-hydrostatic non-surface-tension pressure, a zero derivative boundary condition is applied on all boundaries other than the outlet boundary, at which a zero value is specified. For hydrostatic and surface-tension pressure, $p_{hs}$, a zero normal-derivative boundary condition is applied on the lateral (side) boundaries and a zero value is applied at the top boundary of the domain. For the bottom boundary, the pressure gradient $\frac{\partial p_{hs}}{\partial z}=-\rho g$ is enforced, in which the density~$\rho$ from the nearest node within the domain is used. For the hydrostatic and surface-tension pressure, and the non-hydrostatic non-surface-tension pressure correction steps, the boundary conditions just described are applied at each of the levels within the multigrid method, see~\citet{Phillips2022-progress}.

\subsubsection{Isotropic non-linear Petrov-Galerkin method}\label{Petrov-Galerkin}
We apply an isotropic non-linear Petrov-Galerkin method to both the velocity components and the volume fraction field in order to reduce the oscillations that would otherwise occur, as a result of applying a high-order discretisation, see Godunov's theorem~\citep{Wesseling}. 
Introducing non-linearity enables us to control the oscillations. Based on \citet[page 185]{donea2003finite}, the approach outlined in this section will calculate diffusion coefficients to be used for volume fraction and velocities that are calculated in Equations~\eqref{Petrov-C}, \eqref{Petrov-u}, \eqref{Petrov-v} and~\eqref{Petrov-w}. Using three methods, we calculate three diffusion coefficients, and select the one with the lowest value at each grid point. Suppose that $\bm{\Psi}^n$ is either the volume fraction field~$\bm{C}^n$ or a velocity component (either $\bm{u}^n,\bm{v}^n$ or $\bm{w}^n$). Then, according to \citet{Codina1993}, if the discretised derivatives of ${\bm{\Psi}}^n$ with respect to $x$, $y$ and $z$ are written as $\bm{\Psi_{x}}^n =  \bm{f}(\bm{\Psi}^n;{\bf{w_{x}}}) $, $\bm{\Psi_{y}}^n =  \bm{f}(\bm{\Psi}^n;{\bf{w_{y}}})$ and $\bm{\Psi_{z}}^n =  \bm{f}(\bm{\Psi}^n;{\bf{w_{z}}})$, the first residual-based diffusion coefficient is given by:
\begin{equation}\label{k-abs}
   \bm{k_{abs}}^n = \alpha_{kabs}|\bm{a_\Psi}^n| \odot \bm{h_\Psi}^n  \oslash (\epsilon_k \bm{1}+\frac{1}{3}(|\bm{\Psi_{x}}^n|+|\bm{\Psi_{y}}^n|+|\bm{\Psi_{z}}^n|)) \,,
\end{equation}
where $\epsilon_k$ is a small positive number added to the term in the divisor, so that the diffusion coefficients do not become too large (in this paper, \num{e-7} is used); and $\bm{a_\Psi}^n$ represents the residual of the advection part of the governing equation. For the general case where the grid spacing is different in different directions, the lengthscale associated with the diffusion is given by
\begin{equation} \label{h_Psi}
\bm{h_\Psi}^n = 
  \left( \Delta x|\bm{\Psi_{x}}^n|+\Delta y|\bm{\Psi_{y}}^n|+\Delta z|\bm{\Psi_{z}}^n|\right)  
  \oslash 
  \left(\epsilon_k \bm{1}+|{\bm{\Psi_{x}}^n|+|\bm{\Psi_{y}}^n|+|\bm{\Psi_{z}}^n|}\right) \, ,
\end{equation}  
and the expression for the first diffusion coefficient becomes
\begin{equation}\label{efficient-kabs}
\bm{k_{abs}}^n = \alpha_{kabs}|\bm{a_\Psi}^n| \odot (\Delta x|\bm{\Psi_{x}}^n|+\Delta y|\bm{\Psi_{y}}^n|+\Delta z|\bm{\Psi_{z}}^n|)  \oslash (\epsilon_k \bm{1}+\frac{1}{3}(|\bm{\Psi_{x}}^n|+|\bm{\Psi_{y}}^n|+|\bm{\Psi_{z}}^n|)^2) \,. 
\end{equation}
For the case when the grid spacing is the same in all three directions, Equation~\eqref{h_Psi} simplifies to $\bm{h_\Psi}^n=\Delta x \bm{1}$ resulting in
\begin{equation}\label{efficient-kabs-special-case}
 \bm{k_{abs}}^n = \alpha_{kabs} \Delta x| \bm{a_\Psi}^n| \oslash (\epsilon_k \bm{1}+\frac{1}{3}(|\bm{\Psi_{x}}^n|+|\bm{\Psi_{y}}^n|+|\bm{\Psi_{z}}^n|)) \,.
\end{equation}

Taking the expression from \citet{Hansbo1991}, the second residual-based diffusion coefficient is given by:
\begin{eqnarray}
\label{k-square}
   \bm{k_{square}}^n & = & \alpha_{ksquare}\,{ ({\bm{a_\Psi}^n})^2}\odot  \bm{h_\Psi}^n \nonumber \\[2mm]
    & & \oslash \left( 
    \epsilon_k \bm{1} +\frac{1}{3} (|\bm{u}^n\odot\bm{\Psi_{x}}^n|+|\bm{v}^n\odot\bm{\Psi_{y}}^n|
    +|\bm{w}^n\odot\bm{\Psi_{z}}^n|
    )(({\bm{\Psi_x}}^n)^2+({\bm{\Psi_{y}}}^n)^2+({\bm{\Psi_{z}}}^n)^2)\right) ,\quad
\end{eqnarray}
where $({\bm{a_\Psi}^n})^2={\bm{a_\Psi}^n}\odot {\bm{a_\Psi}^n} $ and $\vert\bm{\Psi_x}^n\vert$ is a tensor in which the absolute value of each entry of $\bm{\Psi_x}^n$ is taken. The coefficients $\alpha_{kabs}$, $\alpha_{ksquare}$ have been determined by trial and error, and are defined as:
\begin{equation*}
  \alpha_{kabs} =\frac{1}{8\times 2^{\rm p}}  \hspace{0.5cm}\text{ and }\hspace{0.5cm} \alpha_{ksquare} =\frac{8}{2^{\rm p}}\,, 
\end{equation*} 
where $\rm p$ is the polynomial order of the finite element expansion, \ie ${\rm{p}}=1$ for linear $3\times3\times3$ filters, ${\rm p}=2$ for quadratic $5\times5\times5$ filters and ${\rm p}=3$ for cubic filters.  
  
To calculate the advection residual, $\bm{a_\Psi}^n$, which appears in Equations~\eqref{k-abs} and~\eqref{k-square}, a simple mixed mass approach is 
used
\begin{equation}
    \bm{a_\Psi}^n =  \alpha_a 
    \bm{f}\left( \; \bm{u}^n \odot \bm{f}(\bm{\Psi}^n;{\bf{w_{x}}}) +\bm{v}^n \odot\bm{f}(\bm{\Psi}^n;{\bf{w_{y}}})+\bm{w}^n \odot\bm{f}(\bm{\Psi}^n;{\bf{w_{z}}}) ; \; \frac{1}{m_l} {\bf{w_m}} - {\bf{E}} \right),
    \label{a_psi} 
\end{equation}
where $m_l=\Delta x\Delta y\Delta z$ is the lumped mass term, $\bf{w_m}$ filter representing the consistent mass matrix and $\bf{E}$ is the ``identity filter'' (on convolving this with a tensor produces the same tensor). A scalar coefficient, $\alpha_a$, is introduced, for which a value of $\alpha_a=2$ has been found effective. The underlying assumption, in using Equation~\eqref{a_psi} for calculating the residual tensor, is that most of the errors in the discretisation are within the spatial gradients associated with the transport terms, which, given the dominance of these terms in the applications presented here, is a reasonable assumption. 

Finally, the rate of diffusion should not exceed the rate of advection, so we introduce a maximum diffusion coefficient calculated as follows:
\begin{equation}
\label{max-k}
    \bm{k_{max}}^n = \Delta x\vert\bm{u}^n\vert + \Delta y\vert\bm{v}^n\vert + \Delta z\vert\bm{w}^n\vert \; . 
\end{equation}
Assuming that all the diffusion coefficients given by Equations~\eqref{k-abs},  \eqref{k-square}, \eqref{max-k} are conservatively large, one can determine the diffusion coefficient using: 
\begin{eqnarray}  \label{k_Psi-n}
     \bm{k_\Psi}^n &=& \bm{\rho_\Psi}^n \odot \min \{ \bm{k_{max}}^n,\, \bm{k_{abs}}^n, \,\bm{k_{square}}^n\} +\mu \bm{1}, \\
     \text{or} \quad {k_\Psi}^n_{i,j,k} &=& {\rho_\Psi}^n_{i,j,k} \min\{ k_{max\, i,j,k}^n, k_{abs\, i,j,k}^n, k_{square\, i,j,k}^n \} +\mu\,.
\end{eqnarray}
At each grid point (\ie for each $i$, $j$, $k$) the minimum of the three diffusion coefficients is taken. When solving for the velocity components $\bm{\rho_\Psi}^n=\bm{\rho}^n$ and when solving for the volume fraction field, $\bm{\rho_\Psi}^n=\bm{1}$. Also, if $\bm{\Psi}^n$ is a velocity component, $\mu$ is the dynamic viscosity and if $\bm{\Psi}^n$ is the volume fraction field, $\mu=0$ is the diffusivity. By taking the minimum value of these diffusion coefficients (Equation~\eqref{k_Psi-n}), it is hoped that the scheme will minimally intervene in the discretisation. 

\subsubsection{Compressive advection for interface-capturing with Petrov-Galerkin and extrema detecting for the volume fraction field} 
\label{compress-adv-k}
The positive Petrov-Galerkin diffusion can result in a smeared interface, so to counteract this, we introduce negative diffusion through another Petrov-Galerkin-based formulation. We use the approach outlined in~\citet{Pavlidis2016} (Equation~(33) in that paper) to sharpen up interfaces by adding negative diffusion to the advection equation for the volume fraction when this field has no extrema in a particular direction, which is determined by an oscillatory detecting variable. 

We now outline the new approach used to detect oscillations. Variables detecting oscillations in each direction with increasing~$x$, $y$ and~$z$, are calculated for linear $3\times3\times3$ filters $S_{i,j,k}^{x\, n}$, $S_{i,j,k}^{y\, n}$, $S_{i,j,k}^{z\, n}$ as follows: 
\begin{subequations}
\begin{eqnarray} 
    S_{i,j,k}^{x\, n} &=& 
(C_{i,j,k}^n - C_{i-1,j,k}^n)(C_{i+1,j,k}^n - C_{i,j,k}^n), \label{Sx-lin} \\
    S_{i,j,k}^{y\, n} &=& (C_{i,j,k}^n - C_{i,j-1,k}^n)(C_{i,j+1,k}^n - C_{i,j,k}^n)\,, \label{Sy-lin} \\
    S_{i,j,k}^{z\, n} &=& (C_{i,j,k}^n - C_{i,j,k-1}^n)(C_{i,j,k+1}^n - C_{i,j,k}^n)\,. \label{Sz-lin} 
\end{eqnarray} 
\end{subequations}
When quadratic and higher-order filters are used, the following oscillatory-detecting variables are calculated: 
\begin{subequations}
\begin{eqnarray}
    S_{i,j,k}^{x\, n} = \min\{ 
&\;&(C_{i-1,j,k}^n - C_{i-2,j,k}^n)(C_{i,j,k}^n - C_{i-1,j,k}^n), \nonumber\\
&\;&(C_{i,j,k}^n - C_{i-1,j,k}^n)(C_{i+1,j,k}^n - C_{i,j,k}^n), \label{Sx-quad}  \\
&\;&(C_{i+1,j,k}^n - C_{i,j,k}^n)(C_{i+2,j,k}^n - C_{i+1,j,k}^n) \;\;\; \}, \nonumber
\\
    S_{i,j,k}^{y\, n} = \min\{ 
&\;&(C_{i,j-1,k}^n - C_{i,j-2,k}^n)(C_{i,j,k}^n - C_{i,j-1,k}^n), \nonumber \\
&\;&(C_{i,j,k}^n - C_{i,j-1,k}^n)(C_{i,j+1,k}^n - C_{i,j,k}^n),  \label{Sy-quad}  \\
&\;&(C_{i,j+1,k}^n - C_{i,j,k}^n)(C_{i,j+2,k}^n - C_{i,j+1,k}^n) \;\;\; \},\nonumber
\\
    S_{i,j,k}^{z\, n} = \min\{ 
&\;&(C_{i,j,k-1}^n - C_{i,j,k-2}^n)(C_{i,j,k}^n - C_{i,j,k-1}^n), \nonumber \\
&\;&(C_{i,j,k}^n - C_{i,j,k-1}^n)(C_{i,j,k+1}^n - C_{i,j,k}^n), \label{Sz-quad} \\
&\;&(C_{i,j,k+1}^n - C_{i,j,k}^n)(C_{i,j,k+2}^n - C_{i,j,k+1}^n) \;\;\; \}\,. \nonumber
\end{eqnarray}
\end{subequations}
The condition $S_{i,j,k}^{x\, n}\geqslant 0$ indicates that there is no change of sign in gradient and thus no oscillation; $S_{i,j,k}^{x\, n}<0$ indicates that there is an oscillation.  Similar statements can be made for $S_{i,j,k}^{y\, n}$ and $S_{i,j,k}^{z\, n}$. Equations~\eqref{Sx-quad}, \eqref{Sy-quad} and~\eqref{Sz-quad} provide a much more rigorous approach to detecting an oscillation, locally, in a particular direction than the approach used in Equations~\eqref{Sx-lin}, \eqref{Sy-lin} and \eqref{Sz-lin}. In fact, the use of the latter can lead to significant oscillations when diffusion is introduced. 
In addition to using $S_{i,j,k}^{x\, n}<0$ for detecting an oscillation (with similar terms for~$y$ and~$z$), if $C^{n}_{i,j,k}>0.95$ or $C^{n}_{i,j,k}<0.05$, then an oscillation is also deemed to have been detected and thus the algorithm will apply positive Petrov-Galerkin diffusion and does this by setting $S_{i,j,k}^{x\, n}=-1$ $S_{i,j,k}^{y\, n}=-1$ $,S_{i,j,k}^{z\, n}=-1$. These  modifications to $S_{i,j,k}^{x\, n}$ $S_{i,j,k}^{y\, n}$ $,S_{i,j,k}^{z\, n}$ reduce the likelihood of the volume fraction field becoming greater than one or less than zero. The weights that determine how much positive (\eg $w_{i,j,k}^{x\, n+}$) and how much negative (\eg $w_{i,j,k}^{x\, n-}$) diffusion is applied, are calculated from:   
\begin{subequations}
\begin{align}
    w_{i,j,k}^{x\, n+} &= -\max\{-1, \min\{0,{\cal K}^+ S_{i,j,k}^{x\, n}\}\}, \;\;\; &
    w_{i,j,k}^{x\, n-} &= \min\{1, \max\{0,{\cal K}^- S_{i,j,k}^{x\, n} \}\}, \\
    w_{i,j,k}^{y\, n+} &= -\max\{-1, \min\{0,{\cal K}^+ S_{i,j,k}^{y\, n}\}\}, \;\;\; &
    w_{i,j,k}^{y\, n-} &= \min\{1, \max\{0,{\cal K}^- S_{i,j,k}^{y\, n}\}\}, \\
    w_{i,j,k}^{z\, n+} &= -\max\{-1, \min\{0,{\cal K}^+ S_{i,j,k}^{z\, n}\}\}, \;\;\; &
    w_{i,j,k}^{z\, n-} &= \min\{1, \max\{0,{\cal K}^- S_{i,j,k}^{z\, n}\}\}\,. 
\end{align}
\end{subequations}
The values of ${\cal K}^+ = 300$ and ${\cal K}^- = 300$ 
are used here which tend to mean that these weightings produce their extreme values of~0 and~1 most of the time. The negative diffusion coefficient used to maintain a sharp interface~\citep{Pavlidis2014,Pavlidis2016} is: 
\begin{eqnarray}
    k_{i,j,k}^{n-} = -\beta \frac{u_{i,j,k}^{2}+v_{i,j,k}^{2}+w_{i,j,k}^{2}}{\epsilon_{x}+((C_{x\,i,j,k}^n)^2+(C_{y\,i,j,k}^n)^2+(C_{z\,i,j,k}^n)^2)(|u_{i,j,k}^{n*}|+|v_{i,j,k}^{n*}|+|w_{i,j,k}^{n*}|)/3}, 
    \label{negativek}
\end{eqnarray}
where $\beta=0.1$; $\epsilon_{x} = 10^{-4}$; $C_{x\,i,j,k}^n$, $C_{y\,i,j,k}^n$ and $C_{z\,i,j,k}^n$ represent the gradient of volume fraction field in each direction at time level~$n$ (\eg $C_{xi,j,k}^n = \bm{C_x}^n\vert_{i,j,k} $); and the $*$ superscript denotes the normalisation operator for each velocity component which is defined by
\begin{eqnarray}
(u_{i,j,k}^{n*},v_{i,j,k,j,k}^{n*},w_{i,j,k}^{n*}) = (C_{xi,j,k}^n,C_{yi,j,k}^n,C_{zi,j,k}^n)
    \left( 
    \frac{u_{i,j,k}^n C_{xi,j,k}^n+v_{i,j,k}^n C_{yi,j,k}^n+w_{i,j,k}^n C_{zi,j,k}^n}{\epsilon_{x}+(C_{xi,j,k}^n)^2+(C_{yi,j,k}^n)^2+(C_{zi,j,k}^n)^2}
    \right)\,.
    \label{normalisation}
\end{eqnarray}
This negative diffusion coefficient is used to generate negative diffusion coefficients in~$x$, $y$ and~$z$ as follows: 
\begin{equation}
k_{i,j,k}^{x\, n-} = \frac{k_{i,j,k}^{n-} }{\Delta x}, \;\;\; \;\;
k_{i,j,k}^{y\, n-} = \frac{k_{i,j,k}^{n-} }{\Delta y}, \;\;\; \;\;
k_{i,j,k}^{z\, n-} = \frac{k_{i,j,k}^{n-} }{\Delta z} \,. 
\end{equation} 
An isotropic positive diffusion coefficient is also applied to reduce oscillations. To do this, Equation~\eqref{efficient-kabs} is used to obtain $k_{absi,j,k}^{n}$ which is multiplied by $\mathcal{K}$ to form:  
\begin{eqnarray}
    k_{i,j,k}^{x\, n+} =k_{i,j,k}^{y\, n+} =k_{i,j,k}^{z\, n+} = {\cal K} \, k_{absi,j,k}^{n}  \,,
    \label{positivek}
\end{eqnarray}
in which $\mathcal{K}=3$ (determined by trial and error) artificially elevates the Petrov-Galerkin diffusivity in order to control oscillations in the volume fraction field generated by negative diffusion, Equation~\eqref{negativek}. The diffusion coefficient for interface capturing becomes:
\begin{equation}
    k_{i,j,k}^{x\, n} = w_{i,j,k}^{x\, n+} k_{i,j,k}^{x\, n+} + w_{i,j,k}^{x\, n-} k_{i,j,k}^{x\, n-},
\end{equation}
and similarly for $k_{i,j,k}^{y\, n}$, $k_{i,j,k}^{z\, n}$. After stabilising the overall diffusion coefficient, the final anisotropic diffusion coefficients become:
\begin{equation}
    k_{i,j,k}^{x\, n} \leftarrow 
    \max\{-k_{min}, \min\{k_{max},k_{i,j,k}^{x\, n}\}\},
\end{equation}
where $k_{min}$ and $k_{max}$ are  determined based on the stability condition of explicit time-stepping schemes:
\begin{equation}\label{alpha_min_and_max}
k_{min} = \frac{\alpha_{min}}{\Delta t} \max\{\Delta x, \Delta y, \Delta z\}^{2} \,,  \qquad  k_{max} = \frac{\alpha_{max}}{\Delta t}\min\{\Delta x, \Delta y, \Delta z\}^{2} \,, 
\end{equation}
in which $\alpha_{min}=0.0001$, $\alpha_{max}=0.05$ has been determined by trial and error. If there are issues with dissipation of the interface (not the case within this paper), we have seen that reducing $\alpha_{max}$ to 0.005 will improve the results. Expressions similar to Equation~\eqref{alpha_min_and_max} are developed for $k_{i,j,k}^{y\, n}$ and $k_{i,j,k}^{z\, n}$.

\subsection{Solution method for the multiphase flow equations}\label{sec:solution_algorithm}
\begin{figure}[htp]
 \centering
 \includegraphics[scale=0.55]{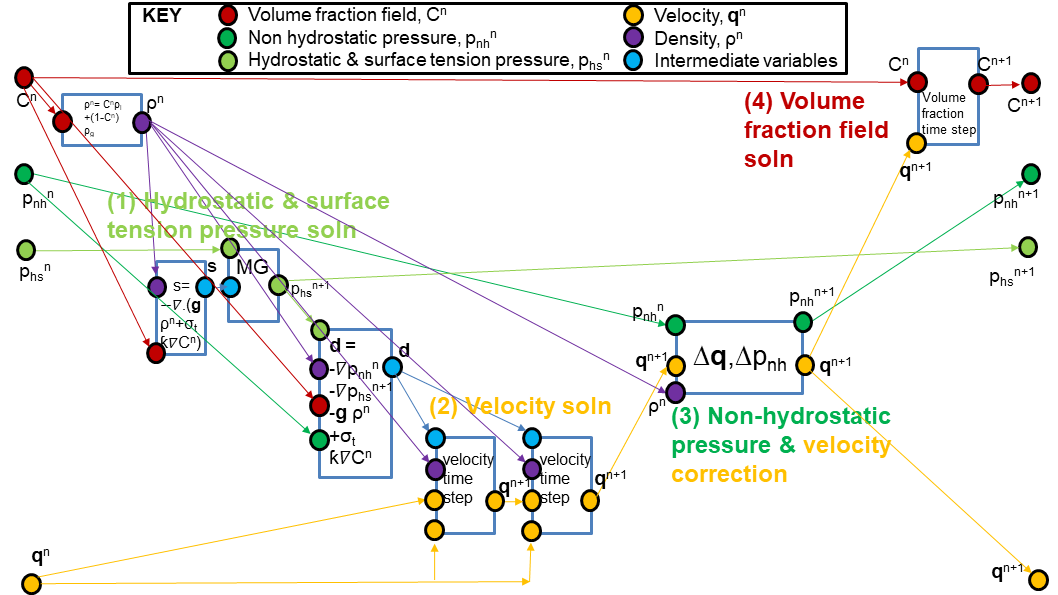}
 \caption{Schematic of the multiphase flow solver as a neural network containing the four stages of the overall solution method. See Figure~\ref{fig:NN-NS2} for a schematic of the velocity and the non-hydrostatic pressure correction.}
 \label{fig:NN-NS}
\end{figure}

This section describes the overall solution method for solving the Navier-Stokes and volume fraction field equations. Although these equations are highly coupled, a segregated solution method is applied in four stages. First, the hydrostatic and surface-tension pressure is found; second, a velocity solution is calculated; third, non-hydrostatic and non-surface-tension pressure is calculated and the velocity solution is corrected; and finally, an advection equation is solved for the volume fraction. Figure~\ref{fig:NN-NS} shows a diagrammatic representation of the multiphase solution method implemented as a neural network, where the four stages in the solution process are identified as detailed below. 

\begin{figure}
 \centering
 \includegraphics[scale=0.57,trim=0mm 7mm 0mm 0mm, clip]{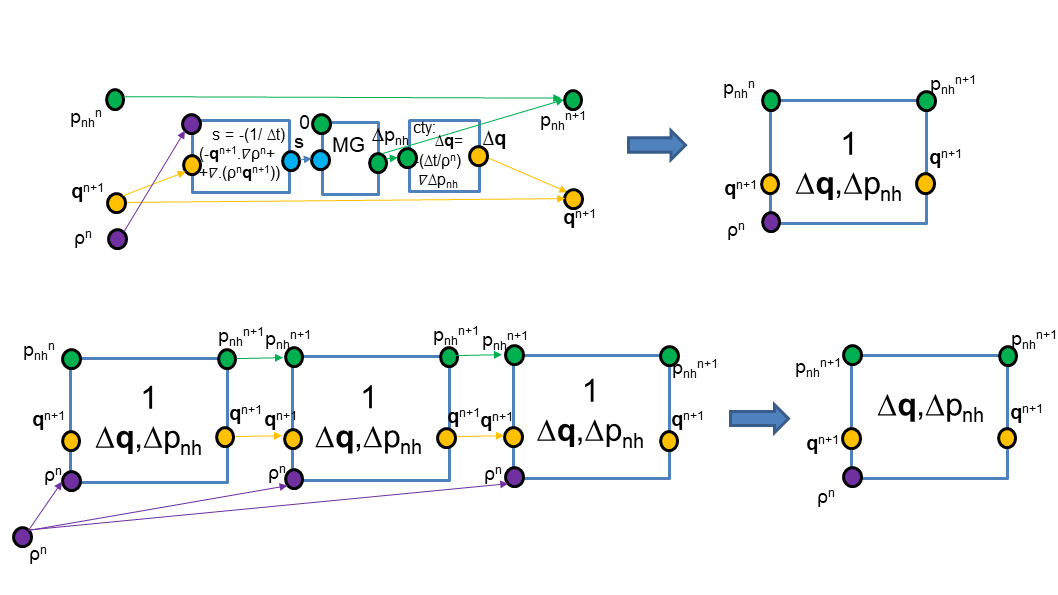}
 \caption{Schematic of the velocity and the non-hydrostatic pressure correction necessary to satisfy the continuity equation. Details of a single iteration of the pressure and velocity correction are shown in top left and top right, respectively. 
 Three of these iterations are combined to form an overall pressure and velocity correction to satisfy the continuity equation, the schematic of which is shown bottom right. (See Figure~\ref{fig:NN-NS} for a key of the colours). }
 \label{fig:NN-NS2}
\end{figure}

\begin{figure}[htp]
 \centering
\includegraphics[width=0.95\textwidth,trim=0mm 0mm 0mm 30mm, clip]
{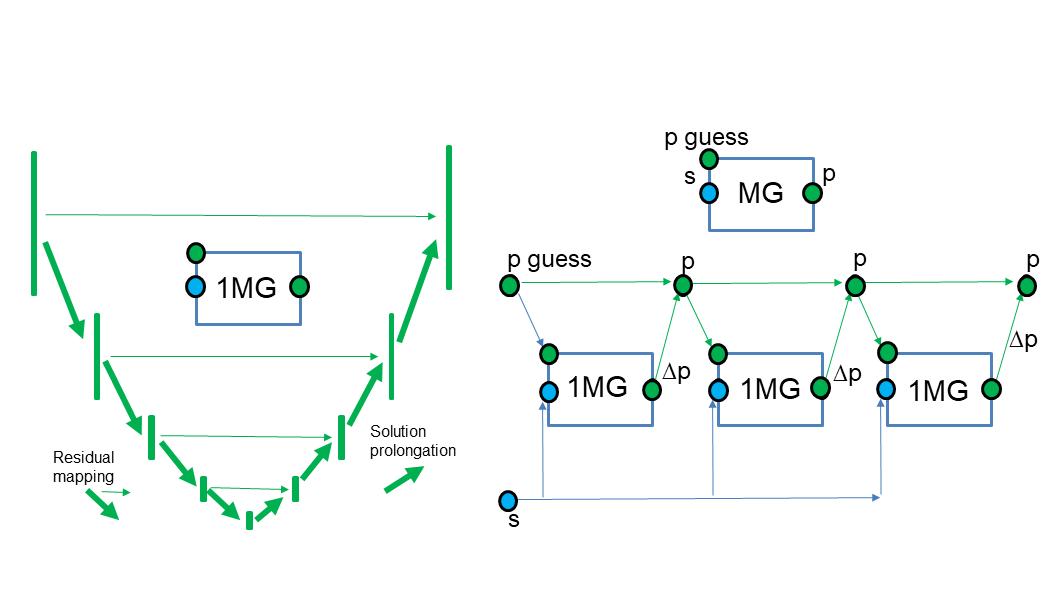}
\vspace{-1.0cm} 
 \caption{Saw-tooth multigrid method, based on a U-Net architecture, used to solve for the hydrostatic and surface-tension pressure and the non-hydrostatic non-surface-tension pressure correction.}
 \label{fig:multi-grid}
\end{figure}

\textbf{Stage 1: Hydrostatic and Surface-Tension Pressure Solution.} 
By taking the divergence of the terms on the right-hand side of Equation~\eqref{NS-eqn-disc-guess} (excluding the term involving $p_{nh}$) and setting this to zero, we obtain an equation for the hydrostatic and surface-tension pressure as follows: 
\begin{equation}\label{hydro-pressure}
    -\nabla \cdot  \nabla p_{hs}^{n+1} + \nabla \cdot  (\bm{s}_{q}^{n} + \bm{s}_{t}^{n}) = 0\,,
\end{equation}
where the buoyancy term is given in Equation~\eqref{eq:gravity} and the surface tension term is given by Equation~\eqref{surface-tension-uvw}. After discretising Equation~\eqref{hydro-pressure} in space, we obtain an expression for the residual of the pressure  
\begin{equation}\label{ph_disc}
\bm{r_{hs}}^{n+\frac{1}{2}}=-\bm{f}(\bm{{p_{hs}}}^{n+1} ;\wdiff) 
+ \bm{f}( \bm{s}_{\bm{t} u}^{n} ;{\bf{w_{x}}})
+ \bm{f}( \bm{s}_{\bm{t} v}^{n} ;{\bf{w_{y}}})+ \bm{f}( \bm{s}_{\bm{t} w}^{n}+ g\bm{\rho}^{n} ;{\bf{w_{z}}})\,, 
\end{equation}
which is forced to zero using the U-Net multigrid solver in order to calculate the hydrostatic pressure tensor $\bm{{p_{hs}}}^{n+1}$. Figure~\ref{fig:multi-grid} shows a schematic diagram that describes how this and the (non-hydrostatic non-surface-tension) pressure correction equation are solved using a saw tooth multigrid method using the U-Net architecture, see~\citep{Phillips2022-progress}. A fixed number of 20~multigrid cycles for each pressure equation solution is used in this paper, except where otherwise stated.  

\textbf{Stage 2: Velocity Solution.} 
With a surface tension coefficient for water of $\sigma_t=\SI{7.28e-2}{\newton\per\meter}$, the tensors representing surface tension, see Equation~\eqref{surface-tension-uvw}, are substituted in the right-hand side of the discrete residuals of the velocities (see Equations~\eqref{Petrov-u}, \eqref{Petrov-v} and~\eqref{Petrov-w}), together with the pressure gradients and buoyancy source. The solution for the velocities requires a two-step approach using the discrete sources 
${\bm{\tilde s_u}}^{n+\frac{1}{2}}$, 
${\bm{\tilde s_v}}^{n+\frac{1}{2}}$, 
${\bm{\tilde s_w}}^{n+\frac{1}{2}}$ (see Equations~\eqref{Petrov}) and starting with the best approximation to $\bm{u}^{n+\frac{1}{2}}$, $\bm{v}^{n+\frac{1}{2}}$, $\bm{w}^{n+\frac{1}{2}}$, that is $\bm{u}^{n+\frac{1}{2}}=\bm{u}^{n}$, $\bm{v}^{n+\frac{1}{2}}=\bm{v}^{n}$, $\bm{w}^{n+\frac{1}{2}}=\bm{w}^{n}$. By setting the residual to zero, applying the inverse identity filter and applying Hadamard entry-wise division, the approximate velocities at $n+1$ are given by: 
\begin{eqnarray}
        \bm{\tilde u}^{n+1} &=& \left( -\bm{f}( \bm{\tilde s_u}^{n+\frac{1}{2}} ; {\bf{w}_{ml}^{-1}} ) +\frac{1}{\Delta t} {\bm \rho}^n \odot {\bm{u}}^n \right) \oslash \left(\frac{1}{\Delta t}\bm{\rho}^n +\bm{\sigma}
        \right), \nonumber \\ 
        \bm{\tilde v}^{n+1} &=& \left( -\bm{f}( \bm{\tilde s_v}^{n+\frac{1}{2}} ; {\bf{w}_{ml}^{-1}} )+\frac{1}{\Delta t} {\bm \rho}^n \odot {\bm{v}}^n\right) \oslash \left(\frac{1}{\Delta t}\bm{\rho}^n +\bm{\sigma}
        \right) , \nonumber \\ 
        \bm{\tilde w}^{n+1} &=& \left( -\bm{f}( \bm{\tilde s_w}^{n+\frac{1}{2}} ; {\bf{w}_{ml}^{-1}} )+\frac{1}{\Delta t} {\bm \rho}^n \odot {\bm{w}}^n \right) \oslash \left(\frac{1}{\Delta t}\bm{\rho}^n +\bm{\sigma}
        \right) .
        \label{velocity-1}
\end{eqnarray} 
    The velocities at time level $n+\half$ are updated: 
    \begin{equation}
            \bm{u}^{n+\frac{1}{2}}=\frac{1}{2}(\bm{u}^{n}+\bm{\tilde u}^{n+1}), \quad 
            \bm{v}^{n+\frac{1}{2}}=\frac{1}{2}(\bm{v}^{n}+\bm{\tilde v}^{n+1}), \quad 
            \bm{w}^{n+\frac{1}{2}}=\frac{1}{2}(\bm{w}^{n}+\bm{\tilde w}^{n+1})\,.
    \end{equation}
A second iteration can be performed by recalculating the sources 
${\bm{\tilde s_u}}^{n+\frac{1}{2}}$, 
${\bm{\tilde s_v}}^{n+\frac{1}{2}}$, 
${\bm{\tilde s_w}}^{n+\frac{1}{2}}$ (Equation~\eqref{Petrov}) with the latest solutions at time levels~$n+1$ and~$n+\frac{1}{2}$ and using Equation~\eqref{velocity-1} again to calculate an improved estimate of the solutions at time level~$n+1$. In this work we use two iterations, although more could be done if desired. 

\textbf{Stage 3: Non-Hydrostatic Pressure and Velocity Correction.} 
In the following, the non-hydrostatic pressure and velocity correction equations are formed, see Figure~\ref{fig:NN-NS2}. 
To do this, Equation~\eqref{NS-eqn-disc-guess} is modified by replacing ${\bm{\tilde{q}}}^{n+1}$ and $\tilde{p}_{nh}^n$ with the variables that satisfy mass conservation at time level~$n+1$ (${\bm{q}}^{n+1}$ and ${p}_{nh}^{n+1}$) resulting in:
\begin{equation}\label{NS-eqn-disc}
\begin{split}
\rho^{n} \left(\frac{\bm{q}^{n+1} -\bm{q}^{n} }{\Delta t} 
+ 
u^{n+\frac{1}{2}}\frac{\partial \bm{q}^{n+\frac{1}{2}} }{\partial x} +  v^{n+\frac{1}{2}}\frac{\partial \bm{q}^{n+\frac{1}{2}} }{\partial y} +  w^{n+\frac{1}{2}}\frac{\partial \bm{q}^{n+\frac{1}{2}} }{\partial z} 
\right) + \sigma \bm{\tilde q}^{n+1} 
- \nabla \cdot({\mu}\nabla \bm{q}^{n+\frac{1}{2}}) \\ = -\nabla p_{hs}^{n+1} -\nabla p_{nh}^{n+1} + \bm{s}_{q}^{n} 
+ \bm{s}_{t}^{n}.  
\end{split}
\end{equation}
On subtracting Equation~\eqref{NS-eqn-disc-guess} from Equation~\eqref{NS-eqn-disc}, the following velocity correction equation is formed: 
\begin{equation}
     \frac{\rho^{n}}{\Delta t} \Delta \bm{q} = - \nabla \Delta p_{nh}, 
\label{vel-correction} 
\end{equation}
in which $\Delta \bm{q}=\bm{q}^{n+1} -\bm{\tilde{q}}^{n+1}$ 
and $\Delta p_{nh}=p_{nh}^{n+1}-\tilde{p}_{nh}^{n+1}$. By combining the equation for the volume fraction discretised in time (Equation~\eqref{transport-indicator-disc-in-time}) with the expression for density (Equation~\eqref{rho}), we obtain an equation for conservation of mass:
\begin{equation}\label{eq:density_derivation--}
    \frac{\rho^{n+1}-\rho^{n}}{\Delta t} +  \nabla \cdot \left( \bm{ q}^{n+1} \rho^{n}\right)  = 
    \frac{\rho^{n+1}-\rho^{n}}{\Delta t} + \bm{ q}^{n+1} \cdot \nabla \rho^{n}
    =0\, .
\end{equation}
Taking the divergence of Equation~\eqref{vel-correction} and substituting in Equation~\eqref{eq:density_derivation--} enables us to form the correction for non-hydrostatic non-surface-tension pressure (see \citet{Pavlidis2014,Pavlidis2016}):  
\begin{equation}
    - \nabla \cdot \nabla \Delta p_{nh} 
    = 
    \frac{1}{\Delta t} 
    \nabla\cdot (\rho^{n} \bm{q}^{n+1}-\rho^{n} \bm{\tilde q}^{n+1})
    = -\frac{1}{\Delta t} \left(\frac{\rho^{n+1}-\rho^{n}}{\Delta t}  + \nabla \cdot (\rho^{n} \bm{\tilde q}^{n+1})  \right). 
    \label{nonhydro-pressure1} 
\end{equation}
Approximating the mass conservation equation by using the best estimate of velocity ($\bm{\tilde q}^{n+1}$), 
\begin{equation}\label{eq:density_derivation}
    \frac{\rho^{n+1}-\rho^{n}}{\Delta t} =  - \bm{\tilde q}^{n+1}\cdot \nabla  \rho^{n} \, ,
\end{equation} 
means that we can replace the unknown density~$\rho^{n+1}$ in Equation~\eqref{nonhydro-pressure1}, resulting in the non-hydrostatic non-surface-tension pressure correction of 
\begin{equation}
   - \nabla \cdot \left(\nabla \Delta p_{nh}\right) 
   = -\frac{1}{\Delta t} \left(\nabla \cdot (\rho^{n}\bm{\tilde q}^{n+1}) - \bm{\tilde q}^{n+1}\cdot\nabla \rho^{n}   \right). 
    \label{nonhydro-pressure2}
\end{equation}

The fully discretised form of Equation~\eqref{nonhydro-pressure2} is:
\begin{eqnarray}
\bm{r_{nh}}^{n+\frac{1}{2}}
&=& -\bm{f}(\bm{\Delta {p_{nh}}} ;\wdiff) \nonumber \\
&-& \frac{1}{\Delta t}\left( \bm{f}( \bm{\rho}^{n} \odot \bm{\tilde u}^{n+1} ;{\bf{w_{x}}})+ 
\bm{f}( \bm{\rho}^{n} \odot \bm{\tilde v}^{n+1} ;{\bf{w_{y}}}) 
+ \bm{f}( \bm{\rho}^{n} \odot \bm{\tilde w}^{n+1} ;{\bf{w_{z}}}) \;\right) 
\nonumber \\
&+& \frac{1}{\Delta t}\left( \bm{\tilde u}^{n+1} \odot \bm{f}( \bm{\rho}^{n} ;{\bf{w_{x}}})+ 
\bm{\tilde v}^{n+1} \odot \bm{f}( \bm{\rho}^{n} ;{\bf{w_{y}}})
+\bm{\tilde w}^{n+1} \odot \bm{f}( \bm{\rho}^{n} ;{\bf{w_{z}}})
\;\right), 
  \label{ph_disc2}
\end{eqnarray}
in which $\bm{\tilde u}^{n+1}$,  $\bm{\tilde v}^{n+1}$, $\bm{\tilde w}^{n+1}$
are the estimates for $\bm{ u}^{n+1}$, $\bm{v}^{n+1}$, $\bm{w}^{n+1}$ obtained at the end of Stage~2.  
The residual in Equation~\eqref{ph_disc2} is forced to zero using the U-Net multigrid solver to calculate the non-hydrostatic pressure correction tensor $\bm{\Delta {p_{nh}}}$. Although the right-hand sides of Equations~\eqref{nonhydro-pressure1} and~\eqref{nonhydro-pressure2} are identical in the continuum, at the discrete level, they are very different. For instance, the sign of the implied diffusion (realised through the Petrov-Galerkin diffusion introduced into the equations for $\bm{C}^{n+1}$) is reversed. 

Equation~\eqref{vel-correction} is then discretised in space to form an equation for the velocity corrections: 
\begin{eqnarray}
    \bm{\Delta u} &=& -\Delta t \left( \bm{f}\left( \bm{f}( \bm{\Delta p_{nh}} ;{\bf{w_{x}}}); {\bf{w_{ml}^{-1}}} \right) \oslash \bm{\rho}^{n}\right),\nonumber\\
    \bm{\Delta v} &=& -\Delta t \left( \bm{f}\left( \bm{f}( \bm{\Delta p_{nh}} ;{\bf{w_{y}}}); {\bf{w_{ml}^{-1}}} \right) \oslash \bm{\rho}^{n} \right), \label{vel-cor}\\
    \bm{\Delta w} & =& -\Delta t \left( \bm{f}\left( \bm{f}( \bm{\Delta p_{nh}} ;{\bf{w_{z}}}); {\bf{w_{ml}^{-1}}} \right) \oslash \bm{\rho}^{n} \right). \nonumber
\end{eqnarray}
Finally, the non-hydrostatic non-surface-tension pressure and velocities are updated:
\begin{equation*}
    \bm{p_{nh}}^{n+1} = \bm{p_{nh}}^n + \bm{\Delta p_{nh}} \quad\text{ and } \quad
    \begin{array}{lcl}
      \bm{u}^{n+1} &\leftarrow& \bm{u}^{n+1} + \bm{\Delta u}\,, \\
      \bm{v}^{n+1} &\leftarrow& \bm{v}^{n+1} + \bm{\Delta v}\,, \\
      \bm{w}^{n+1} &\leftarrow& \bm{w}^{n+1} + \bm{\Delta w}\,.
      \end{array}
\end{equation*}
To obtain improved mass conservation, one can repeat this process of solving Equations~\eqref{ph_disc2} and~\eqref{vel-cor}, and updating the non-hydrostatic non-surface-tension pressure and velocities at time level~$n+1$. Here this process is carried out just twice, see Figure~\ref{fig:Multigrid_iteration} . 

\textbf{Stage 4: Volume Fraction Solution.} Solve for volume fraction field based on interface tracking with compressive advection. Setting the residual in Equation~\eqref{Petrov-C} to zero and rearranging, gives 
  \begin{equation}
    \bm{C}^{n+1} = -\Delta t\bm{f}( 
    \bm{\tilde{s}_C}^{n+\frac{1}{2}} ; {\bf{w}_{ml}^{-1}} ) + \bm{C}^{n}\,,
    \label{transport-indicator-1}
  \end{equation}
  in which ${\bf{w}_{ml}}=m_l {\bf{E}}$ and ${\bf{w}_{ml}^{-1}}={m_l}^{-1} {\bf{E}}$, where ${\bf{E}}$ has values $\text{E}_{i,j,k}=1$ when $i=j=k=0$, zero otherwise. 

\subsection{Computational speed of the approach} 
The following equation for the time it takes to run the multiphase code for a given simulation seems to hold (to within 10\%) for the simulations performed here: 
  \begin{equation}
\text{Wall clock time of simulation in seconds} = 
 \text{p} \mathcal{T} \mathcal{N} \left[  M_s + ( N_h+ N_I N_{nh}) H_s \right], 
    \label{speed}
  \end{equation}
in which $\mathcal{N}=N_x\times N_y\times N_z$ is the number of nodes or grid points, $\mathcal{T}$ is the number of time-steps in the simulation, $\text{p}$ is the ConvFEM polynomial order, $N_h$ is the number of multigrid iterations for the hydrostatic pressure, $N_{nh}$ is the number of multigrid iterations for the non-hydrostatic pressure, and $N_I$ is the number of pressure-velocity correction iterations. The coefficients $M_s$ and $H_s$ depend on the computer being used: $M_s$~is associated with the cost of solving for momentum and volume fraction, and $H_s$ is associated with the cost of a multigrid solve. For the NVIDIA A10 Tensor Core GPU and NVIDIA RTX A5000, the value of these coefficients are $M_s=\SI{1.2e-8}{\second}$ and $H_s=\SI{1.1e-10}{\second}$, and $M_s=\SI{1.37e-8}{\second}$ and $H_s=\SI{1.3e-10}{\second}$, respectively. Also when the surface tension calculation is used, $M_s$ decreases by 1.12\% and $H_s$ increases by 14\%. For example, model~5 from Table~\ref{tab:numerical_detail} uses quadratic elements ($\text{p}=2$), $\mathcal{N}=\num{67e6}$ nodes, $\mathcal{T}=\num{4e4}$ time-steps, $N_h=N_{nh}=20$ multigrid iterations, $N_I=2$ correction iterations and, thus, according to Equation~\eqref{speed}, will take \SI{24.42}{hours} on the NVIDIA A10 Tensor Core GPU, which compares well with the actual run time for this simulation of \SI{23.04}{hours}.  

Table~\ref{tab:compare_speed} reports computational speeds (wall time per time step) of multiphase flow codes quoted in the literature for a range of computer systems and compares this with estimated timings of NN4PDEs when running on one NVIDIA A10 Tensor Core GPU. The reference and computer architecture that was used is given in the first column (where $8 \times GPU$ indicates that 8 GPUs were used). The second column gives the number of nodes in the simulations carried out in each reference. The wall time obtained from each reference is shown in the third column. In the fourth column we adjust the wall times to take account of when the references used more than one CPU or GPU. We assume perfect scaling and multiply the quoted time by~512 in the case where 512~CPUs were used, for example. In order to compare the speed of our code with these references without setting up each of their cases and methods, we use Equation~\eqref{speed} to adjust our computational speed to the size of problem (number of nodes) given in each reference. In this calculation, we assume linear elements ($\text{p}=1$) and $N_h=N_{nh}=20$ multigrid iterations. Hence, we have a way of estimating how long our code would take to run the problems in the cited references. In comparison to the reference studies \citep{greaves2006simulation,nguyen2020efficient,nguyen2018novel} (running on one CPU) and \citep{Bryngelson2021} (running on 100 CPUs), our code demonstrates a significant speed-up when executed on one GPU, running at least \num{2000} times faster than the other codes. Moreover, we notice that NN4PDEs has a comparable performance ($\SI{1.625}{\second}$) to the reference work \citep{Crialesi-Esposito2023} (\SI{1.528}{\second}), despite using a less powerful GPU (the A10) than their GPU (the A100). It has been reported that the A100 is at least twice as capable as the A10 for inference tasks~\citep{baseten_blog}.

\begin{table}[htbp]
\centering
\begin{tabular}{lllll}
\toprule
Computer architectures  & Number of & \multicolumn{2}{c}{Wall time per time step} & Estimated wall time per \\
\cmidrule{3-4}
 & nodes/cells & as quoted & estimate for single CPU/GPU & time step for NN4PDEs \\
\toprule
1 $\times$ CPU~\citep{greaves2006simulation}  & \num{16384}  &   \SI{10.50}{\second} & \phantom{estimate }\SI{10.50}{\second} & \SI{0.609}{\milli\second} \\
1 $\times$ CPU~\citep{nguyen2020efficient}    & \num{108960}  & \SI{11.12}{\second} & \phantom{estimate }\SI{11.12}{\second} & \SI{4.02}{\milli\second} \\
1 $\times$ CPU~\citep{nguyen2018novel}        & \num{180000}   & \SI{14.19}{\second} & \phantom{estimate }\SI{14.19}{\second} & \SI{6.71}{\milli\second} \\
8 $\times$ GPU~\citep{Crialesi-Esposito2023}        & $512^{3}$  & $\SI{0.191}{\second}$ & \phantom{estimate }\SI{1.528}{\second}  & $\SI{1.625}{\second}$ \\
512 $\times$ CPU~\citep{Crialesi-Esposito2023}    & $512^{3}$  & $\SI{1.075}{\second}$ &\phantom{estimate } \SI{550.4}{\second}  & $\SI{1.625}{\second}$ \\
100 $\times$ CPU~\citep{Bryngelson2021}        & $500^{3}$  & \SI{220}{\second} & \phantom{estimate }\SI{22 000}{\second}  & $\SI{1.514}{\second}$ \\
8 $\times$ GPU~\citep{Radhakrishnan2023}   & \num{64000000}   & \SI{0.273}{\second} & \phantom{estimate }\SI{2.184}{\second}  & $\SI{0.775}{\second}$ \\
\bottomrule
\end{tabular}
\caption{Comparison of the computational speed of numerical models found in the literature with the estimated computational speed of NN4PDEs. One single CPU is used in~\citep{greaves2006simulation,nguyen2020efficient,nguyen2018novel}; 8~A100 GPUs and 512~Intel Xeon Gold~6130 CPUs are tested in~\citep{Crialesi-Esposito2023}; 100 AMD Opteron CPUs are used in~\citep{Bryngelson2021}; and 8~V100 GPUs are used in~\citep{Radhakrishnan2023}. The wall time for each reference refers to the overall timing for complete architectures (\eg the simulation on 8 GPUs takes $\SI{0.191}{\second}$, and the equivalent time that this is estimated to take on a single GPU is 8$\times$ $\SI{0.191}{\second} = $\SI{1.528}{\second}). The wall time for NN4PDEs to solve the same problem (with one time step) on a single GPU is estimated using Equation~\eqref{speed}.}
\label{tab:compare_speed}
\end{table}

\section{Results}\label{results}
In this section, the proposed multiphase flow solver is validated against a number of classic CFD problems: collapsing water columns (in 2D and 3D), and a rising water bubble. All the models, presented here, were run on two types of single GPU: NVIDIA A10 Tensor Core GPU and NVIDIA RTX A5000. 

\subsection{Collapsing water column}\label{results:cwc}
A column of water is subjected to gravity ($g = \SI{9.81}{m/s^{2}}$), leading to the collapse of the column towards a flat floor. Two fluids (water and air) are considered with different densities, $\rho_{\text{water}} = \SI{1000}{kg/m^{3}}$ and $\rho_{\text{air}} = \SI{1}{kg/m^{3}}$. The problem is initialised by setting the volume fraction of the fluid to be zero for water and one for air. Free-slip boundary conditions are imposed on the sides and bottom of the computational domain. A zero pressure boundary condition is applied to the hydrostatic and non-hydrostatic pressure at the top of the domain. Neumann boundary conditions (\ie zero derivative) are applied for both volume fraction and density fields on all surfaces of the domain. 

Eight different numerical model with various grid sizes and discretisation orders were applied to this problem in both 2D (Model 1) and 3D (Models 2--8), as summarised in Table~\ref{tab:numerical_detail}. For all models, a uniform grid spacing is used to generate structured grids, see grid size in Table~\ref{tab:numerical_detail}. Various orders of discretisation are used here (linear, quadratic and cubic ConvFEM) as shown in Table~\ref{tab:disc_combinations}. Model~7 is similar to Models~2, 4, 5 and~6 as regards grid size (width and height) and time step, but has a depth of the domain that is 4 times larger, in order to investigate 3D behaviour of the collapsing column. To ensure that the Courant number is below 0.1, a fixed time step is applied of $\SI{0.5}{ms}$ (Model 1), $\SI{0.025}{ms}$ (Models 2--7) and $\SI{0.1}{ms}$ (Model 8). 
\begin{table}[htbp]
    \centering
    \begin{tabular}{cc@{$\times$}c@{$\times$}crr@{.}lr@{.}lcll}
    \toprule
    Model & \multicolumn{4}{c}{Number of grid points}  & \multicolumn{2}{c}{Grid size} & \multicolumn{2}{c}{Time step} & Obstacle & \multicolumn{1}{c}{Discretisation} & \multicolumn{1}{c}{Figure}\\
    \cmidrule{2-5}
     & $N_x$ & $N_y$ & $N_z$ & \multicolumn{1}{c}{ nodes} & \multicolumn{2}{c}{(mm)}  &  \multicolumn{2}{c}{(ms)}  & & \\
    \toprule
    1 & 512 & \multicolumn{2}{l}{\hspace{-1mm}512}  & 262,144 & 1&0 & \qquad0&5 & no & linear & Figure~\ref{fig:validation_wc}\\ 
    2 & 1024 & 64 & 1024 & 67,108,864 & 0&5 & 0&025 & no & linear & Figures~\ref{fig:Interface_no_obstacle_L},\ref{fig:Indicator_no_obstacle_L}\\
    3   & 1024 & 64 & 1024 & 67,108,864 & 0&5 & 0&025 & yes & linear & Figure~\ref{fig:water_column_obstacle_1024_64_1024}\\
    4    & 1024 & 64 & 1024 & 67,108,864 & 0&5 & 0&025 & no & quadratic & Figures~\ref{fig:Interface_no_obstacle_Q},\ref{fig:Indicator_no_obstacle_Q}\\
    5   & 1024 & 64 & 1024 & 67,108,864 & 0&5 & 0&025& no & linear$/$quadratic & Figure~\ref{fig:Interface_no_obstacle_Q_LC},\ref{fig:Indicator_no_obstacle_Q_LC}\\
    6    & 1024 & 64 & 1024 & 67,108,864 & 0&5 & 0&025 & no & linear$/$cubic & Figures~\ref{fig:Interface_no_obstacle_C_LC},\ref{fig:Interface_no_obstacle_C_LC}\\    
    7     & 1024 & 256 & 1024 & 268,435,456 & 
    0&5 &  0&025 & no & linear$/$quadratic & Figure~\ref{fig:water_column_1024_256_1024}\\
    8   & 512 & 512 & 512 & 134,217,728 & 1&0 & 0&1 & no & linear$/$quadratic  & Figures~\ref{fig:Iso_surface_cubic_water_column},\ref{fig:Indicator_cubic_water_column}\\
    \bottomrule
    \end{tabular}
    \caption{Details of eight different collapsing water column models simulated by the proposed NN4PDEs approach. The numbers of grid nodes listed do not include the halo nodes through which the boundary conditions were applied. Table~\ref{tab:disc_combinations} explains the mixed formulations (\eg linear/quadratic, linear/cubic etc) that are listed here in the ``ConvFEM discretisation'' column. The computational domain can be calculated by multiplying the number of nodes in each direction by the grid size, or looking at Tables~\ref{results:cwc:section:3d:Table:Dimensions} and Figure~\ref{fig:Computation_domain_water_column}.}
    \label{tab:numerical_detail}
\end{table}

\begin{table}[htbp]
\centering
\begin{tabular}{lll}
\toprule
description       & volume fraction (diffusion term) & all other terms \\
\toprule
linear            & linear    & linear \\
quadratic         & quadratic & quadratic \\
linear/quadratic  & linear    & quadratic \\
linear/cubic      & linear    & cubic \\
\bottomrule
\end{tabular}
\caption{Discretisations used in the various models given in Table~\ref{tab:numerical_detail}.} 
\label{tab:disc_combinations}
\end{table}
\subsubsection{2D collapsing water column}\label{results:cwc:section:2d}
\begin{figure}[htp]
    \centering
    \subfloat[Evolution of the non-dimensional front $x^{*}$ in time]{\includegraphics[scale=0.45]{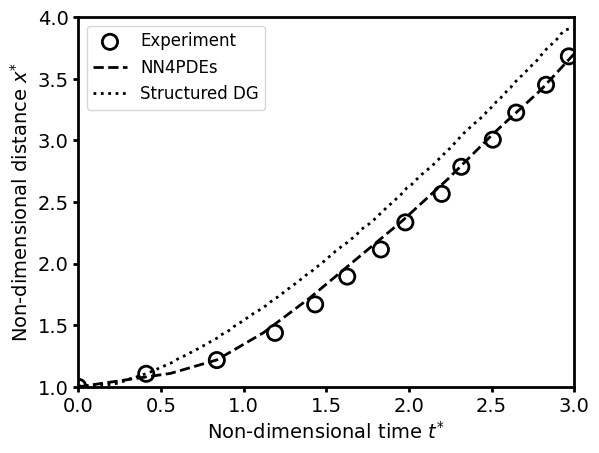}\label{fig:val_width}}
    \quad
    \subfloat[Evolution of the non-dimensional height of the water column $y^{*}$ in time]{\includegraphics[scale=0.45]{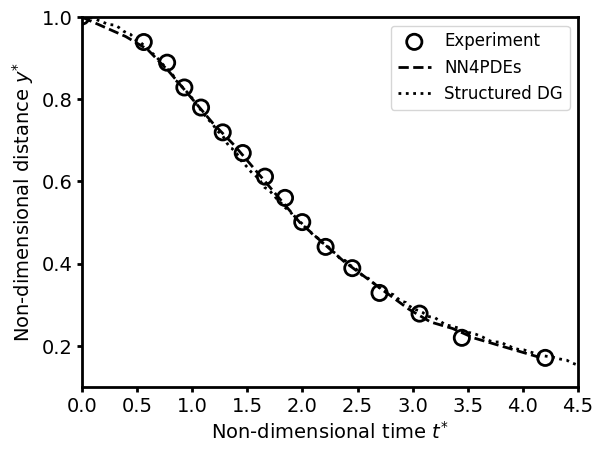}\label{fig:val_height} }
    \caption{\label{fig:validation_wc} Results for the collapsing water column from NN4PDEs predicted using Model~1 (see Table~\ref{tab:numerical_detail}). For comparison,  experimental data~\citep{yeoh2009assessment} and numerical results~\citep{Pavlidis2016} for a 2D version of this test case are also shown.}
\end{figure}

In order to validate the accuracy of our simulator when predicting the evolution of an initial column of water in 2D, Model~1 is adopted (see Table~\ref{tab:numerical_detail}). The results are compared with the experimental measurements by \citet{yeoh2009assessment}, see also \citet{martin1952part}. The dimensions of the 2D domain are $\SI{0.512}{m} \times \SI{0.512}{m}$ in the $x$~and $y$~directions, respectively. A column of water is initialised with width of $w_{0}= \SI{0.064}{m}$ ($x$ direction) and height of $h_{0}=\SI{0.128}{m}$ ($y$ direction). As the water column collapses due to gravity (imposed in the negative $y$-direction), the water front advances over the flat floor and the height of the column decreases in time. The initial column width ($w_{0}$) and height ($h_{0}$) are applied to normalise distances in the $x$~and $y$~directions ($x^{*}=w/w_{0}$ and $y^{*}=h/h_{0}$). The evolution of the initial stages of the water column collapse is shown by plotting the progress of the normalised front ($x^{*}$) and height ($y^{*}$) of the column through time ($t^{*}$) in Figure~\ref{fig:validation_wc}, demonstrating excellent agreement with experimental data (taken from Figure~7 within~\citet{yeoh2009assessment}). In Figure~\ref{fig:validation_wc} we also plot numerical results from a Discontinuous Galerkin (DG) formulation~\citep{Pavlidis2016} solved on a 2D structured mesh. The normalised distance $x^{*}$ predicted by NN4PDEs is closer to the experimental data than the DG model. The predicted normalised height of the water column, $y^{*}$, is similar for both numerical models. For the column front (Figure~\ref{fig:val_width}), time is normalised by $\sqrt{2g/w_{0}}$ and for the column height (Figure~\ref{fig:val_height}), time is normalised by $\sqrt{2g/h_{0}}$, as done by~\citet{yeoh2009assessment}. 

\subsubsection{3D collapsing water column}\label{results:cwc:section:3d}
\begin{figure}[htbp]
    \centering
    \subfloat[rectangular domain without obstacle]{\includegraphics[scale=0.52]{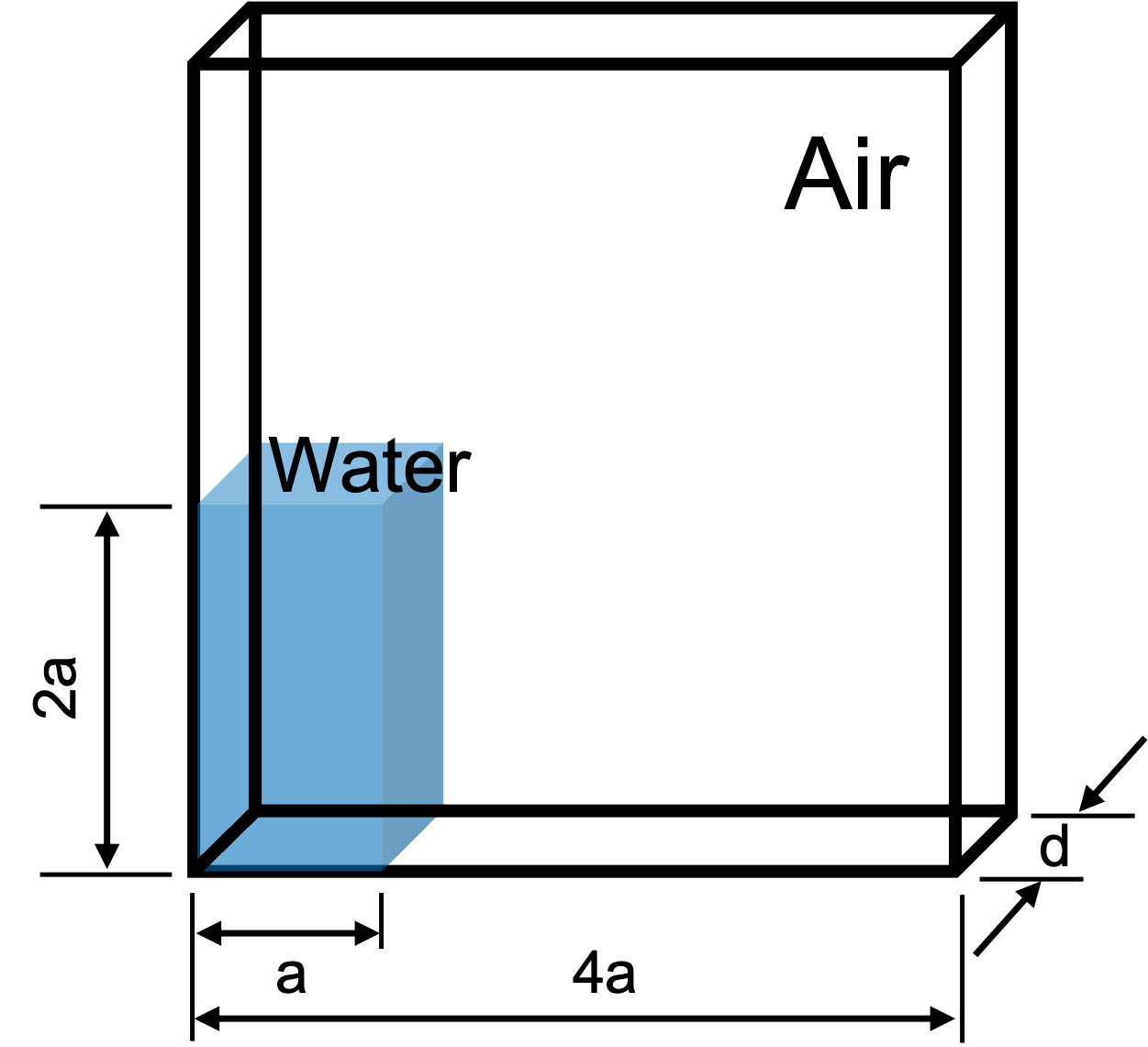}\label{fig:domain2}}
    \quad
    \subfloat[rectangular domain with obstacle]{\includegraphics[scale=0.52]{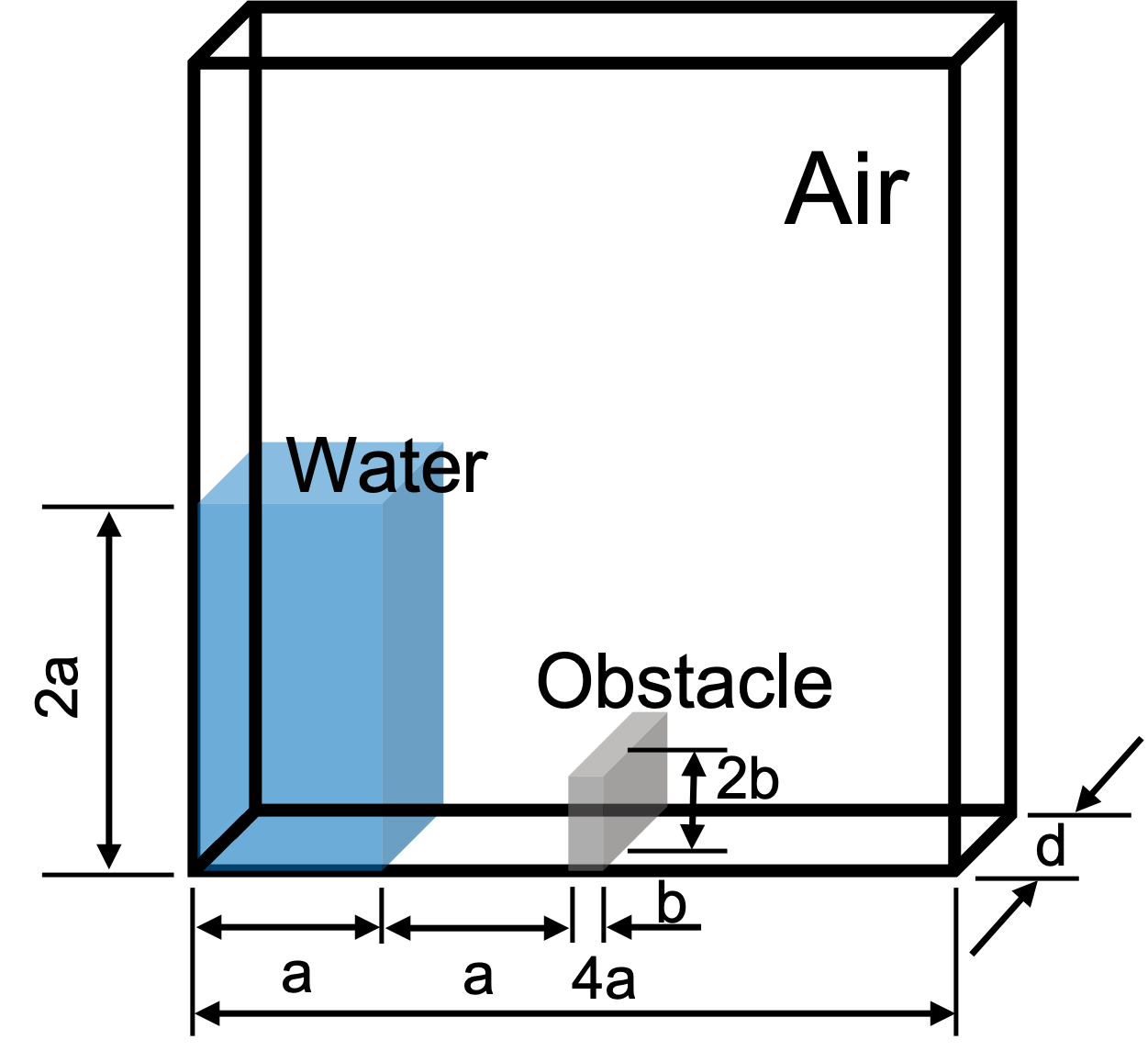}\label{fig:domain3}}
    \quad
    \subfloat[cubic domain without obstacle]{\includegraphics[scale=0.56]{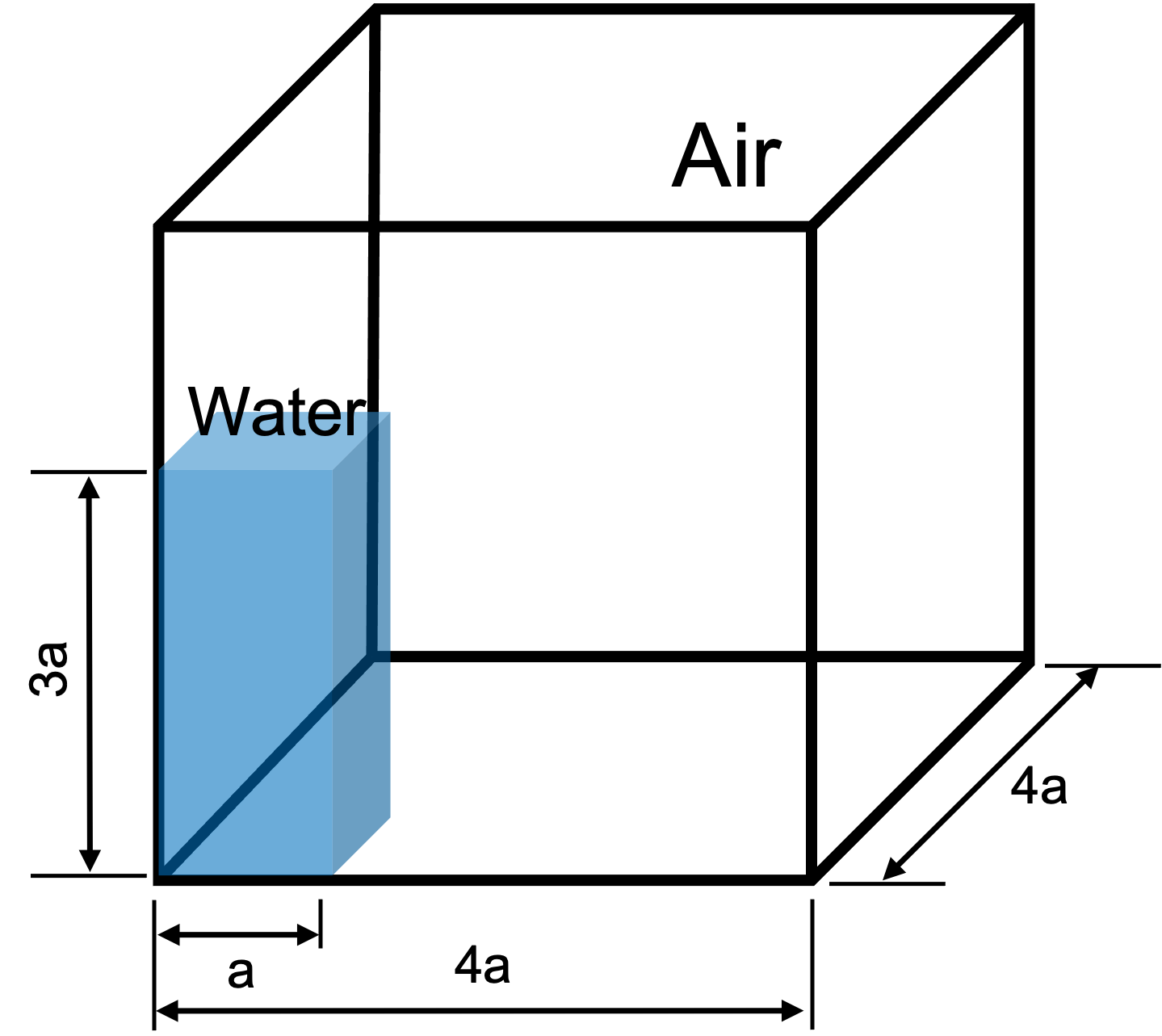}\label{fig:domain1}}
    \caption{Schematic diagrams of three different scenarios used in the collapsing water column tests. The values of \textsf{a}, \textsf{b} and \textsf{d} are \SI{0.128}{m}, \SI{0.024}{m} and \SI{0.032}{m}, respectively. For all models, the height of the domains is 4\textsf{a}. For Model~7, not shown here, the length of the edges in~$x$ and~$y$ are 4\textsf{a} and 4\textsf{d}. See also Table~\ref{results:cwc:section:3d:Table:Dimensions}.}
    \label{fig:Computation_domain_water_column}
\end{figure}

In this section, three collapsing water column scenarios are investigated, for which the dimensions of both the computational domain and the water column vary (see Table~\ref{results:cwc:section:3d:Table:Dimensions} and Figure~\ref{fig:Computation_domain_water_column}). The force due to gravity is imposed in the negative $z$ direction (vertical), the $x$ direction is horizontal and the $y$ direction is into the page.
\begin{table}[htbp]
\centering
    \begin{tabular}{lccc}
        \toprule
Test case &  Figure~\ref{fig:domain2}   &  Figure~\ref{fig:domain3}     & Figure~\ref{fig:domain1} \\
\midrule
 Domain & $\SI{0.512}{m} \times \SI{0.032}{m} \times \SI{0.512}{m}$ & $\SI{0.512}{m} \times \SI{0.032}{m} \times \SI{0.512}{m}$ & $\SI{0.512}{m} \times \SI{0.512}{m} \times \SI{0.512}{m}$\\
Water         & $\SI{0.128}{m} \times \SI{0.032}{m} \times \SI{0.256}{m}$ & $\SI{0.128}{m} \times \SI{0.032}{m} \times \SI{0.256}{m}$ & $\SI{0.128}{m} \times \SI{0.128}{m} \times \SI{0.384}{m}$ \\
Obstacle              &   ---                   & $\SI{0.024}{m} \times \SI{0.032}{m} \times \SI{0.048}{m}$ &   ---               \\
\bottomrule
\end{tabular}
    \caption{Collapsing water column: size of computational domains and water columns used in three scenarios of the 3D collapsing water columns. See Figure~\ref{fig:Computation_domain_water_column}.}\label{results:cwc:section:3d:Table:Dimensions}
\end{table}

\begin{figure}[htbp]
    \centering
     \subfloat[Linear (convolutional) filters (Model~2).]{\includegraphics[scale=0.32]{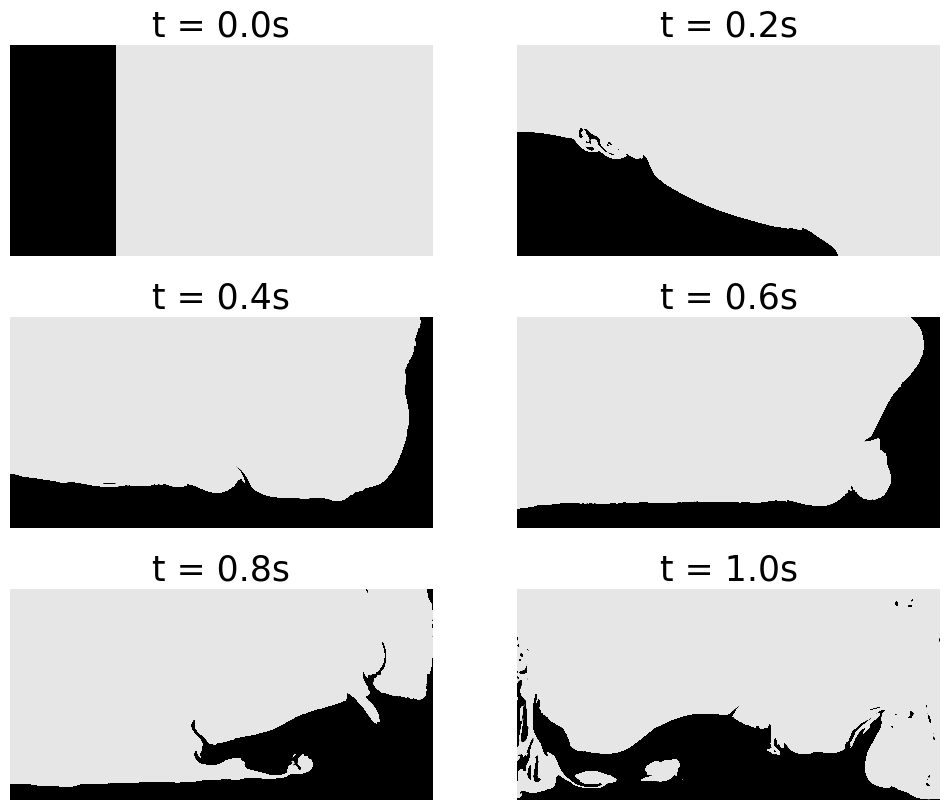}\label{fig:Interface_no_obstacle_L} }
     \quad
     \subfloat[Quadratic convolutional filters (Model~4).]{\includegraphics[scale=0.345]{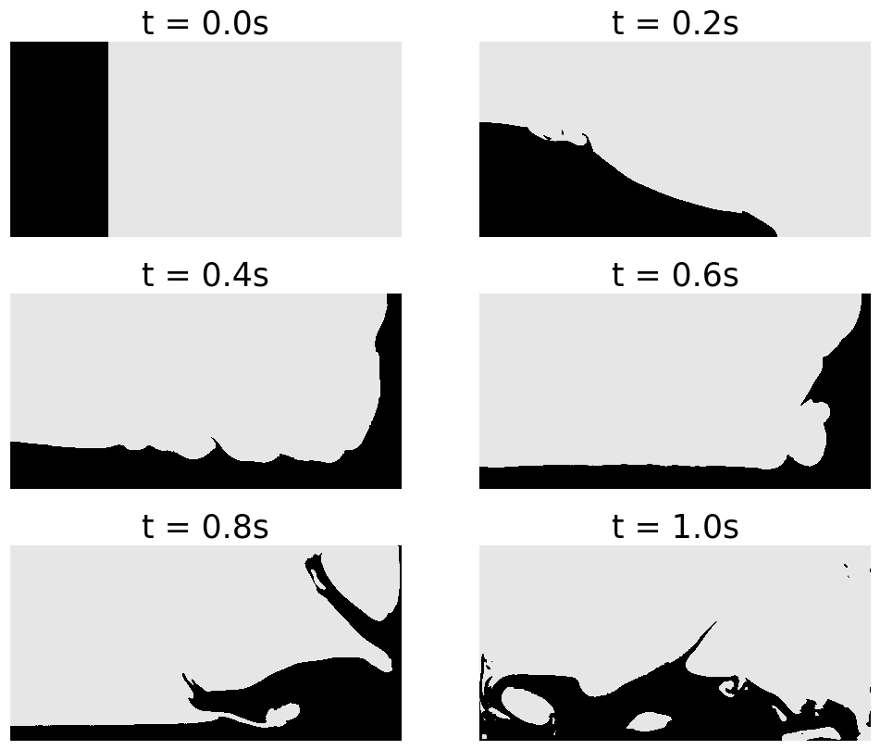}\label{fig:Interface_no_obstacle_Q} }
     \quad
    \subfloat[Linear filters for volume fraction diffusion, quadratic filters for all other terms (Model~5).]{\includegraphics[scale=0.32]{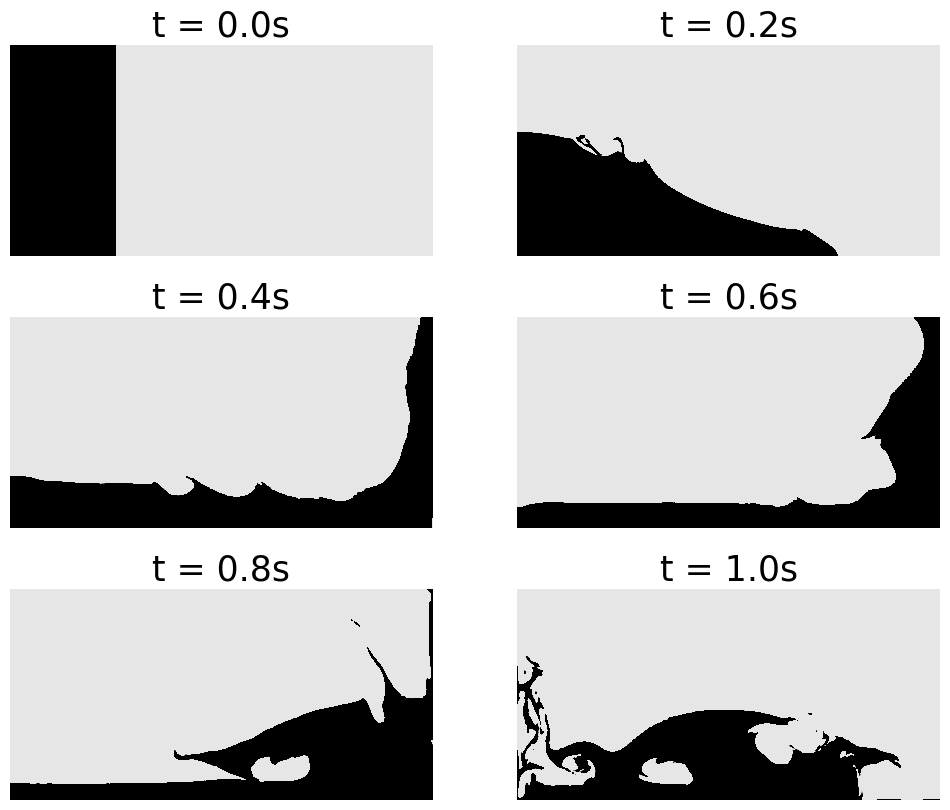}\label{fig:Interface_no_obstacle_Q_LC} }
     \quad
     \subfloat[Linear filters for volume fraction diffusion, cubic filters for all other terms (Model~6).]{\includegraphics[scale=0.32]{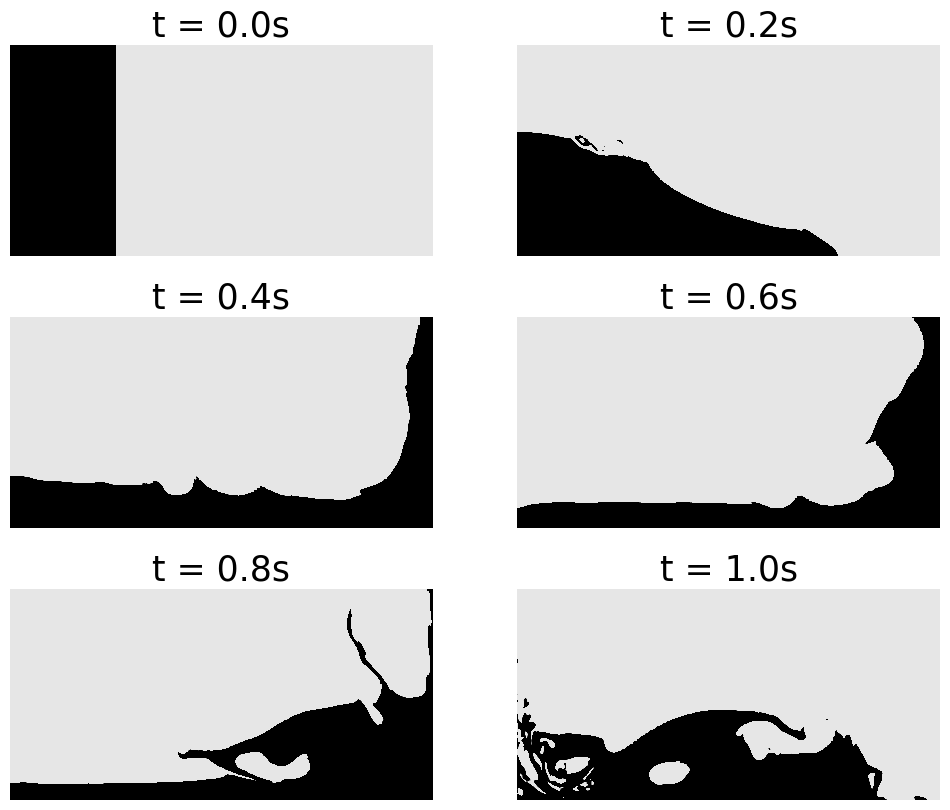}\label{fig:Interface_no_obstacle_C_LC}}
    \caption{Snapshots of water-air interface for the 3D collapsing water column without an obstacle from $t = \SI{0}{s}$ to $t = \SI{1.0}{s}$ with $\SI{0.2}{s}$ intervals and solved on a grid of 1024$\times$64$\times$1024 nodes. Volume fraction values greater than 0.5 are shown in black and where less than 0.5 is coloured grey. The resulting fields are shown on an~$xz$ plane at the midpoint of the extent of the domain in the $y$~direction. See Table~\ref{tab:numerical_detail} for details.}
    \label{fig:density_water_column_1024_64_1024}
\end{figure}

\begin{figure}
    \centering
     \subfloat[Linear convolutional filters (Model~2).]{\includegraphics[scale=0.32]{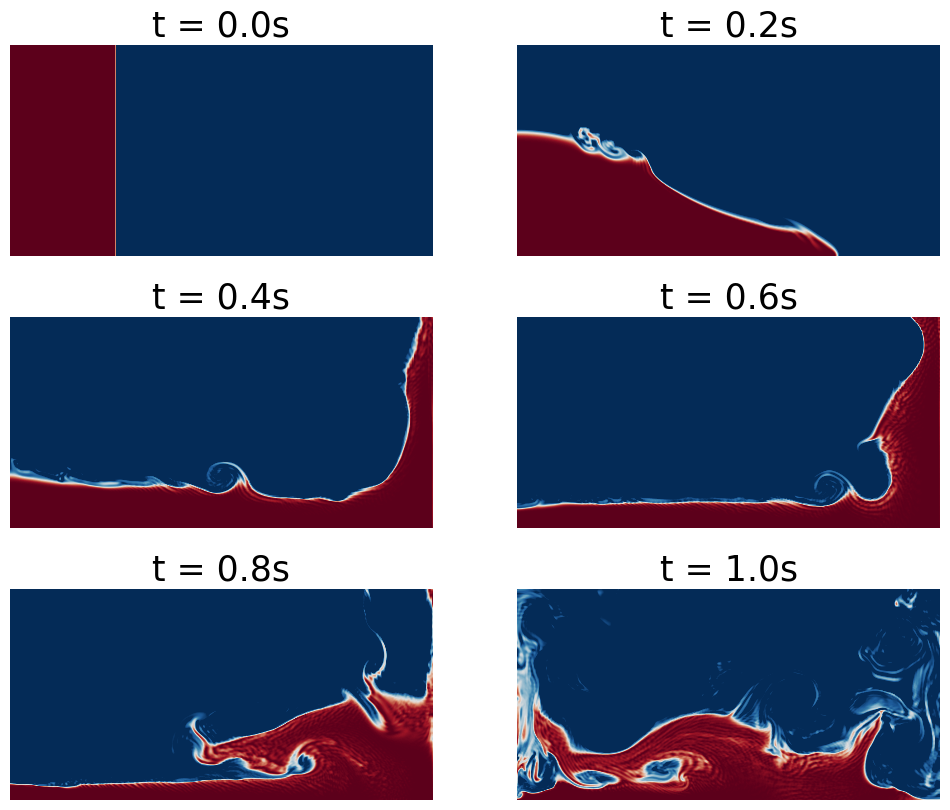}\label{fig:Indicator_no_obstacle_L}}
     \quad
     \subfloat[Quadratic convolutional filters (Model~4).]{\includegraphics[scale=0.345]{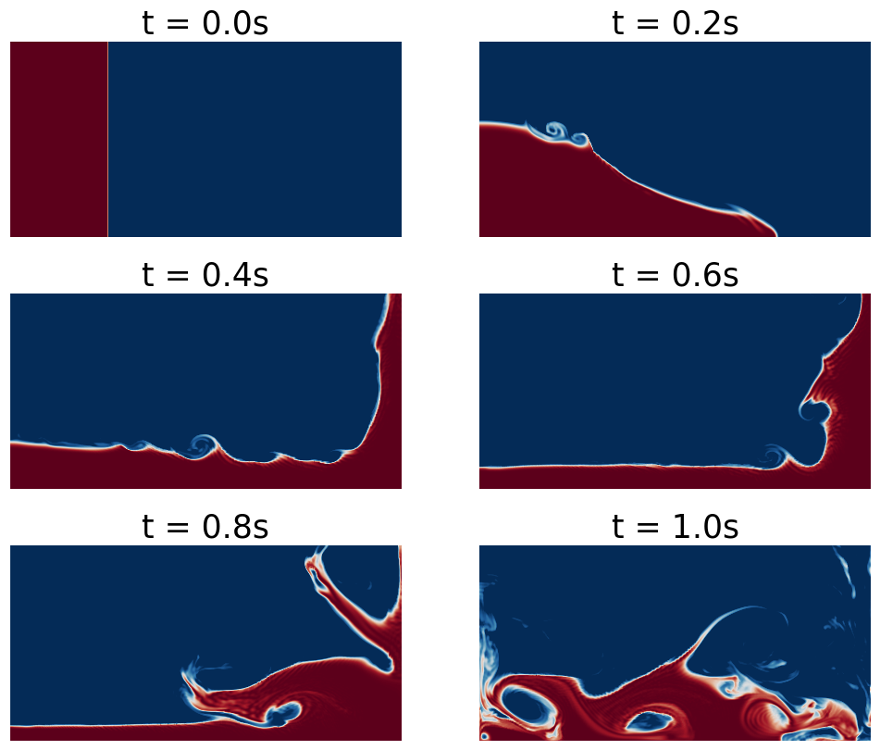}\label{fig:Indicator_no_obstacle_Q}}
     \quad
     \subfloat[Linear filters for volume fraction diffusion, quadratic filters for all other terms (Model~5).]{\includegraphics[scale=0.32]{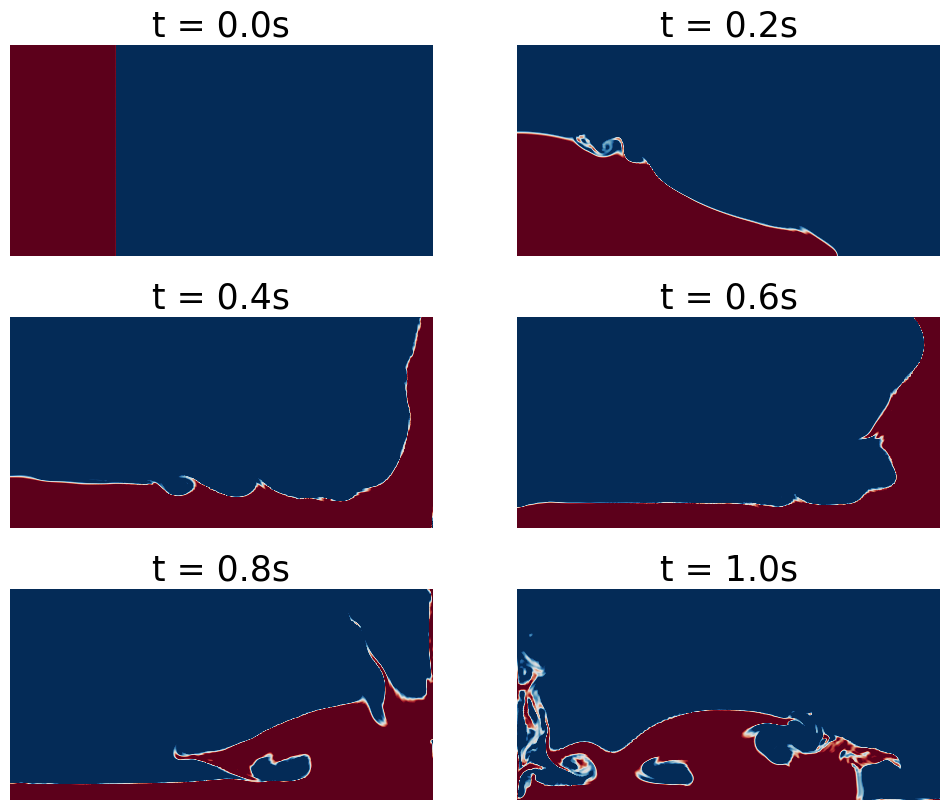}\label{fig:Indicator_no_obstacle_Q_LC}}
     \quad
     \subfloat[Linear filters for volume fraction diffusion, cubic filters for all other terms (Model~6).]{\includegraphics[scale=0.32]{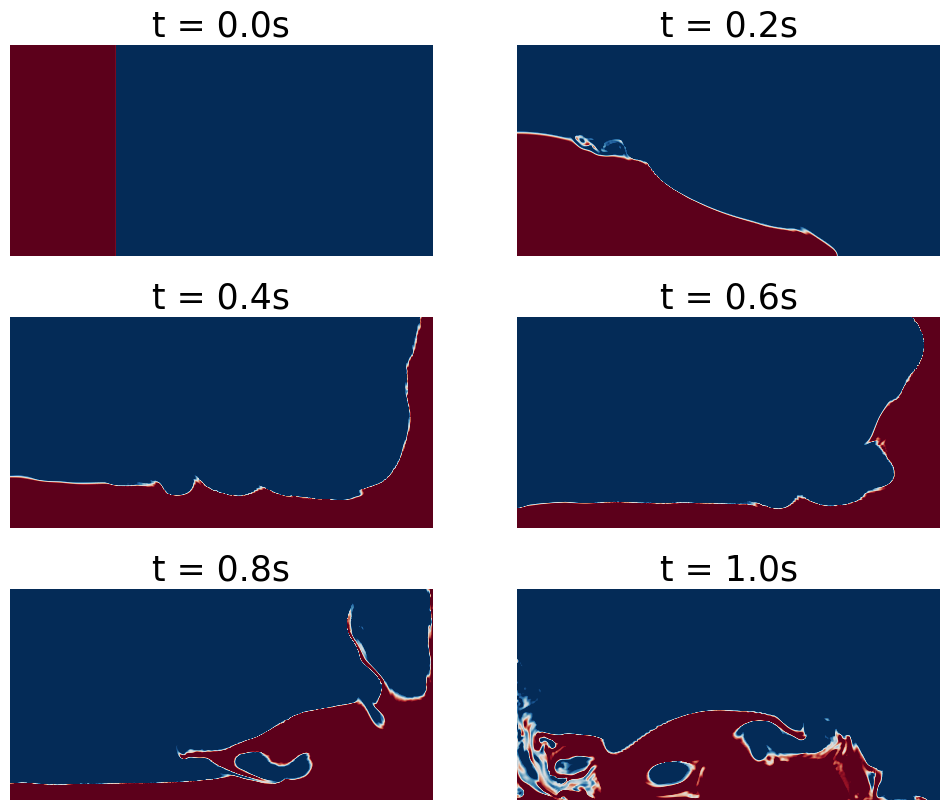}\label{fig:Indicator_no_obstacle_C_LC}}
    \caption{Snapshots of volume fraction field for the 3D collapsing water column without an obstacle from $t = \SI{0}{s}$ to $t = \SI{1.0}{s}$ at intervals of $\SI{0.2}{s}$, solved on a grid of 1024$\times$64$\times$1024 nodes. The volume fraction is shown on an $xz$~plane at the midpoint of the extent of the domain in the $y$~direction.} 
    \label{fig:indicator_water_column_1024_64_1024}
\end{figure}

Figures~\ref{fig:density_water_column_1024_64_1024} and~\ref{fig:indicator_water_column_1024_64_1024} show transient results in the range \SI{0}{s} $\leqslant t \leqslant$\SI{1}{s} for a  grid of 67~million ConvFEM nodes from Models~2, 4, 5 and~6, which have the same domain and grid size, but different orders and combinations of FE discretisations (see Table~\ref{tab:numerical_detail} for details). Figure~\ref{fig:density_water_column_1024_64_1024} shows the water-air interface, chosen here to be where the volume fraction field has a value of~0.5, and Figure~\ref{fig:indicator_water_column_1024_64_1024} shows the volume fraction field, which ranges from~0 to~1. Overall, numerical results indicate similar time-dependent spatial distributions of the water volume fraction field to the numerical and experimental investigations of~\citet{yeoh2009assessment}. The evolution of the water columns is shown through the volume fraction field in Figures~\ref{fig:density_water_column_1024_64_1024} and~\ref{fig:indicator_water_column_1024_64_1024}, and the results display good  qualitative agreement with those shown experimentally and computationally by~\citet{yeoh2009assessment} (see their Figure~2, and Figures~4 and 5 respectively).

Comparing the approaches implemented in these four models, a better performance in terms of interface sharpness between phases was obtained from Models~5 and~6 (as demonstrated in Figures~\ref{fig:Indicator_no_obstacle_Q_LC} and~\ref{fig:Indicator_no_obstacle_C_LC}) due to the compressive-advection interface-capturing formulation (see Section~\ref{compress-adv-k}). The reason is that higher-order (quadratic and cubic FE) diffusion schemes introduce more oscillations than lower-order (linear FE) schemes, which leads to a more diffusive volume fraction field as shown in Figure~\ref{fig:Indicator_no_obstacle_Q}. In order to mitigate the additional diffusion of the volume fraction field caused by higher-order discretisations, Models~5 and~6 combine advantages of lower- and higher-order FE discretisations: a lower-order FE discretisation (linear ConvFEM) is used to discretise the diffusion of the volume fraction field and all other terms are discretised using higher-order FE schemes (quadratic or cubic ConvFEM).

\begin{figure}[htbp]
    \centering
    \subfloat[Water-air interface.]{\includegraphics[scale=0.3]{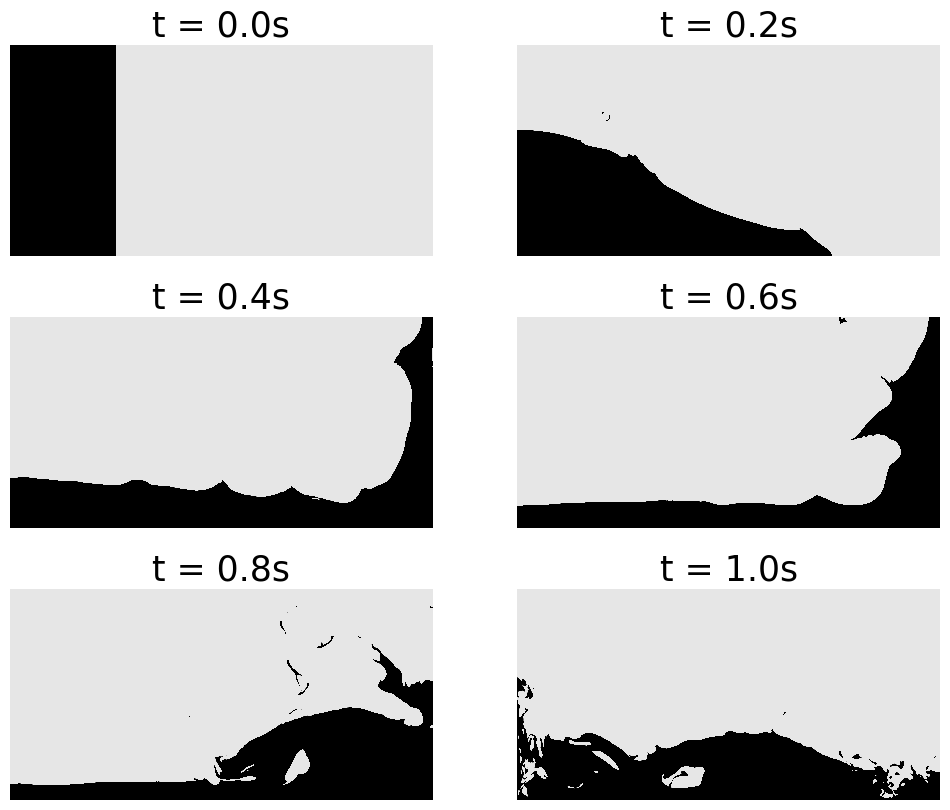}\label{fig:Interface_no_obstacle_256M} }
    \quad
    \subfloat[Volume fraction field.]{\includegraphics[scale=0.3]{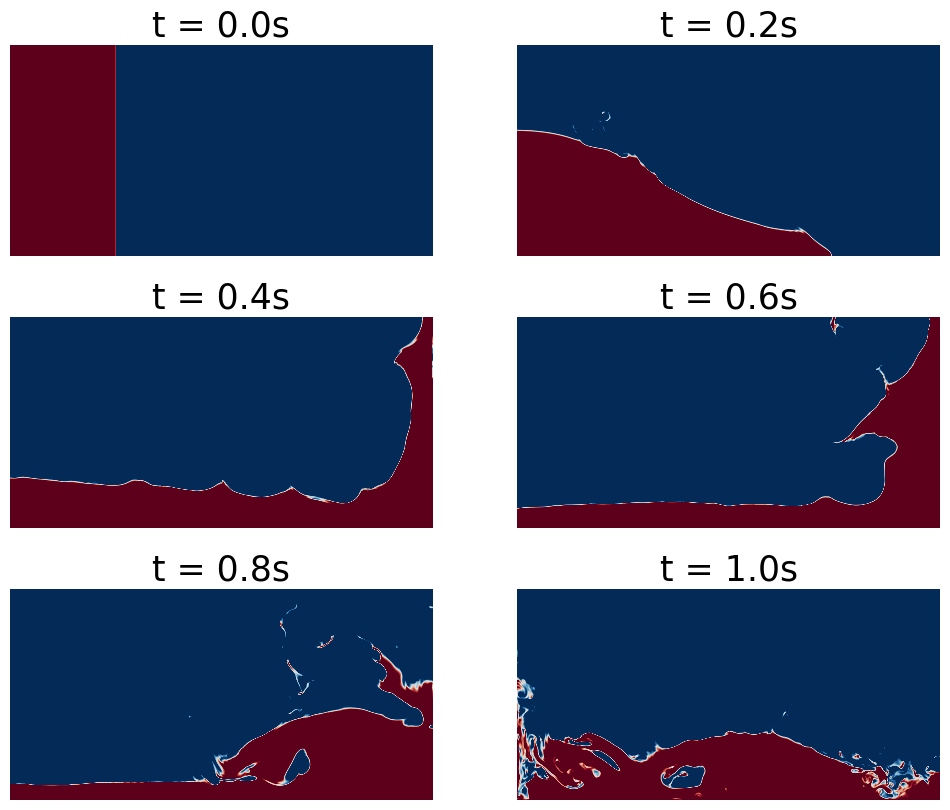}\label{fig:Indicator_no_obstacle_256M}}
    \quad
    \subfloat[Isosurface at $t = \SI{0.4}{s}$. ]{\includegraphics[scale=0.08,trim=200mm 120mm 120mm 250mm, clip]{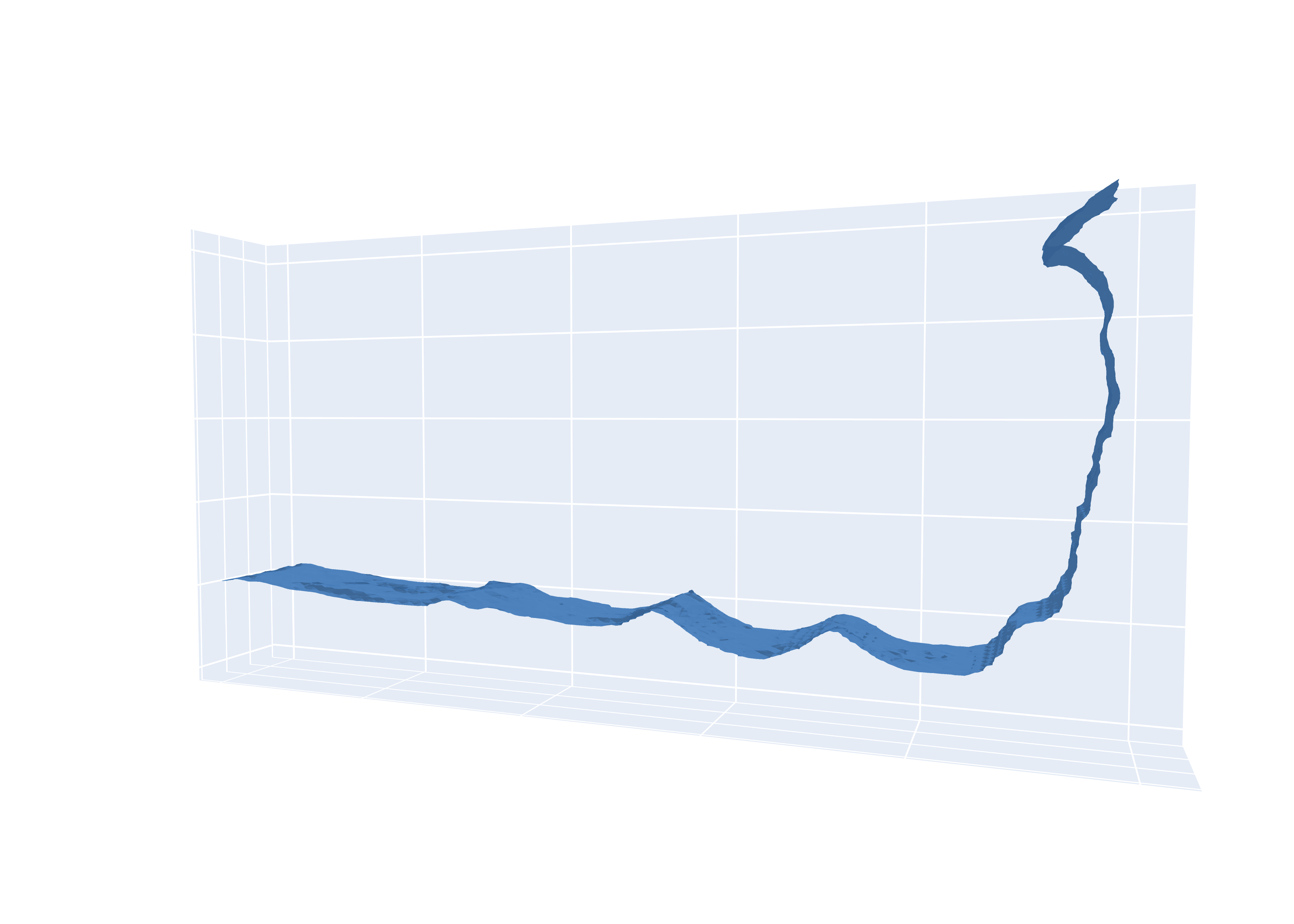}\label{fig:256M_iso_surface1}}
    \quad
    \subfloat[Isosurface at $t = \SI{0.6}{s}$.]{\includegraphics[scale=0.08,trim=200mm 120mm 120mm 250mm, clip]{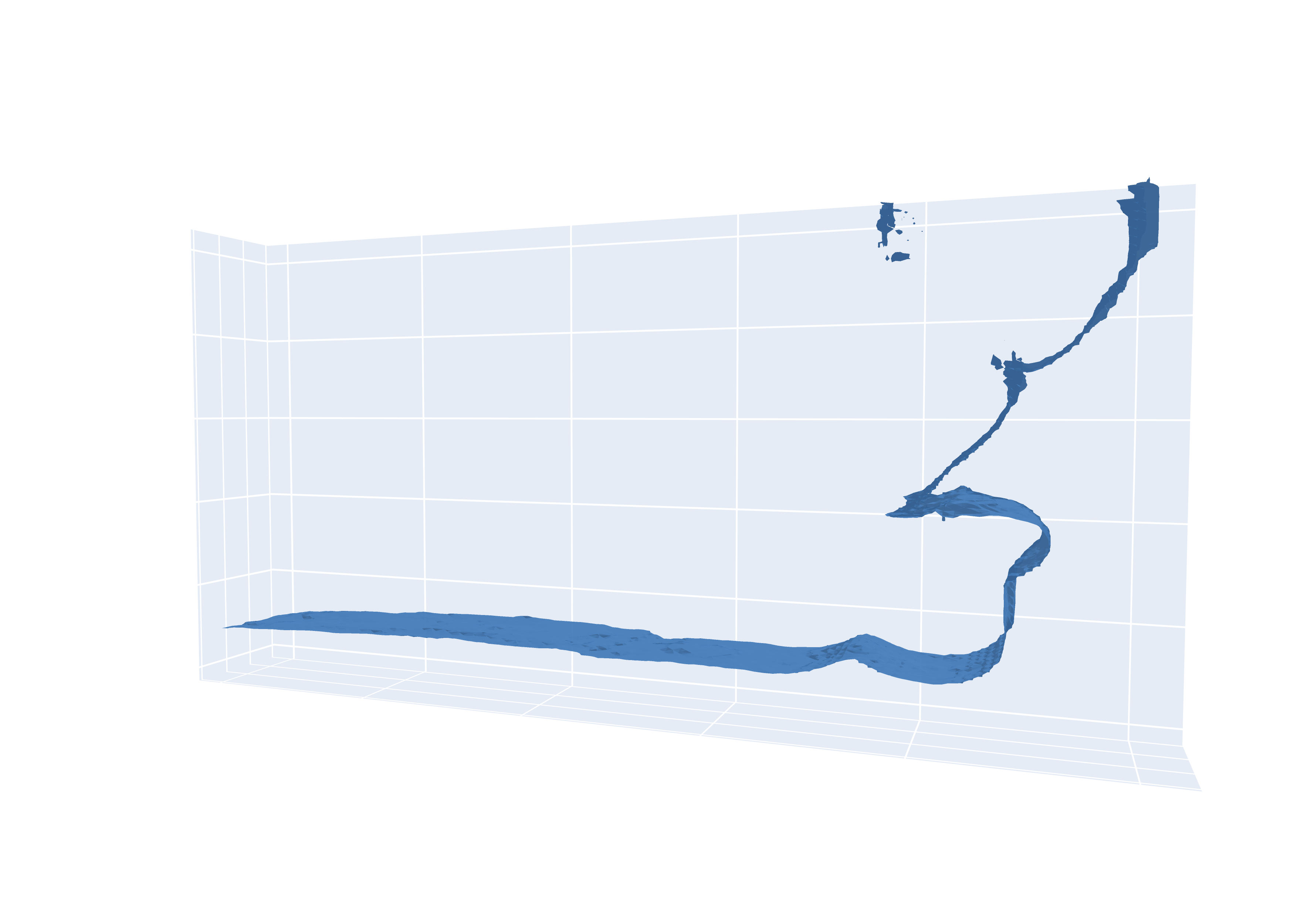}\label{fig:256M_iso_surface2}}
    \quad
    \subfloat[Isosurface at $t = \SI{0.8}{s}$.]{\includegraphics[scale=0.08,trim=200mm 120mm 120mm 250mm, clip]{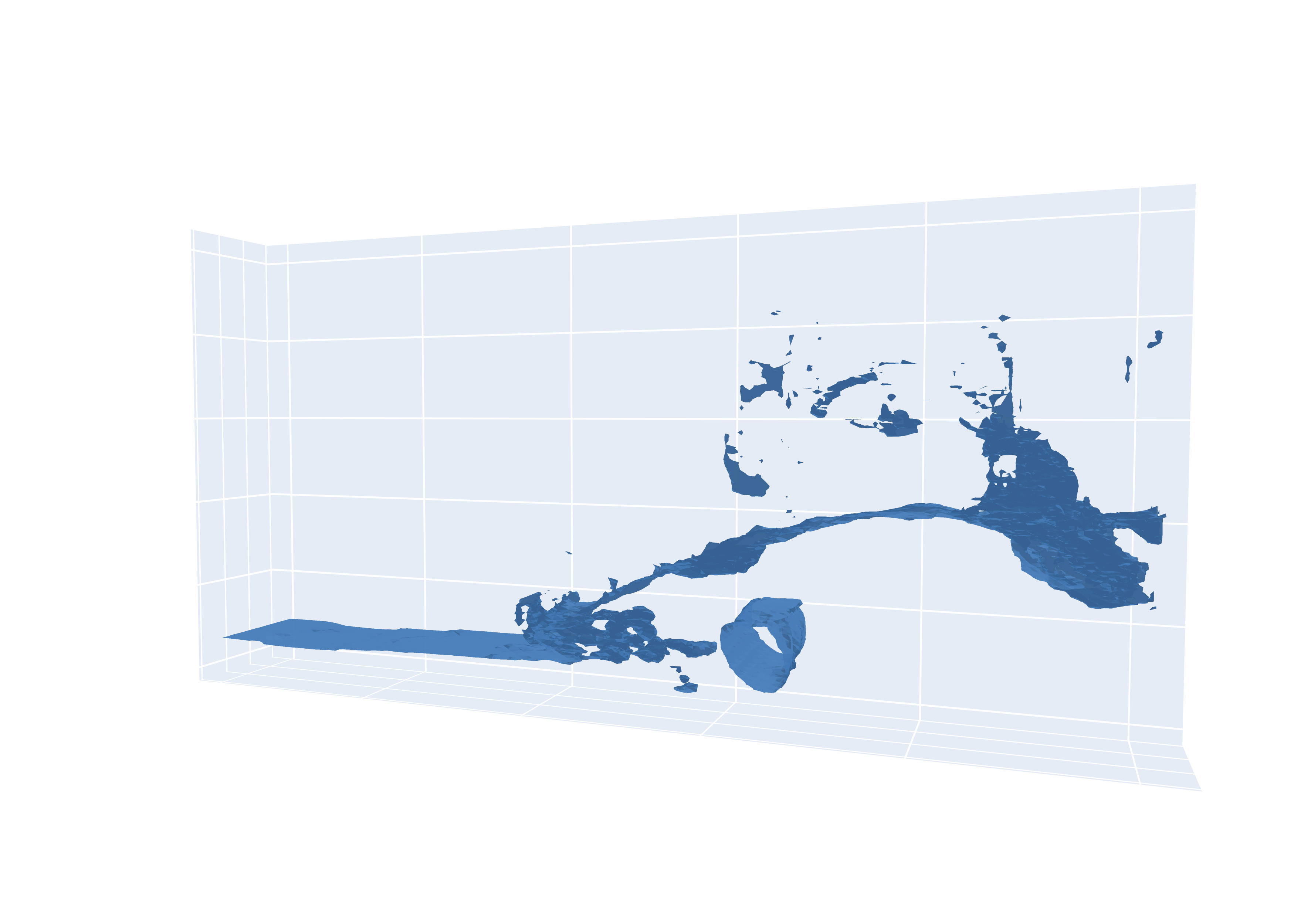}\label{fig:256M_iso_surface3}}
    \quad
    \subfloat[Isosurface at $t = \SI{1.0}{s}$.]{\includegraphics[scale=0.08,trim=200mm 120mm 120mm 250mm, clip]{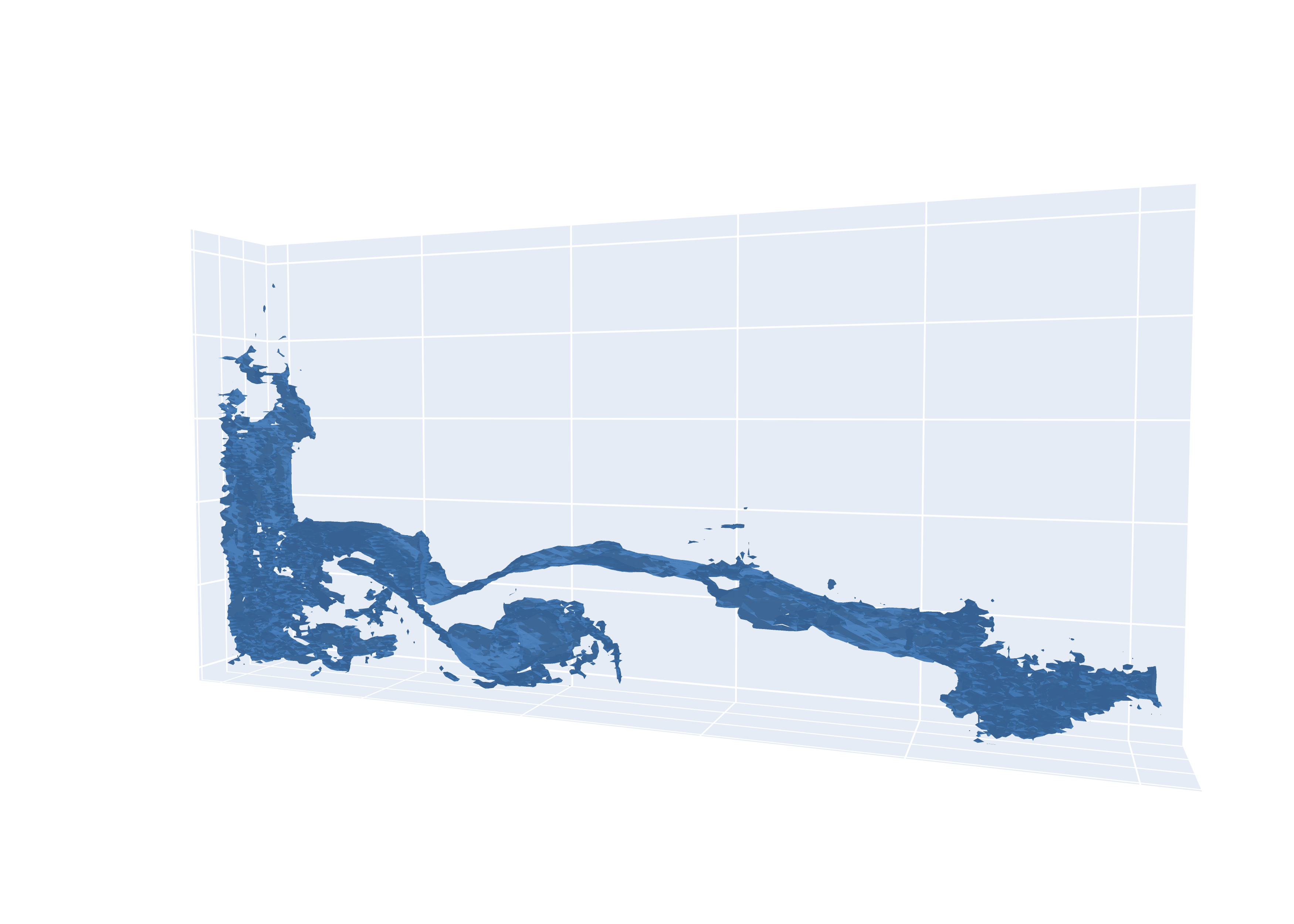}\label{fig:256M_iso_surface4}}
    \caption{Snapshots of three-dimensional numerical predictions of collapsing water column without an obstacle from $t = \SI{0}{s}$ to $t = \SI{1.0}{s}$ solved on a grid of 1024$\times$256$\times$1024 nodes. The results are obtained by Model~7 using linear filters for the volume fraction diffusion and quadratic convolutional filters for all other terms.  Plots~(a) and~(b) are shown on an $xz$ plane located at the midpoint of the domain on the $y$-axis. Plots~(c) to~(f) are isosurfaces showing the water-air interface. }
    \label{fig:water_column_1024_256_1024}
\end{figure}

Figure~\ref{fig:water_column_1024_256_1024} shows results for the water-air interface and volume fraction field for Model~7, which has the same time step and uses the same filters (linear filters for the diffusion term of the volume fraction and quadratic filters for all other terms) as the results shown in Figures~\ref{fig:Interface_no_obstacle_Q_LC} and~\ref{fig:Indicator_no_obstacle_Q_LC} for Model~5. The computational domain used to generate the results shown in Figure~\ref{fig:water_column_1024_256_1024} has a depth that is 4~times larger than that of Model~5, using a total of 256~million nodes (the largest number of nodes used used here). The motivation for running this simulation is to excite fully three-dimensional behaviour (due to the extended domain in the $y$~direction) and also explore the limit of the size of the problem that can be simulated on our single GPU. Comparing Figures~\ref{fig:Interface_no_obstacle_Q_LC} and~\ref{fig:Indicator_no_obstacle_Q_LC} with Figures~\ref{fig:Interface_no_obstacle_256M}  and~\ref{fig:Indicator_no_obstacle_256M}, a similar dispersion and advection of the collapsing water column can be seen in both cases. Figures~\ref{fig:256M_iso_surface1}--\ref{fig:256M_iso_surface4} show the water-air interface (an isosurface at a value of volume fraction of 0.5) at four time levels from \SI{0.4}{s} to \SI{1}{s}. They show the initial water column  descending due to gravity, and forming  waves on the interface between the falling water and the surrounding air in the early stages (Figure~\ref{fig:256M_iso_surface1}). As waves propagate throughout the domain, they breakup upon impact with the boundary, as shown in Figures~\ref{fig:256M_iso_surface3} and~\ref{fig:256M_iso_surface4} and become highly three dimensional, aligning with the results shown in Figure~3 of~\citet{yeoh2009assessment}. 

\begin{figure}[htbp]
    \centering
     \subfloat[Water-air interface.]{\includegraphics[scale=0.3]{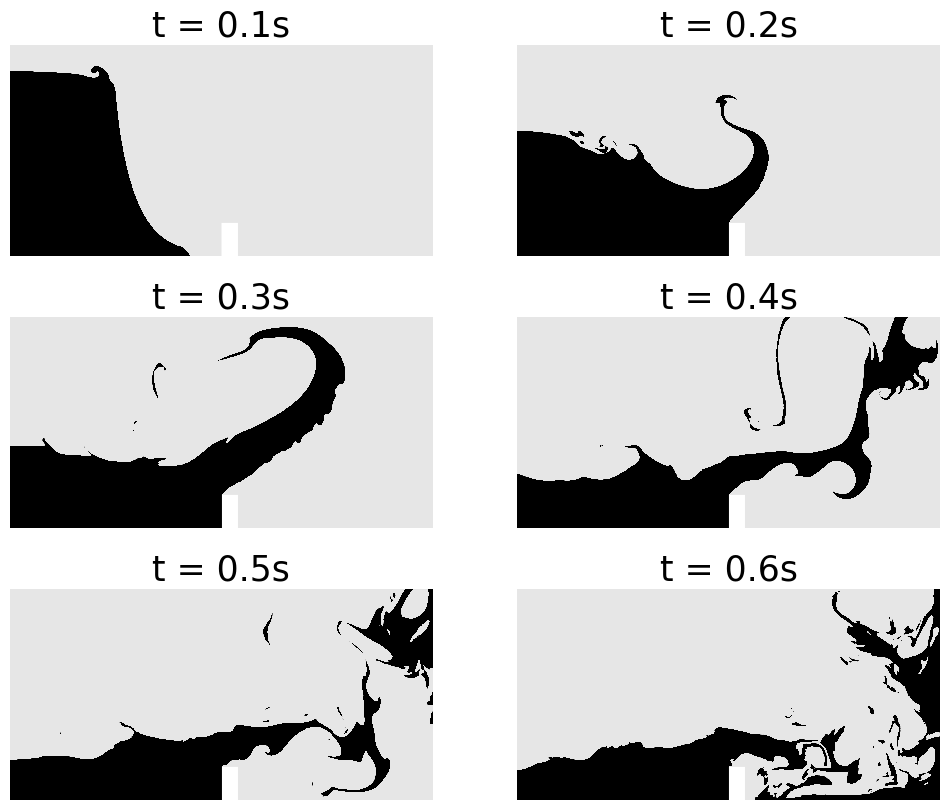}\label{fig:comp_obstacle} }
     \quad
     \subfloat[Volume fraction field.]{\includegraphics[scale=0.3]{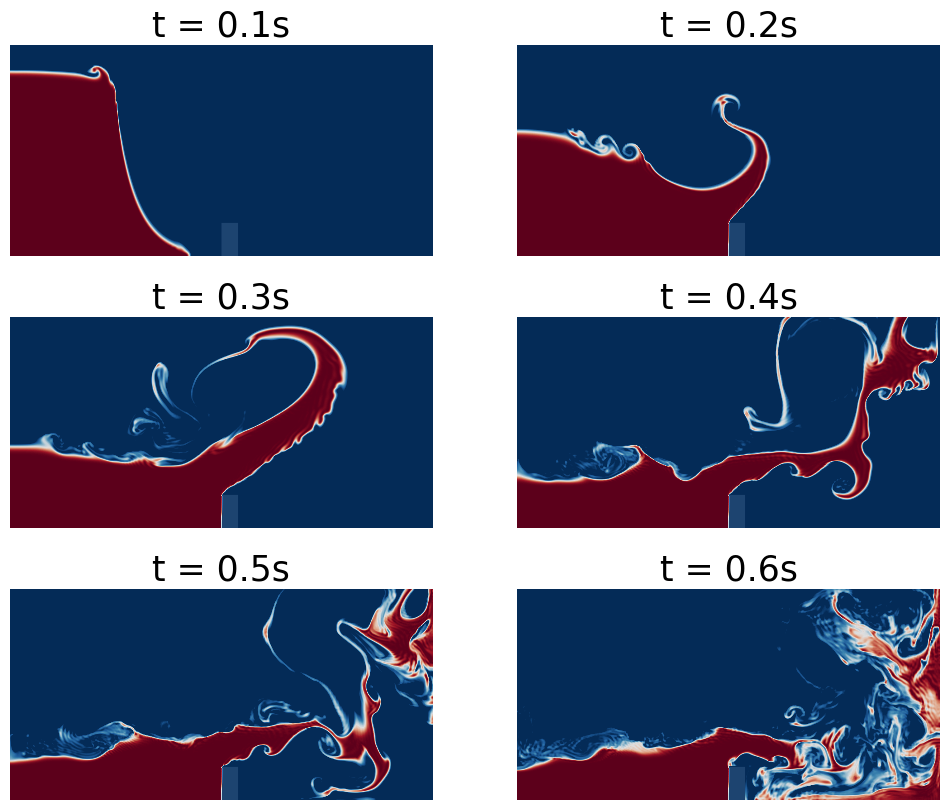}\label{fig:indicator_obstacle}}
    \caption{Snapshots of three-dimensional numerical predictions of the collapsing water column with an obstacle from $t = \SI{0.1}{s}$ to $t = \SI{0.6}{s}$ at an interval of  $\SI{0.1}{s}$. The fields are shown on an $xz$ plane located at the midpoint of the domain on the $y$-axis. The results are obtained using Model~3 (Table~\ref{tab:numerical_detail}) solved on a grid of 1024$\times$64$\times$1024 nodes using linear convolutional filters.}
    \label{fig:water_column_obstacle_1024_64_1024}
\end{figure}

Figure~\ref{fig:water_column_obstacle_1024_64_1024} shows the interaction between the collapsing water and an obstacle (see Figure~\ref{results:cwc:section:3d:Table:Dimensions} for location of the obstacle). Figures~\ref{fig:comp_obstacle} and~\ref{fig:indicator_obstacle} show the water-air interface and volume fraction, respectively. The results indicate that the solution obtained from the NN4PDEs approach captures instabilities induced by the obstacle, including the generation of waves and ripples due to the sudden change in the momentum of the water. Qualitative comparisons can be made between the evolution of the volume fraction field shown here, in Figure~\ref{fig:indicator_obstacle}, with the results shown in~\citet{yeoh2009assessment}, experimentally in their Figure~3 and computationally in their Figures~8 and~9. A close match in terms of wave propagation and interaction between collapsing water and obstacle is seen, with our results and those of~\citet{yeoh2009assessment} producing similar spatial variation of the water field at each time level.

\begin{figure}
    \centering
    \begin{subfigure}[t]{0.45\textwidth}
    \includegraphics[width=\textwidth,trim=200mm 120mm 220mm 250mm, clip]{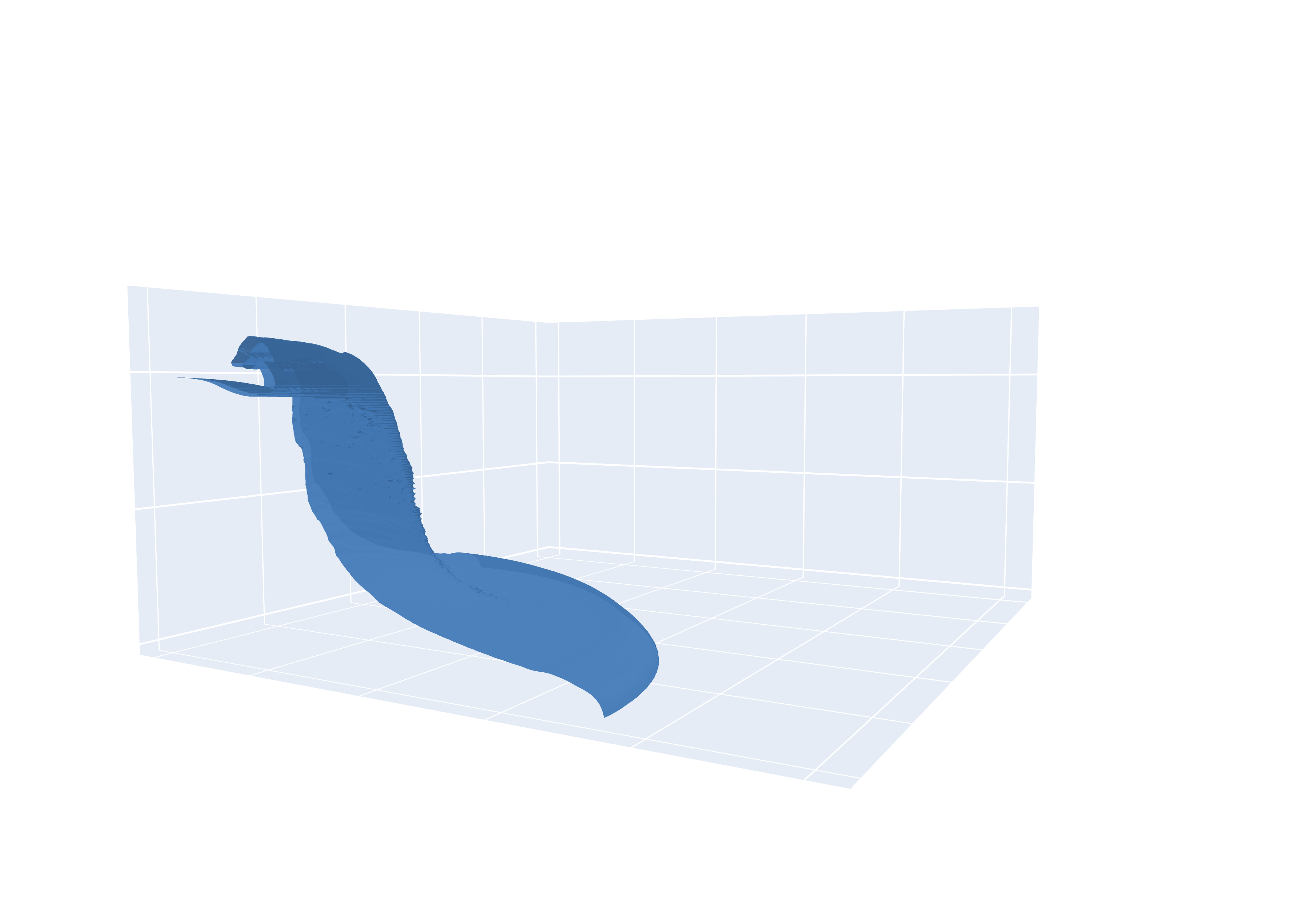}
    \caption*{$t$=\SI{0.15}{s}}
    \end{subfigure}
    \begin{subfigure}[t]{0.45\textwidth}
    \includegraphics[width=\textwidth,trim=200mm 120mm 220mm 250mm, clip]{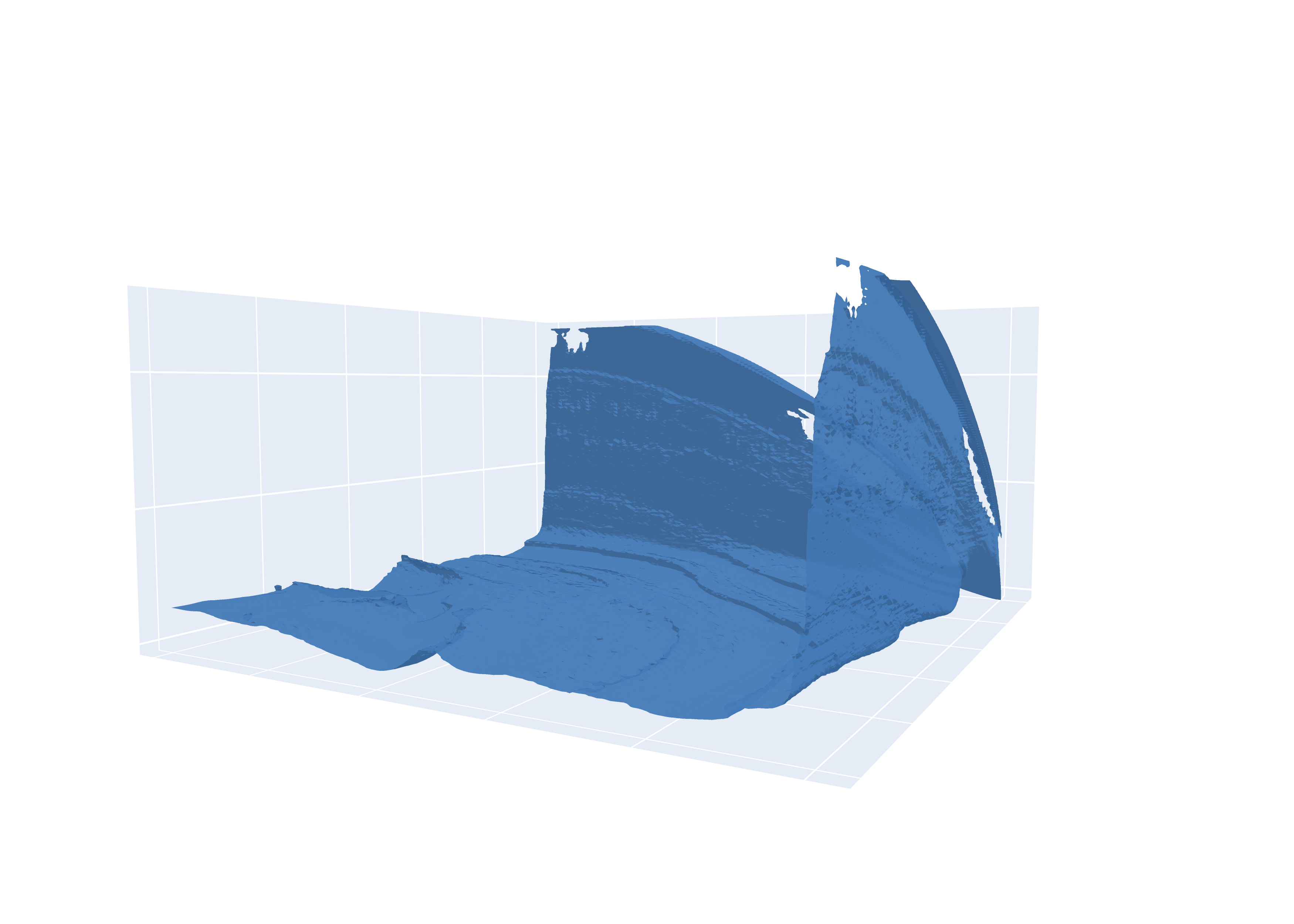}
    \caption*{$t$=\SI{0.3}{s}}
    \end{subfigure}
    \begin{subfigure}[t]{0.45\textwidth}
    \includegraphics[width=\textwidth,trim=200mm 120mm 220mm 250mm, clip]{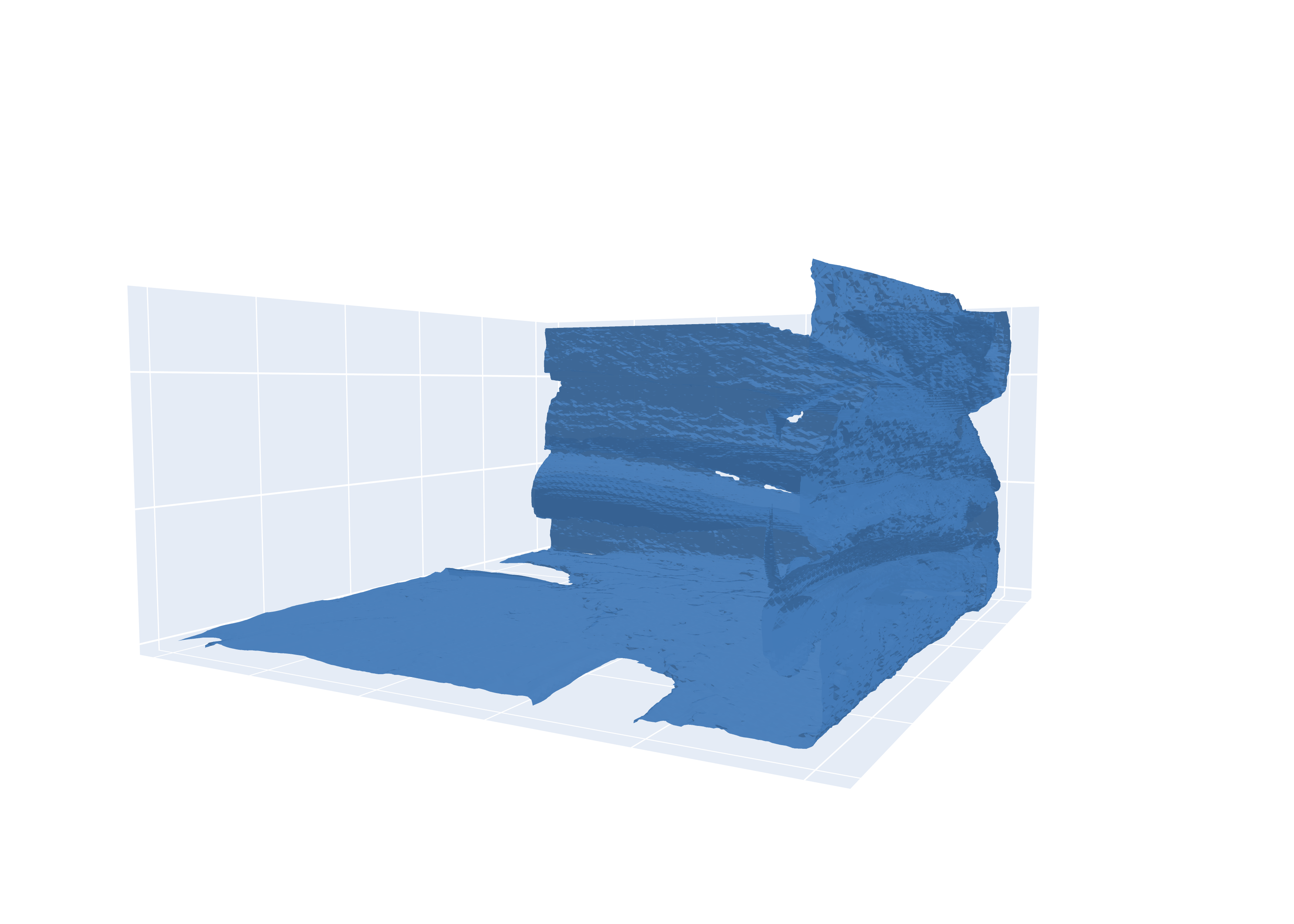}
    \caption*{$t$=\SI{0.45}{s}}
    \end{subfigure}
    \begin{subfigure}[t]{0.45\textwidth}
    \includegraphics[width=\textwidth,trim=200mm 120mm 220mm 250mm, clip]{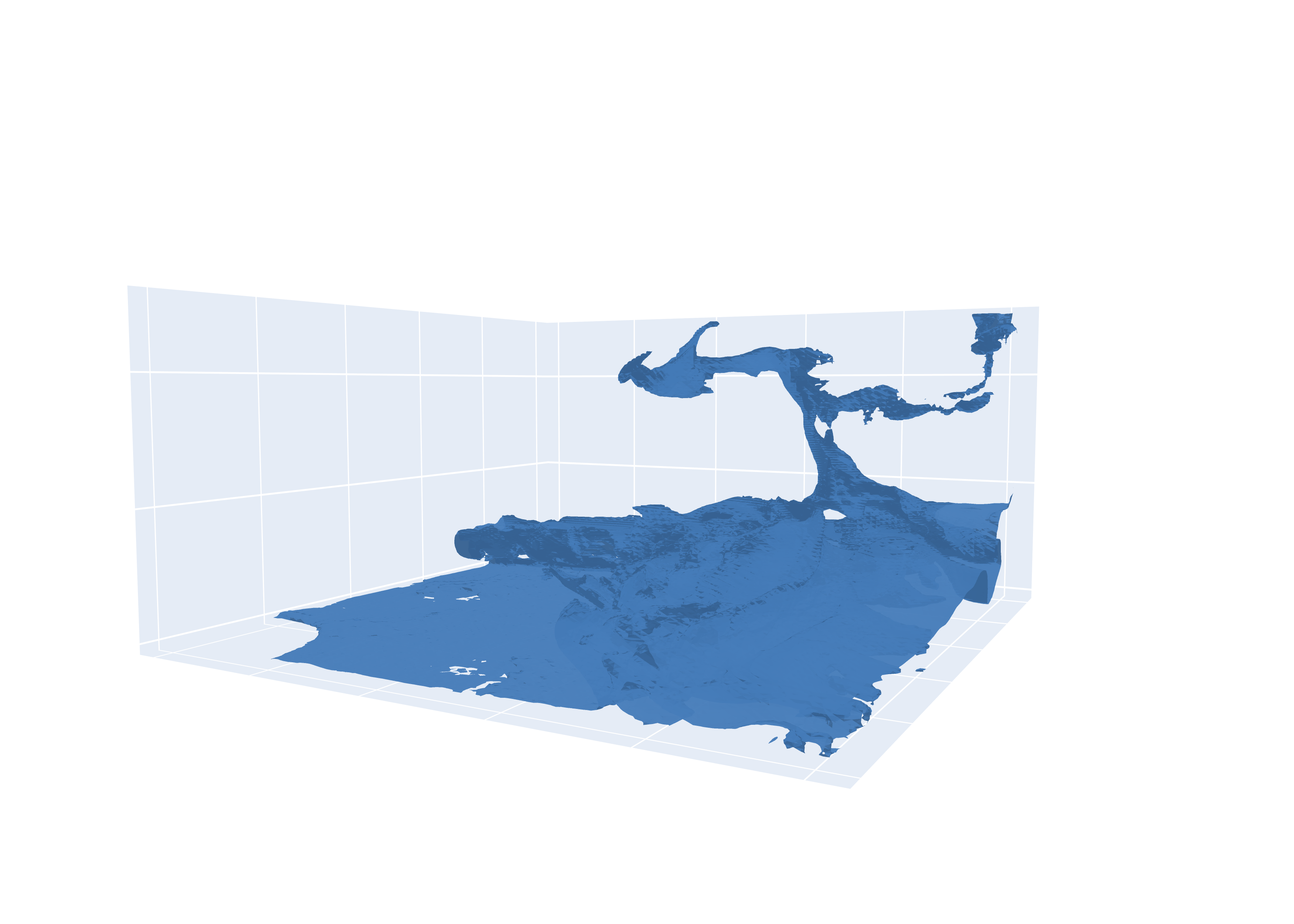}
    \caption*{$t$=\SI{0.6}{s}}
    \end{subfigure}
    \begin{subfigure}[t]{0.45\textwidth}
    \includegraphics[width=\textwidth,trim=200mm 120mm 220mm 250mm, clip]{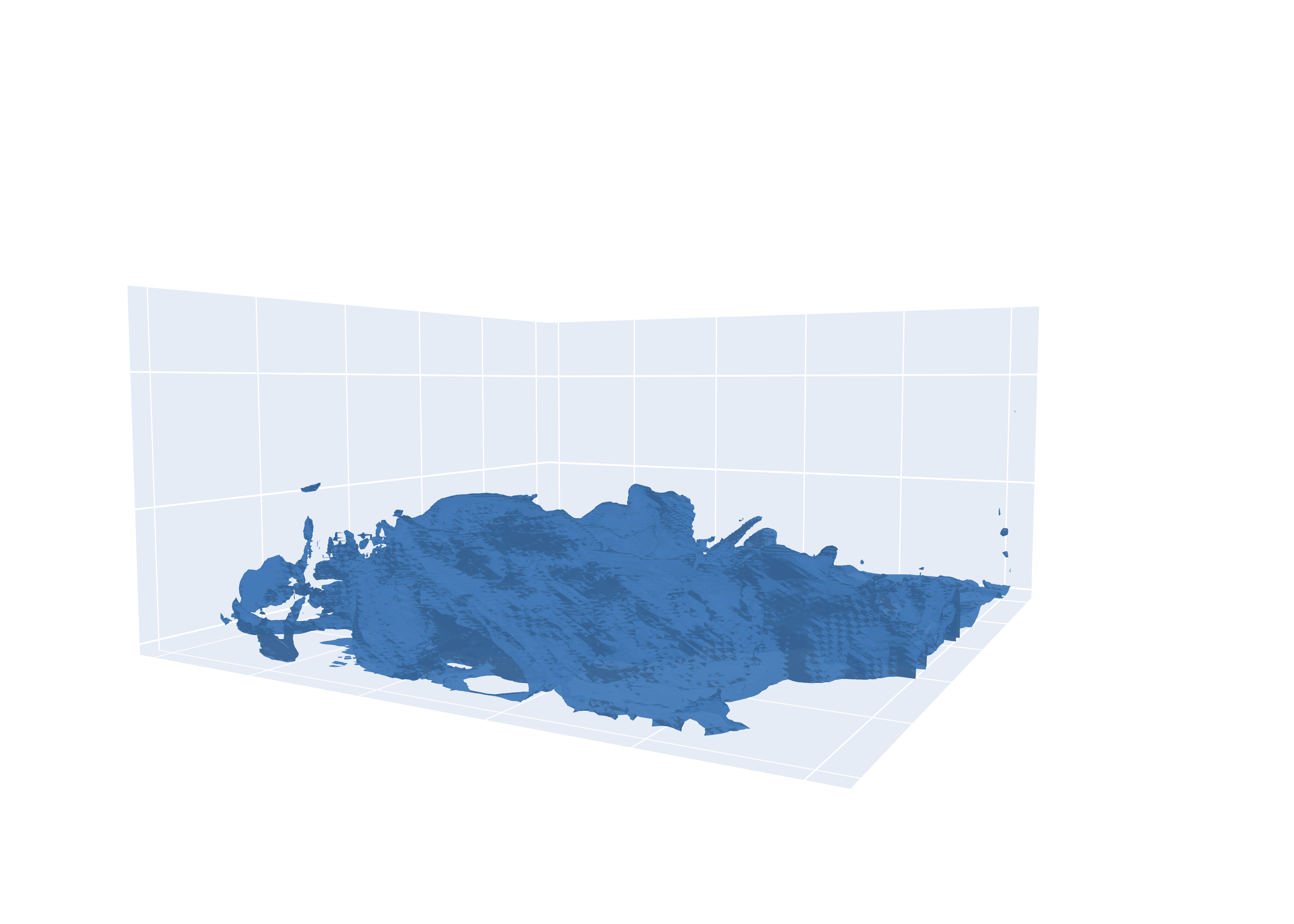}
    \caption*{$t$=\SI{0.75}{s}}
    \end{subfigure}
    \begin{subfigure}[t]{0.45\textwidth}
    \includegraphics[width=\textwidth,trim=200mm 120mm 220mm 250mm, clip]{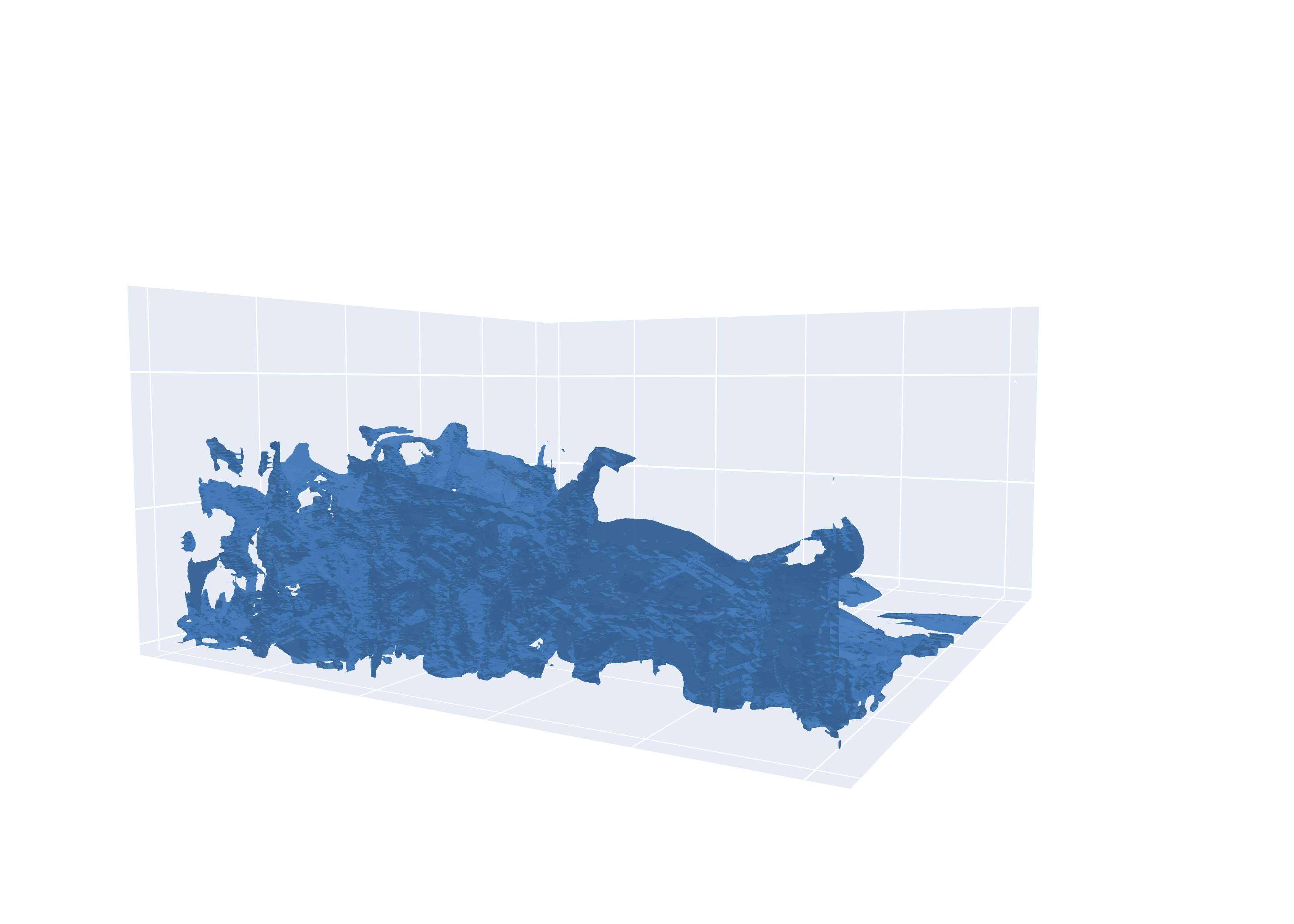}
    \caption*{$t$=\SI{0.9}{s}}
    \end{subfigure}
    \begin{subfigure}[t]{0.45\textwidth}
    \includegraphics[width=\textwidth,trim=200mm 120mm 220mm 250mm, clip]{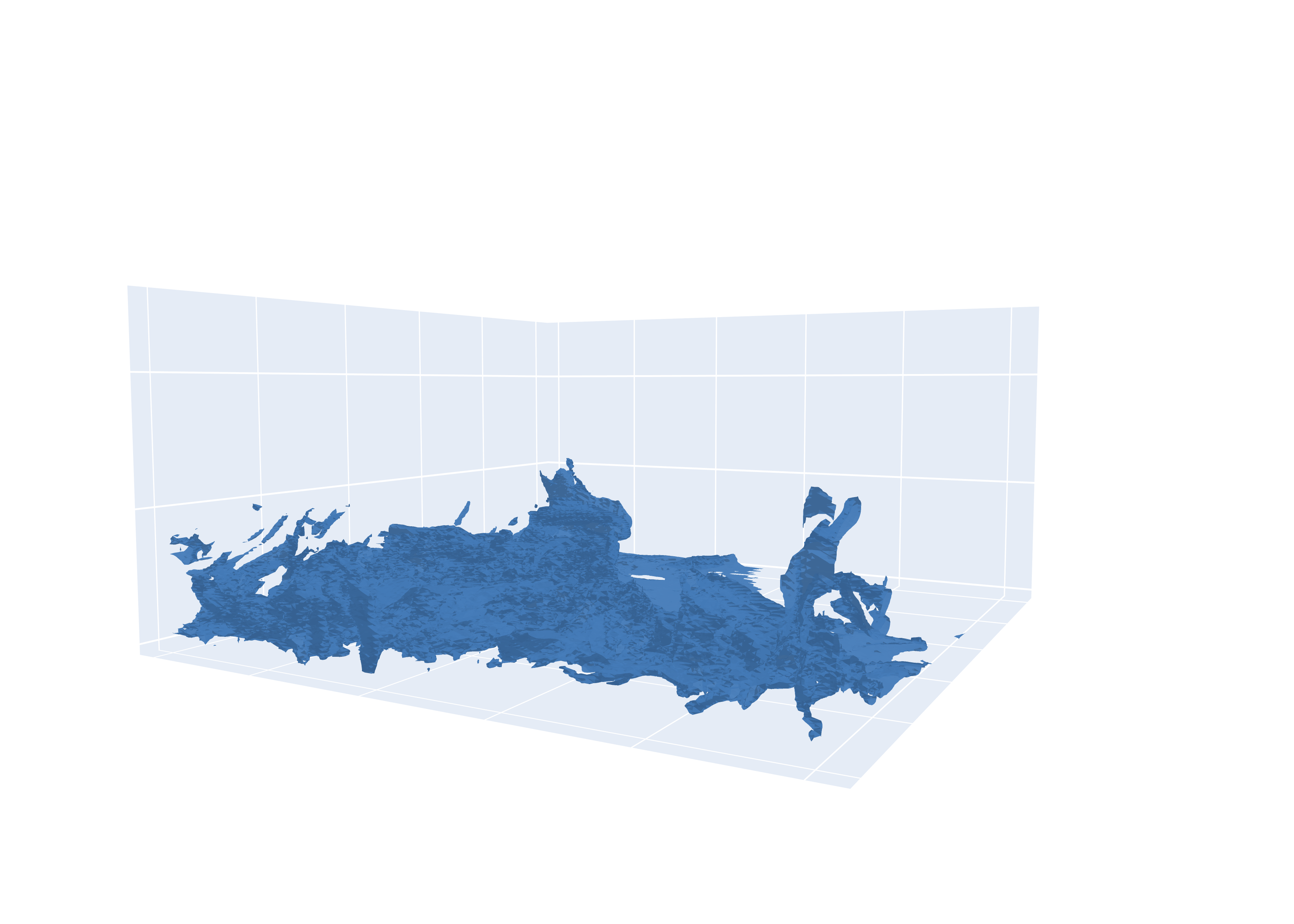}
    \caption*{$t$=\SI{1.05}{s}}
    \end{subfigure}
    \begin{subfigure}[t]{0.45\textwidth}
    \includegraphics[width=\textwidth,trim=200mm 120mm 220mm 250mm, clip]{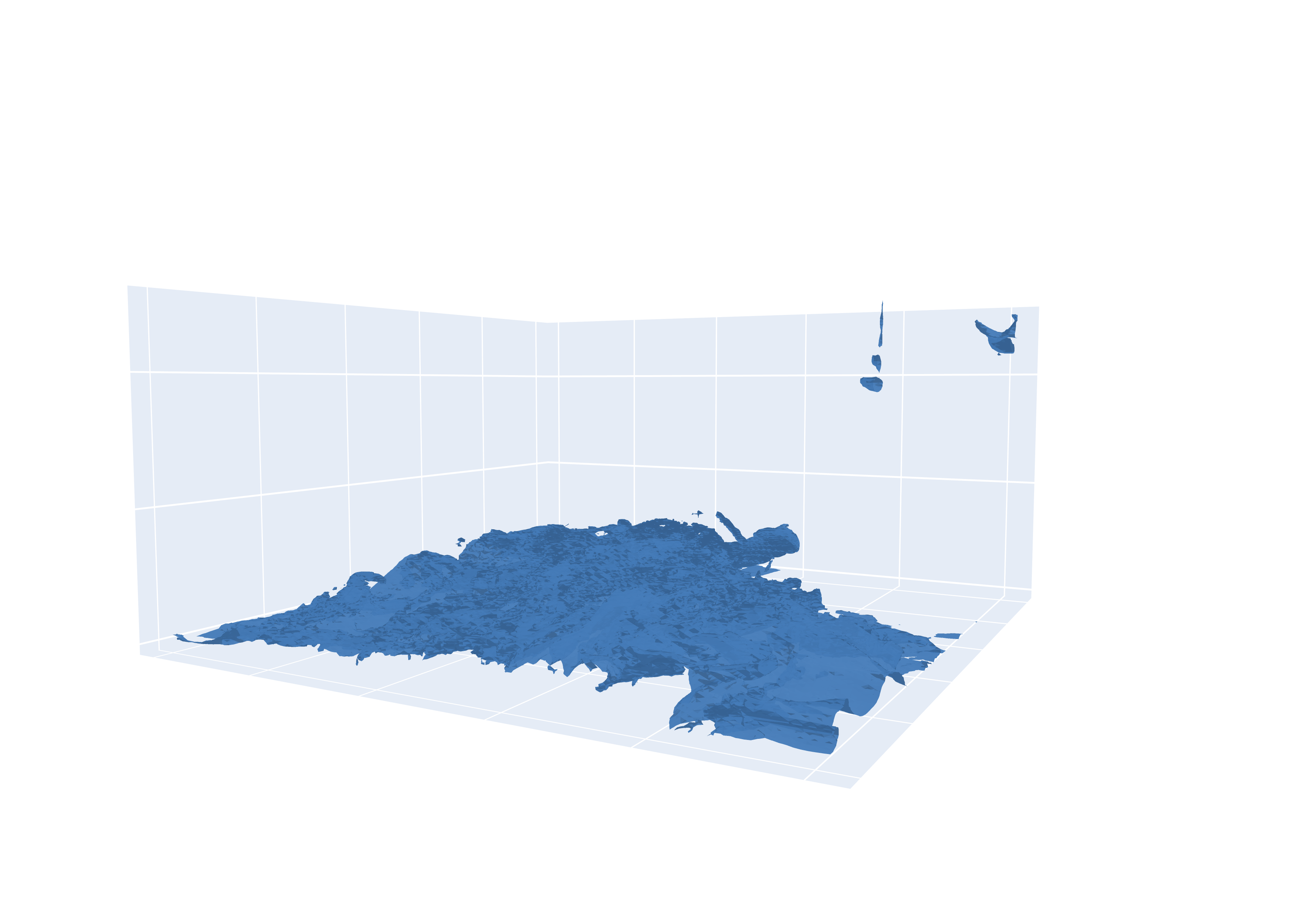}
    \caption*{$t$=\SI{1.2}{s}}
    \end{subfigure}
    \caption{Snapshots of the volume fraction isosurface drawn at a value of 0.5 from $t = \SI{0.15}{s}$ to $t = \SI{1.2}{s}$ at an interval of $\SI{0.15}{s}$. The results are obtained using Model~8 shown in Table~\ref{tab:numerical_detail} and solved on a grid of 512$\times$512$\times$512 nodes and using linear filters for the volume fraction diffusion and quadratic convolutional filters for all other terms.}
    \label{fig:Iso_surface_cubic_water_column}
\end{figure}

\begin{figure}
    \centering
    \begin{minipage}[t]{1\textwidth}
    \centering
    \hspace{15em}
    \vspace{-20em}
    \includegraphics[width=0.6\textwidth]{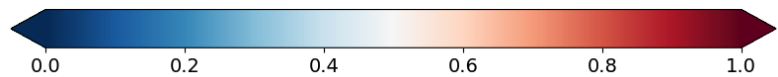}\\[1mm]
    \end{minipage}
    \begin{subfigure}[t]{0.32\textwidth}
    \hspace{5mm}~\includegraphics[width=0.82\textwidth]{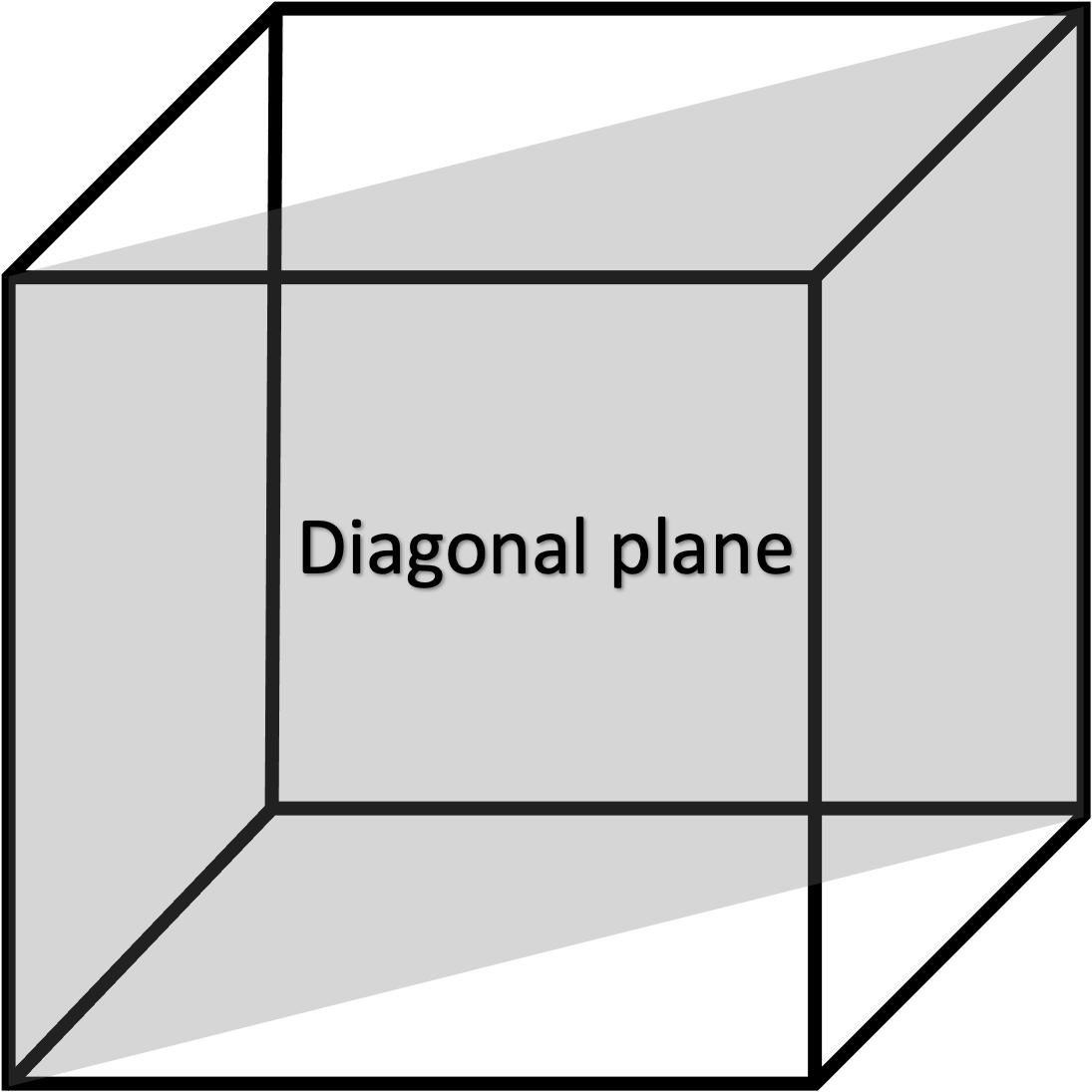}
    \caption*{plotting plane}
    \end{subfigure}
    \begin{subfigure}[t]{0.32\textwidth}
    \includegraphics[width=\textwidth,trim=25mm 15mm 25mm 15mm, clip]{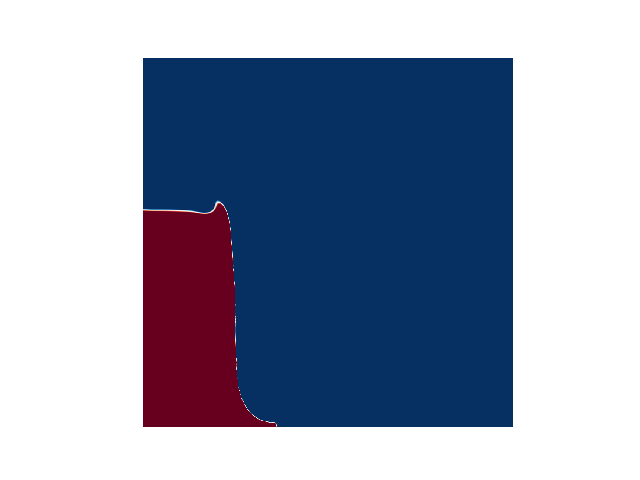}
    \caption*{$t$ = \SI{0.1}{s}}
    \end{subfigure}
    \begin{subfigure}[t]{0.32\textwidth}
    \includegraphics[width=\textwidth,trim=25mm 15mm 25mm 15mm, clip]{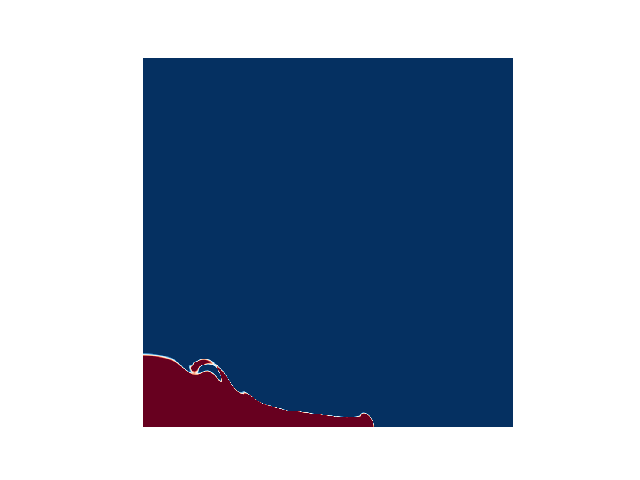}
    \caption*{$t$ = \SI{0.2}{s}}
    \end{subfigure}
    \vspace{0.5em}
    \begin{subfigure}[t]{0.32\textwidth}
    \includegraphics[width=\textwidth,trim=25mm 15mm 25mm 15mm, clip]{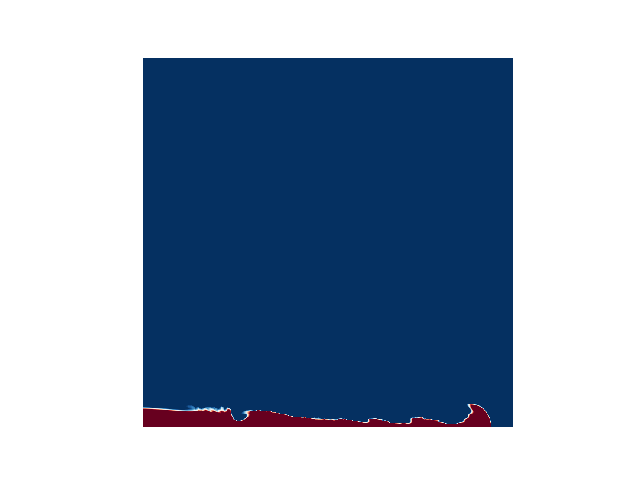}
    \caption*{$t$ = \SI{0.3}{s}}
    \end{subfigure}
    \begin{subfigure}[t]{0.32\textwidth}
    \includegraphics[width=\textwidth,trim=25mm 15mm 25mm 15mm, clip]{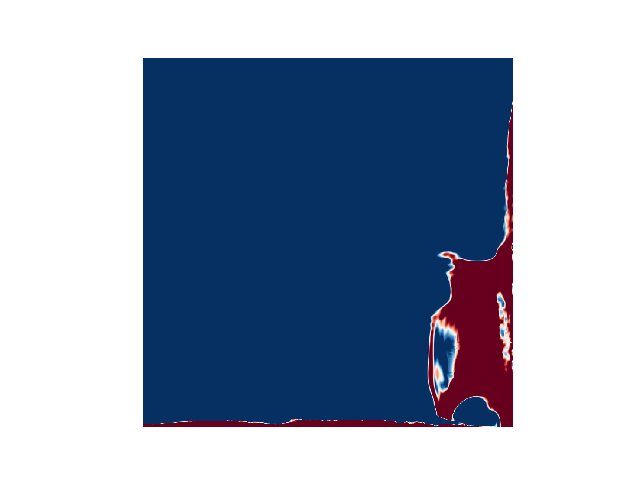}
    \caption*{$t$ = \SI{0.4}{s}}
    \end{subfigure}
    \begin{subfigure}[t]{0.32\textwidth}
    \includegraphics[width=\textwidth,trim=25mm 15mm 25mm 15mm, clip]{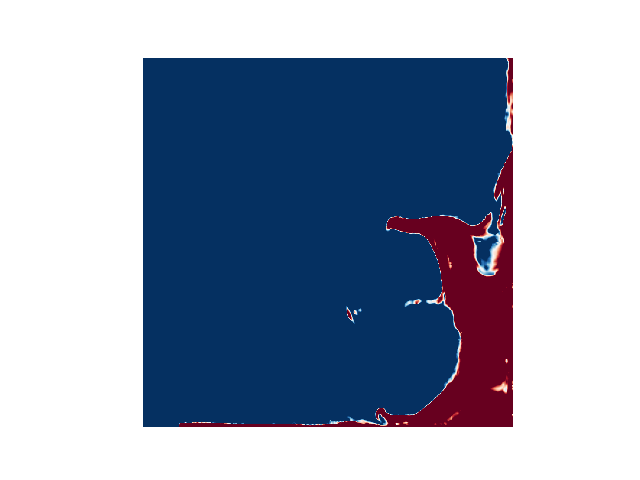}
    \caption*{$t$ = \SI{0.5}{s}}
    \end{subfigure}
    \vspace{0.5em}
    \begin{subfigure}[t]{0.32\textwidth}
    \includegraphics[width=\textwidth,trim=25mm 15mm 25mm 15mm, clip]{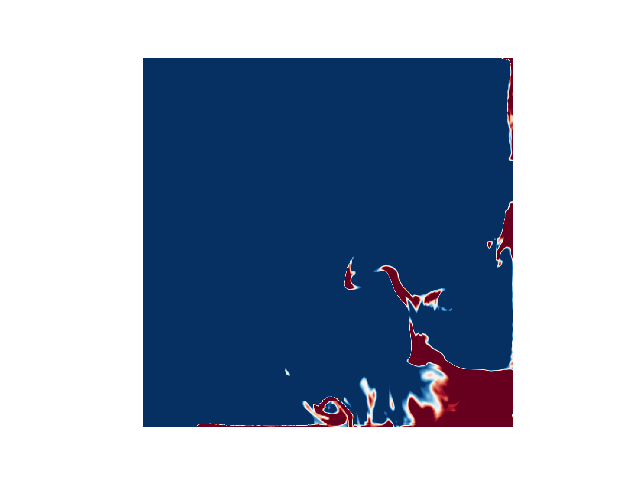}
    \caption*{$t$ = \SI{0.6}{s}}
    \end{subfigure}
    \begin{subfigure}[t]{0.32\textwidth}
    \includegraphics[width=\textwidth,trim=25mm 15mm 25mm 15mm, clip]{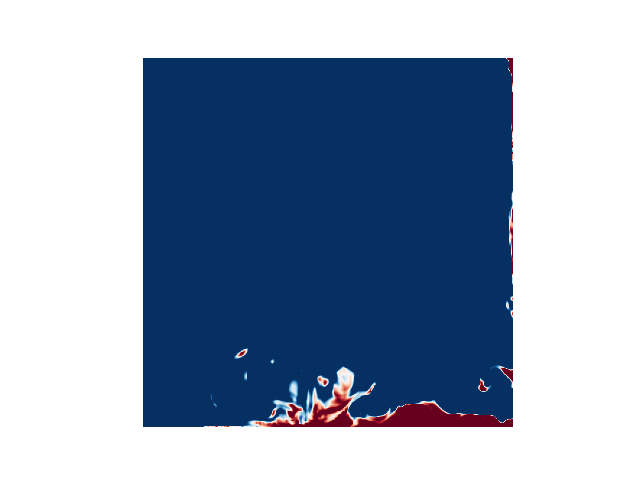}
    \caption*{$t$ = \SI{0.7}{s}}
    \end{subfigure}
    \begin{subfigure}[t]{0.32\textwidth}
    \includegraphics[width=\textwidth,trim=25mm 15mm 25mm 15mm, clip]{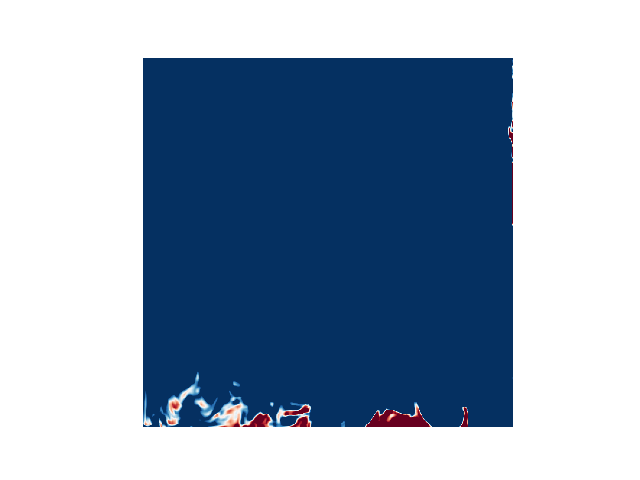}
    \caption*{$t$ = \SI{0.8}{s}}
    \end{subfigure}
    \caption{Snapshots of volume fraction field on a diagonal plane $x=y$ (top left) from $t = \SI{0.1}{s}$ to $t = \SI{0.8}{s}$ with an interval of $\SI{0.1}{s}$ between the snapshots. The results are obtained from Model~8 (for details see Table~\ref{tab:numerical_detail}) and solved on a grid of 512$\times$512$\times$512 nodes using linear convolutional filters for volume fraction diffusion and quadratic convolutional filters for all other terms.}
    \label{fig:Indicator_cubic_water_column}
\end{figure}

The collapsing water column in a cube-shaped computational domain is shown in Figure~\ref{fig:Iso_surface_cubic_water_column}. 
The cube-shaped domain is used here, as it results in highly 3D behaviour (at both large and smaller scales) unlike the previous collapsing water column problems that are solved within a highly anisotropic domain, constrained to have 2D large scale behaviour in which the water column fills the extent of the domain in the $y$ direction (see Models 2--6, and Figures~\ref{fig:domain2} and~\ref{fig:domain3}).  The simulation was conducted with Model~8 and the results plotted at eight time levels from \SI{0.15}{s} to \SI{1.2}{s} with an interval of \SI{0.15}{s}. This figure shows the isosurfaces of the volume fraction field, taken at a value of 0.5 to enable visualisation of the interface between the descending water and surrounding air. At $t = \SI{0.15}{s},\,\SI{0.30}{s}$ and $\SI{0.45}{s}$, the interface shows that the waves are symmetric diagonally across the cube-shaped domain. Following the initial impact with the opposing wall, the water field undergoes a transformation into an asymmetrical shape, leading to intensified splashing and spraying, that is particularly  noticeable at $t = \SI{0.60}{s},\,\SI{0.75}{s}$ and $\SI{0.90}{s}$. Gravity causes the water column to drop and spread across the flat floor. This behaviour is most pronounced around $t = \SI{1.20}{s}$, showcasing the ongoing dynamic evolution of the water field. Figure~\ref{fig:Indicator_cubic_water_column} shows results of volume fraction for Model~8 on a grid of size 512$\times$512$\times$512, plotted on a diagonal plane at eight times from \SI{0.1}{s} to \SI{0.8}{s} at intervals of \SI{0.1}{s}. 

\subsection{Rising bubble in water}\label{results:rising_bubble}
\begin{figure}[htp]
\centering
\subfloat[Schematic of the computational domain on the central plane.]{\includegraphics[scale=0.28]{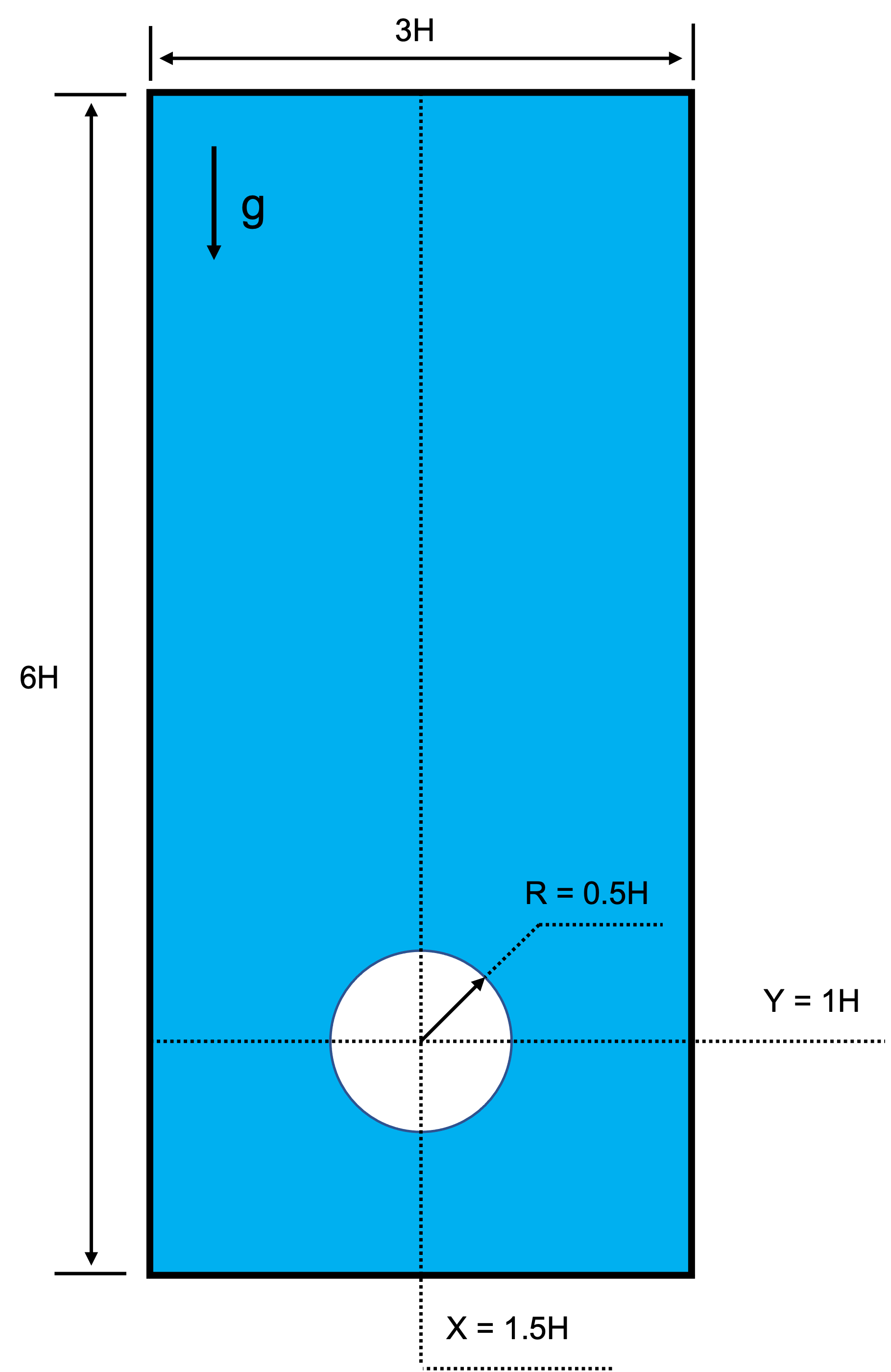}\label{fig:Domain_bubble}}
\qquad
\subfloat[Temporal evolution of bubble along a central plane.]{\includegraphics[scale=0.25]{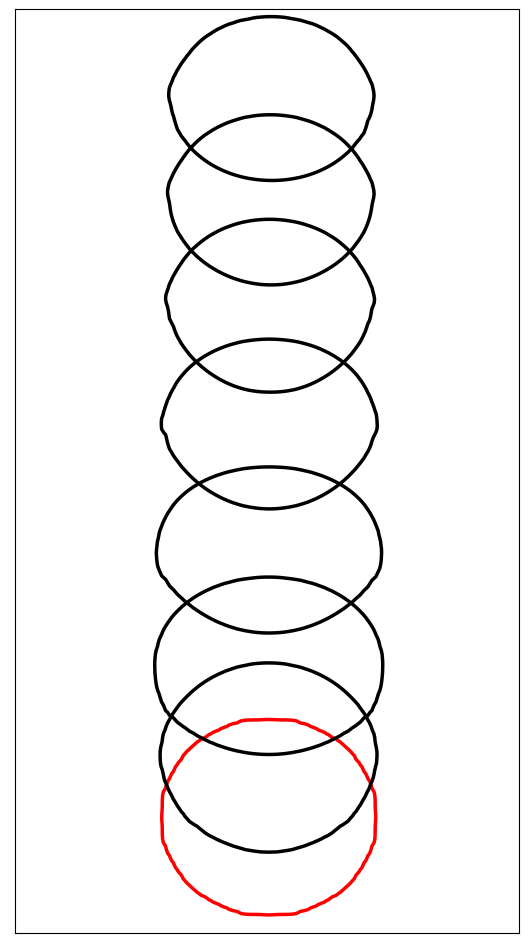}\label{fig:bubble_shape}}
\caption{\label{fig:rising_bubble} Numerical prediction of a single air bubble rising in water with equivalent diameter $D_{b} = \SI{2}{mm}$ using NN4PDEs.} 
\end{figure}

The NN4PDEs approach is used to simulate a single, rising 3D air bubble in water. The computational domain (Figure~\ref{fig:Domain_bubble}) has a cuboidal shape with dimensions of $3H\times 3H \times 6H$ (length in $x$ direction, width in $y$ direction and height in $z$ direction), where $H$ is set to \SI{2}{mm}. An air bubble is centred at $(1.5H,\,1.5H,\,H)$. The boundary conditions used in the simulations within this section are the same as those described for the collapsing water column test-cases  (Section~\ref{results:cwc:section:3d}). One of the best performing discretisations from the previous experiments is used here, namely, linear filters (\ie filters that represent a linear FEM discretisation) for the diffusion term of the volume fraction and quadratic filters (\ie filters that represent a quadratic ConvFEM discretisation) for all other terms. The surface tension model described in Equations~\eqref{surface-tension-uvw} and~\eqref{surface-tension-k} is incorporated for this test case in order to model the rising bubble. 

\begin{figure}[htbp]
\centering 
\subfloat[]{\includegraphics[scale=0.28]{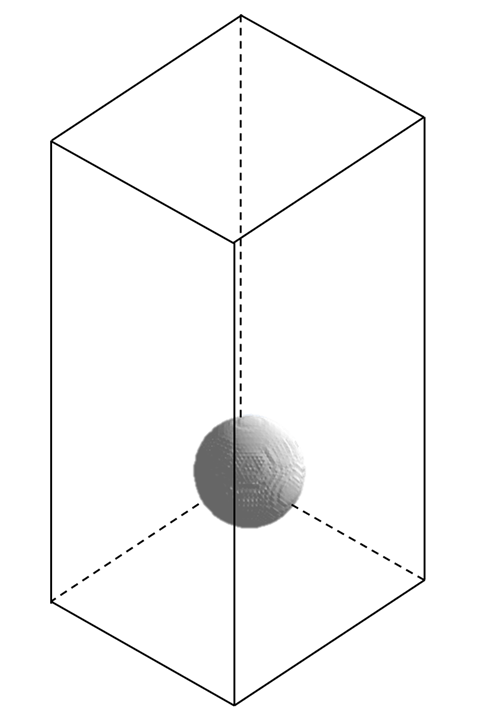}}
\subfloat[]{\includegraphics[scale=0.28]{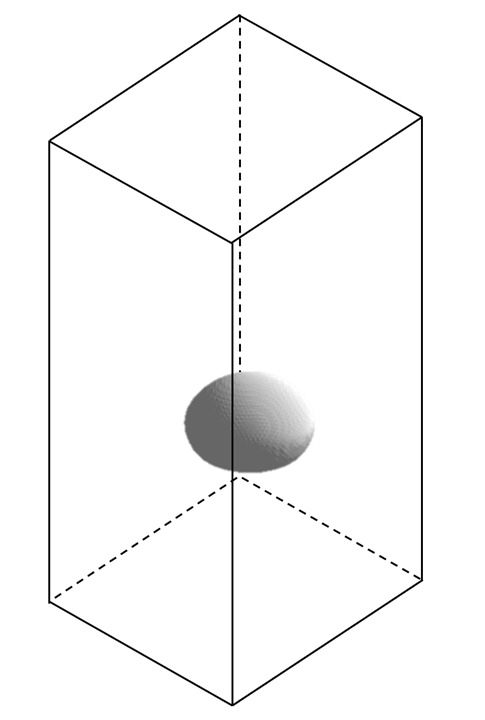}}
\subfloat[]{\includegraphics[scale=0.28]{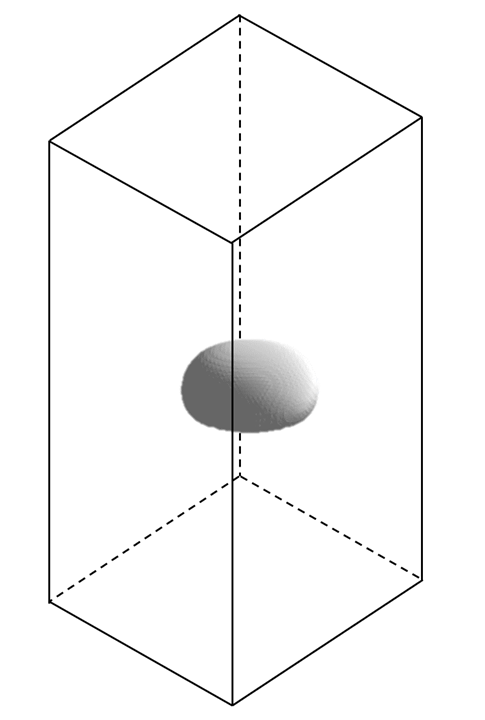}}
\subfloat[]{\includegraphics[scale=0.28]{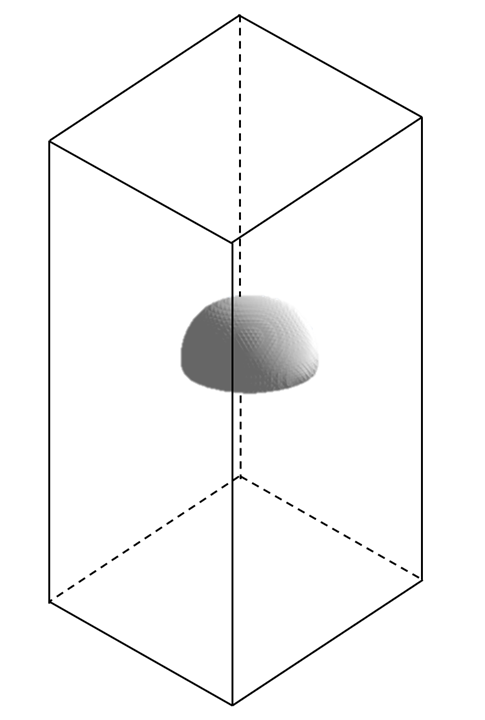}}
\caption{\label{fig:bubble_shape_3d}Numerical prediction of a single air bubble rising in water with $D_{b} = \SI{2}{mm}$ using NN4PDEs. Temporal evolution of bubble visualised using an isosurface for a value of~0.5 of the volume fraction field at four dimensionless time levels, from 0 to 3 with an interval of 1. (a), 0, (b), 1, (c), 2, (d), 3.}
\end{figure}

\begin{figure}
    \centering
    \includegraphics[scale=0.25]{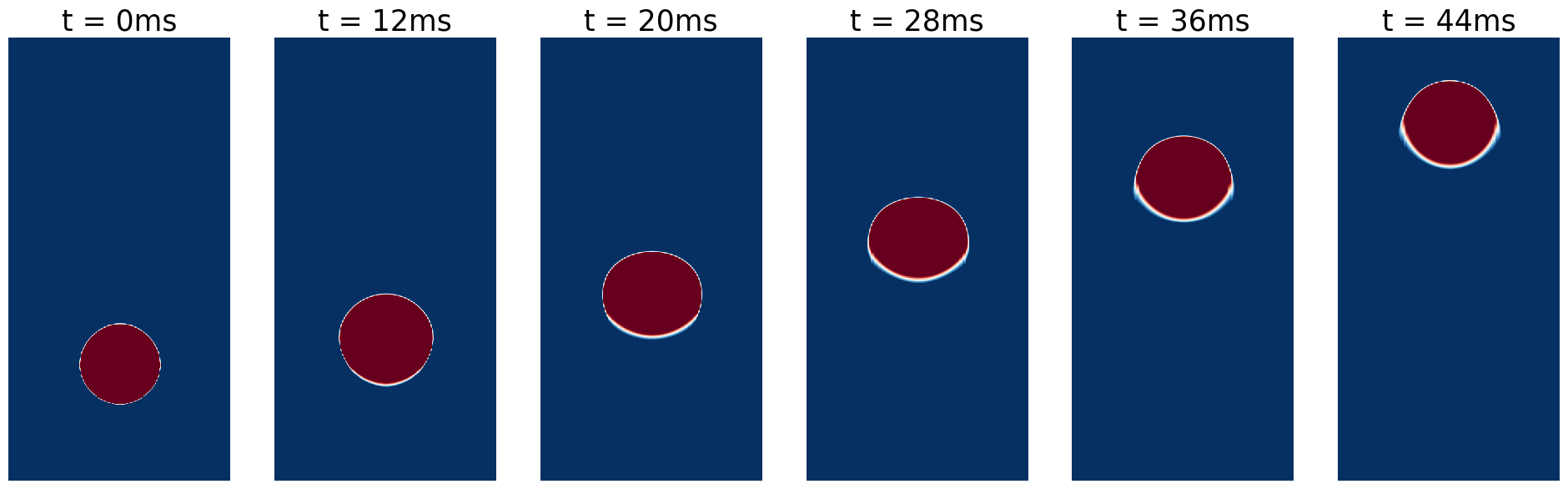}
    \quad
    \includegraphics[scale=0.3]{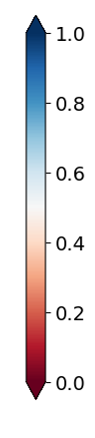}
    \caption{Snapshots of the volume fraction field for a single rising bubble in water, plotted along the central plane in middle of the domain with equivalent diameter $D_b = \SI{2}{mm}$ and at times from $t = \SI{0}{s}$ to $t = \SI{0.044}{s}$.}
    \label{fig:indicator_field_bubble}
\end{figure}

To investigate the ability of the proposed NN4PDEs approach to forecast the dynamics of a rising bubble, a uniform grid is used with 67~million (256$\times$256$\times$512) structured nodes and fixed time step ($\Delta t = \SI{0.1}{\mu s}$). A single, initially spherical, bubble with equivalent diameter of $D_{b} = H = \SI{2}{mm}$ is modelled in this problem. To obtain the results shown, two iterative corrections to velocity and non-hydrostatic pressure were made, and 20~multigrid iterations were used to solve the pressure equations. Figure~\ref{fig:bubble_shape} shows the evolution of the rising bubble through an isosurface, which represents the interface between the bubble and water. The interface is taken to be a volume fraction value of~0.5. The isosurface is plotted on the central plane, and the red circle represents the initial location and shape of the air bubble. Figure~\ref{fig:bubble_shape_3d} shows a 3D visualisation of the shape of the bubble using an isosurface at four different time levels. The shape of the rising bubble simulated with NN4PDEs matches that reported by~\citet{Crialesi-Esposito2023} for a density ratio ($\lambda_{\rho} = \rho_{l} / \rho_{g}$) of~10. Figure~\ref{fig:indicator_field_bubble} illustrates the transient volume fraction field (\SI{0}{s} $\leqslant t \leqslant $\SI{0.044}{s}) of the bubble on the central (diagonal) plane. This plot displays the shape of the bubble more precisely than Figure~\ref{fig:bubble_shape_3d}, revealing that the lower part of the interface of the bubble is more diffuse than the upper part. 

\begin{figure}[htp]
 \centering
 \subfloat[Bubble shape with an equivalent diameter of $\SI{1.92}{mm}$.]{\includegraphics[scale=0.3]{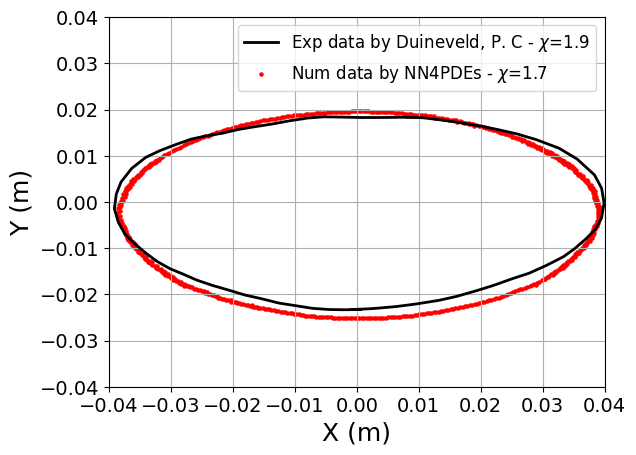}\label{fig:bubble_shape_Val}}
 \quad
 \subfloat[Terminal velocities with different bubble sizes.]{\includegraphics[scale=0.3]{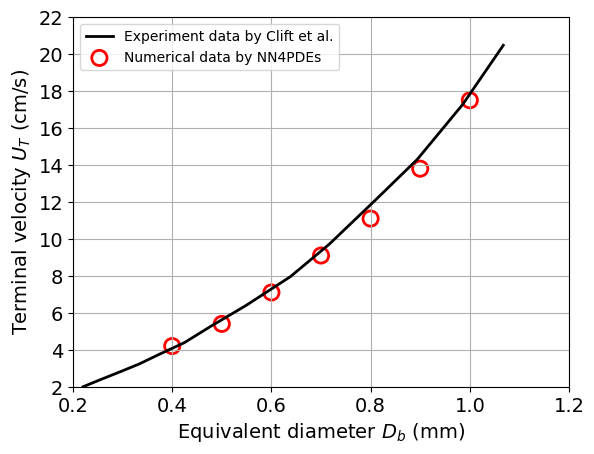}\label{fig:rising_velocity}}
 \quad 
 \subfloat[Bubble rising velocity as a function of time.]{\includegraphics[scale=0.3]{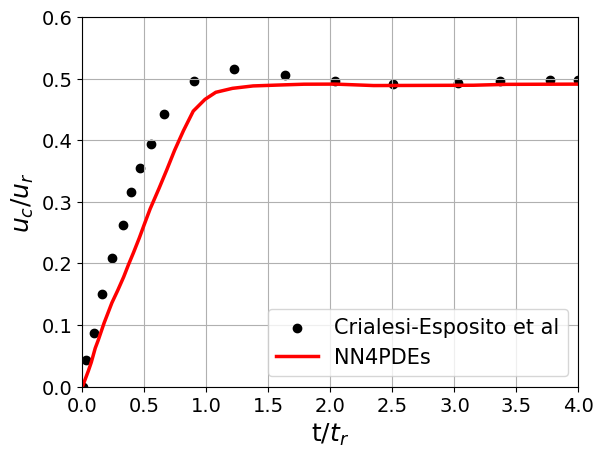}\label{fig:rising_velocity_time_val}}
  \caption{\label{fig:validation_bubble} Comparison of experimental data from~\citet{duineveld1995rise} and numerical data from~\citet{Crialesi-Esposito2023} with three-dimensional numerical results from NN4PDEs for a single bubble rising in water.}
\end{figure}

Figure~\ref{fig:validation_bubble} helps validate the proposed solver by comparing the values relating to bubble shape and terminal velocity from the NN4PDEs approach with experimental measurements and other numerical results. Figure~\ref{fig:bubble_shape_Val} compares the shape of a single rising bubble with an initial diameter of $D_{b} = \SI{1.92}{mm}$, where the red line represents the numerical results and the black line denotes the experimental data of~\citet{duineveld1995rise}. The results indicate that both the bubble shape and aspect ratio ($\chi$) predicted by NN4PDEs match, closely, the experimental data. The terminal velocities of different bubble sizes are also compared with experimental results from~\citet{clift2005bubbles} in Figure~\ref{fig:rising_velocity}. Seven distinct bubble sizes are modelled with initial diameters ranging from $\SI{0.4}{mm}$ to $\SI{1.0}{mm}$. The comparison shows that the velocities of the bubble computed by NN4PDEs always fall within the range of accuracy of the experimental results and are consistently good across the range of diameter sizes. Figure~\ref{fig:rising_velocity_time_val} compares the time evolution of bubble rising velocity with numerical results from~\citet{Crialesi-Esposito2023}. The velocity and time axes are non-dimensionalised using the reference velocity $u_{r} = \sqrt{g d_{0}}$ and reference time $t_{r} = \sqrt{d_{0}/g}$ ($d_{0}$ refers to the diameter of bubble). The results demonstrate the capability of the solver within the NN4PDEs approach to forecast accurately the water-air interface. 

\begin{figure}[htp]
    \centering
    \includegraphics[scale=0.8]{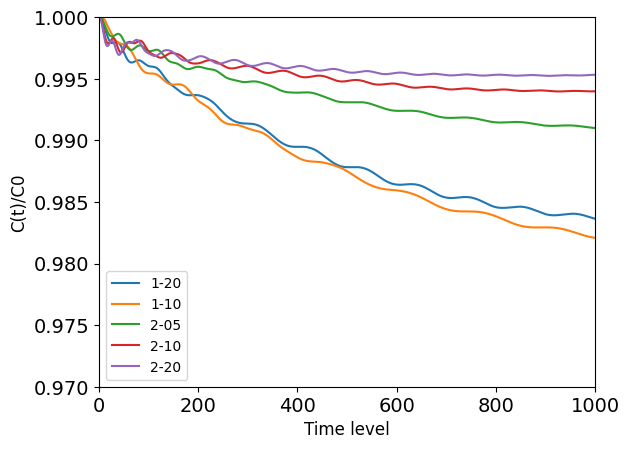}
    \caption{Mass conservation analysis over 1000 time-steps for rising bubble case comparing different iterative approaches for solving non-hydrostatic pressure and velocity corrections. $C(t)/C0$ represents normalised volume fraction integrated over the domain, relative to the initial value $C0=C(t=0)$, where $C(t=n\Delta t)=\int_V C^n dV$. In the legend $i$-$j$ is shown, in which $i$~represents the number of iterations for the pressure and velocity correction within a time step and $j$~represents the number of multigrid iterations used to solve for the non-hydrostatic pressure correction.}
    \label{fig:Multigrid_iteration}
\end{figure}

Figure~\ref{fig:Multigrid_iteration} examines the impact of the iterative approaches used for velocity and non-hydrostatic pressure correction (see Figure~\ref{fig:NN-NS2}) on the prediction of a rising bubble, by examining how well mass is conserved. The results are collected over 1000~time-steps and investigate the temporal evolution of normalised volume fraction from the initial time step by five iterative strategies, which vary the number of corrections to velocity and non-hydrostatic pressure, and the number of multigrid iterations used for solving the non-hydrostatic pressure correction equation. The iterative strategies are labelled $i$-$j$, where~$i$ represents the number of pressure-velocity corrections/iterations and~$j$ represents the number of multigrid iterations for pressure correction. The results suggest that employing more iterations to solve the pressure and velocity corrections (see Figure~\ref{fig:NN-NS2} for a schematic of three iterations) yields a significant improvement in performance as regards mass conservation, as evidenced by a comparison of by cases 1-10 and 2-10 as well as 1-20 and 2-20. However, increasing the number of multigrid iterations leads to a smaller enhancement in performance in terms of conserving mass, as evidenced by a comparison of cases 1-10 and 1-20, with 2-10 and 2-20. Thus, we adopted 2-20 as default iterative approach to obtain the results in Section~\ref{results:rising_bubble}. 
It should also be pointed out that further increasing the number of pressure and velocity correction iterations beyond two significantly reduces the stability of the method and is thus not done here. This seems to be because, in this case, one also needs to iterate over the momentum equations, otherwise the approach tends to `forget' about the momentum equations in favour of forcing down the residual of the continuity equation.  

We remark that the only reason why mass is not exactly conserved is that the volume fraction field is solved in non-conservative form (Equation~\eqref{transport-indicator-disc-in-time}). Simply switching to a conservative method --- by taking the discretised divergence of the convolution of velocity and volume fraction field --- would always conserve mass, but results in small sources and sinks of mass across the domain. Thus, the pressure/velocity correction step is used to enforce mass conservation and force the right-hand side (the discretised divergence of velocity) of the pressure correction step (Equation~\eqref{ph_disc2}) to zero. 

\section{Discussion and Conclusions}\label{conclusion}
This paper further extends a new approach that was initially developed for single-phase fluids. By using tools within Artificial Intelligence (AI) software libraries, the process of solving partial differential equations (PDEs) that have been discretised through standard or other numerical methods can be replicated. Written as a convolutional neural network, this solver (referred to as NN4PDEs) gives the same results as if the numerical methods had been implemented in Fortran or C\verb!++! (to within solver tolerances). Whilst applicable to PDEs in general, this article has focused, for the first time, on applying NN4PDEs to multiphase flows with interface capturing. Furthermore, whilst the underlying discretisation methods can be arbitrary, our demonstration focuses on the use of linear, quadratic and cubic convolutional finite elements (ConvFEM), implemented exactly through convolutional neural networks on structured grids. A segregated velocity-pressure solution method is used here to enforce the incompressibility constraint. To solve the resulting matrix equation for pressure, a sawtooth multigrid method was used and implemented using a type of convolutional neural network known as a U-Net. This solution method proved to be highly efficient needing only a few iterations, with the use of standard neural network architectures such as the U-Net greatly speeding up the model development. For the important task of modelling the interface, a new compressive advection scheme was proposed. Designed to fit within the neural network framework, it was derived using a Petrov-Galerkin approach for robustness and is able to introduce diffusion smoothly in response to the variation of the solution variables.

Two classes of test cases are used to demonstrate this approach, collapsing water columns and rising bubble problems. For the collapsing water column problems, excellent agreement is seen between the solutions from the proposed solver and both experimental data and other numerical solutions from the literature. These test cases are also used to demonstrate how linear, quadratic or cubic finite element discretisations may be used in convolutional layers to form multiphase discretisation and solution methods. It is suggested that the quadratic or cubic ConvFEM discretisations provide a good compromise between computational speed and accuracy. It is also suggested that the use of linear ConvFEM filters for the diffusion term of the volume fraction field is important, in order to reduce the tendency to produce oscillations seen when using high-order (ConvFEM) discretisations. 

While it is difficult to make a rigorous comparison between the computational speed of NN4PDEs and other codes reported in the literature, we have been able to estimate how NN4PDEs compares with selected codes. What we have found is that the speed of the fastest GPU-based codes are comparable to our implementation of NN4PDEs. This seems to suggest that the current NN4PDEs implementation is reasonably well optimised even though the approach has not taken considerable time to develop. This does highlight the advantages of using neural networks to help quickly develop rapid running models. NN4PDEs (running on a single GPU) is at least 2000 faster than codes that run on CPUs, broadly in line with the expectation that code can run faster on GPUs than CPUs.

Expressing PDE solvers as neural networks marks a significant step in forming a bridge between physics-based modelling and AI, bringing to standard numerical methods and models some of the developments made in AI. One benefit of using the NN4PDEs approach is that it enables the power of AI libraries and their built-in technologies to be exploited. For example, the codes described here are deployed on an NVIDIA GPU but could have been run, with no modification, on CPUs, TPUs (tensor processing units) or the latest energy-efficient AI accelerators thanks to the writers of the machine learning libraries. Another benefit of the NN4PDEs approach is that the flow solver can be fully integrated into machine learning workflows such as AI-based surrogate models or digital twins. For example, the automatic differentiation procedures found within machine learning libraries can be applied to physics models enabling easier implementation of optimisation processes such as data assimilation, control and inversion.

\section*{CRediT authorship contribution statement}
\textbf{BC:} software, methodology, writing (original draft, review and editing). 
\textbf{CEH:} methodology, supervision, writing (original draft, review and editing).
\textbf{JG:}  methodology, writing (original draft, review and editing).
\textbf{OKM:} writing (review and editing), funding acquisition.
\textbf{CCP:} conceptualisation, methodology, software, writing (original draft, review and editing), supervision, funding acquisition.

\section*{Data availability statement}
The code used to generate these results can be found at the following github repository:\\ \href{https://github.com/bc1chen/AI4PDE}{https://github.com/bc1chen/AI4PDE}.

\section*{Acknowledgements}
The authors would like to acknowledge the following EPSRC grants: the PREMIERE programme grant, ``AI to enhance manufacturing, energy, and healthcare'' (EP/T000414/1); ECO-AI, ``Enabling \ch{CO2} capture and storage using AI'' (EP/Y005732/1); MUFFINS, ``MUltiphase Flow-induced Fluid-flexible structure InteractioN in Subsea'' (EP/P033180/1); WavE-Suite, ``New Generation Modelling Suite for the Survivability of Wave Energy Convertors in Marine Environments'' (EP/V040235/1);  INHALE, ``Health assessment across biological length scales'' (EP/T003189/1); AI-Respire, ``AI for personalised respiratory health and pollution (EP/Y018680/1); RELIANT, ``Risk EvaLuatIon fAst iNtelligent Tool for COVID19'' (EP/V036777/1); and CO-TRACE, ``COvid-19 Transmission Risk Assessment Case Studies --- education Establishments'' (EP/W001411/1). Support from Imperial-X's Eric and Wendy Schmidt Centre for AI in Science (a Schmidt Futures program) is gratefully acknowledged.

The authors state that, for the purpose of open access, a Creative Commons Attribution (CC BY) license will be applied to any Author Accepted Manuscript version relating to this article. 

\bibliography{mybibfile}

\end{document}